\documentclass[aps,prapplied,reprint,superscriptaddress,showkeys,amsmath,amssymb]{revtex4-2}
\usepackage{newtxtext,newtxmath}
\usepackage{graphicx}
\usepackage{booktabs,longtable,array,tabularx,multirow}
\usepackage[table]{xcolor}

\usepackage[hidelinks]{hyperref}
\usepackage{xurl}

\graphicspath{{figures/}}

\makeatletter
\renewcommand\paragraph{\@startsection{paragraph}{4}{\z@}{0.35ex plus 0.15ex}{-0.8em}{\normalfont\normalsize\bfseries}}
\makeatother
\newcommand{\rhostar}{\rho^{\ast}}
\newcommand{\rhopre}{\rho_{\rm pre}}
\newcommand{\rhofam}{\rho_{\rm fam}}
\newcommand{\rhosplit}{\rho_{\rm split}}
\newcommand{\TTS}{\mathrm{TTS}_{99}}

\newcommand{\sref}[2]{\ref{#1}}
\newcommand{\snote}[2]{\hyperref[#1]{#2}}

\newcommand{\papertitle}{Beyond Ground States: Learning Many-Body Hamiltonians for \texorpdfstring{$p$}{p}-Bit Invertible Arithmetic}

\begin{document}

\title{\papertitle}
\author{Tian Feng}
\email{tian.feng@mail.mcgill.ca}
\noaffiliation
\author{Xutang Fan}
\affiliation{Advanced Cryptography and System Security Key Laboratory of Sichuan Province, Chengdu, 610225, China}
\author{Yiyun Chen}
\affiliation{The Hong Kong University of Science and Technology (Guangzhou), Guangzhou, 511453, China}
\date{November 6, 2026}

\label{page:abstract}
\begin{abstract}
Probabilistic-bit ($p$-bit) invertible logic encodes a Boolean relation as an energy whose ground states are its legal
assignments, computing in any direction by sampling. Traditional constructions target ground-state membership and a wide
minimum energy gap, yet finite-time samplers traverse excited states. We cast Hamiltonian design as inverse statistical
mechanics: fit an energy with interactions up to order four whose Gibbs law approaches uniformity on each clamp case's
legal states, under a gap floor keeping every legal state below every illegal one (so every ground state is legal).
Constrained maximum likelihood (barrier Newton, pruning, refitting) fits candidate terms from a fixed pool, and every
delivered structure is audited exactly over all cases. With candidates verified in one forward pass (as in NP), each
energy is scored by time to a verified solution under restarts using its best of nine tuned samplers. In underdetermined
modes, learned multipliers factor in 0.55, 0.44, 0.52 and 0.46 [0.38, 0.56] of the hand-written encodings' updates from
4$\times$4 to 7$\times$7, faster within every family below 7$\times$7 at no resolved work penalty; learned adders solve
the reverse mode in 0.71 of the updates at 8 bits and about a quarter at 24 and 32 bits, at 0.55 of the work at 32 bits.
At 5$\times$5 under equal pool, budget and floor, a gap-maximizing fit reaches 0.79 [0.73, 0.88] in updates and 1.26 in
work against the hand-written encoding, while likelihood fitting takes 0.55 [0.46, 0.58] and 0.58 [0.48, 0.61] of the
gap fit's updates and work. Rounding coefficients to the narrowest exact words of a tested integer ladder, we run both
energies at each pair's common word on one FPGA color-sweep SSA core (bit-exact with software): learned multipliers take
0.41 of the hand-written encoding's cycles at 4$\times$4 and 0.057 [0.045, 0.068] at 7$\times$7, at up to 2.4 times its
lookup tables. The method requires only a circuit's legal assignments; in determined modes, hand-written encodings take
less work, and from 12 bits on the adder fewer updates.
\end{abstract}


\maketitle

\section{Introduction}

Sampling has become a practical way to compute. In generative modeling, diffusion models synthesize an image from
noise through repeated denoising updates, each step steering a stochastic trajectory toward a learned distribution
\citep{sohldickstein2015deep,song2019generative,ho2020denoising,song2021score}. One lesson from that literature
motivates this study: what a fixed step budget produces depends jointly on the model, the sampler and its noise schedule
\citep{karras2022elucidating}. Probabilistic bits ($p$-bits) are binary stochastic units whose states follow sigmoids of
local fields, and hardware networks of them draw Boltzmann samples directly on physical substrates
\citep{camsari2017stochastic,pervaiz2017hardware,borders2019integer,aadit2022massively,chowdhury2023full}. In
invertible logic, a Boolean circuit such as an adder or a multiplier is encoded into an energy whose ground states
coincide with its legal input--output assignments
\citep{camsari2017stochastic,smithson2019efficient,onizawa2021design,nishino2018study}; clamping any subset of nodes and
sampling the rest runs the circuit in whichever direction is left free. Viewed as conditional generation from an
energy-based model, the Hamiltonian supplies the model, the clamped spins fix the condition, the sampler generates the
free variables, and the schedule of stochastic simulated annealing (SSA) serves as the noise schedule
\citep{onizawa2023fast}. Stochastic computing shrank CMOS and FPGA invertible logic \citep{smithson2019efficient} and
three-body Hamiltonians raised its convergence rates \citep{onizawa2021high}, while SSA extends stochastic computing to
general Ising problems \citep{onizawa2023fast,shin2023memory}.

Most existing invertible-logic energies are designed around zero-temperature ground states. Following conventions from
Ising reductions and quantum annealing, hand-written constructions enforce legal ground states and widen the minimum gap
to illegal states
\citep{lucas2014ising,whitfield2012ground,biamonte2008nonperturbative,bian2014discrete,bian2016mapping,onizawa2021design}.
Along this line, \citet{onizawa2021high} introduced three-body spin gates whose linear-programming weights maximize that
gap while collapsing gate excitations to a few levels, cutting annealing shots roughly 4- and 12-fold over pairwise
encodings on 6- and 8-bit adders for about 10\% more lookup tables per gate and pointing to integer factorization and
machine learning next. Because a finite-time trajectory spends its run among excited states, we target the
finite-temperature landscape directly. At a design inverse temperature the Gibbs law both concentrates probability on
the legal set and spreads it across legal completions, two properties that identical ground states and identical minimum
gaps still leave free to vary. We fit couplings up to interaction order four to that finite-temperature target, keep the
minimum gap as a hard feasibility floor and the gap-maximizing construction as a direct comparator, and benchmark
multipliers from 4$\times$4 to 7$\times$7 and adders from 8 to 32 bits under tuned samplers.

\begin{figure*}[!tbp]
\centering
\includegraphics{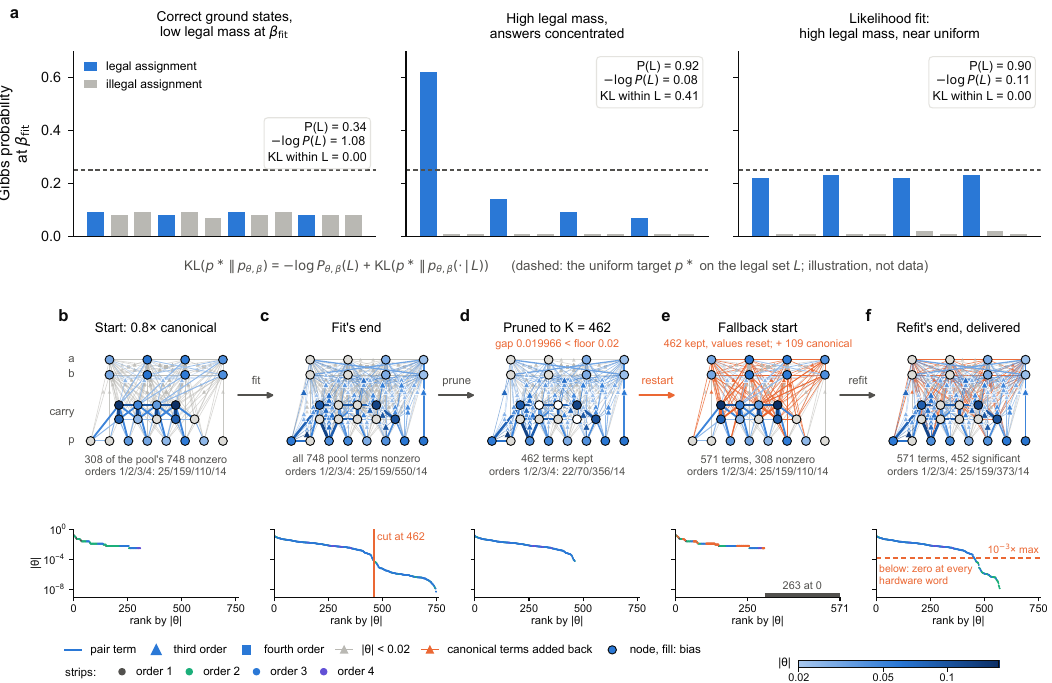}
\caption{\textbf{The design distribution and a learned structure's evolution.} (a) Illustration (not data) of
Eq.~\ref{eq:kl}: correct ground states can hold little legal mass, and a high legal mass can sit on a few answers.
(b--f) The 4$\times$4 structure MUL4-P3: (b) the start, 0.8 times the canonical encoding in P3; (c) the fit's end;
(d) the prune to $K=462$, below the floor (minimum case gap 0.019966 against 0.02); (e) the restart from the canonical
encoding on 571 terms, 263 of them at zero; (f) the delivered structure. Lines, triangles, squares: terms of interaction
order two to four, colored by $|\theta|$ (gray below 0.02; orange, the 109 terms added back); node fill, the bias;
strips, $|\theta|$ by rank.}
\label{fig:concept}
\end{figure*}

Formulated as an inverse problem of statistical mechanics, the goal is to find interactions whose Gibbs law approaches a
target distribution \citep{ackley1985learning,wainwright2008graphical,ravikumar2010high,lokhov2018optimal}. For a
uniform target $p^\ast$ on the legal set $L$ of each clamp case, the Kullback--Leibler divergence to the Gibbs law splits
into two terms (Fig.~\ref{fig:concept}a),
\begin{equation}
\begin{split}
\mathrm{KL}\left(p^\ast \,\|\, p_{\theta,\beta}\right) &= -\log P_{\theta,\beta}(L) \\
&\quad + \mathrm{KL}\left(p^\ast \,\|\, p_{\theta,\beta}(\cdot \mid L)\right),
\end{split}
\label{eq:kl}
\end{equation}
rewarding total probability on valid completions while penalizing unevenness across $L$. On any fixed interaction pool
this divergence is the convex negative log-likelihood of an exponential family \citep{wainwright2008graphical}. Because
ground-state legality alone fixes neither term, we enforce hard linear inequalities placing every legal state at least a
floor $g$ below every illegal state of its case and certify them by exact audit. Even at matched support, node budget
and maximal gap, suppressing higher illegal states raises Gibbs legal mass and can shorten hitting times, as an exact
two-spin family shows (Appendix~\ref{sec:proof-toy}). A single static vector $\theta$ serves all operating modes on the
same hardware substrate as a hand-written encoding.

Boltzmann-machine learning and contrastive approximations
\citep{ackley1985learning,hinton2002training,tieleman2008training} have been demonstrated on stochastic magnetic tunnel
junctions and CMOS $p$-bits \citep{kaiser2022hardware,niazi2024training,jhonsa2026cmos}, where short-run inner updates
can yield effective short-run samplers without a valid steady state \citep{nijkamp2020anatomy}. Other recent approaches
adapt the sampling dynamics rather than the static Hamiltonian \citep{delacour2026probabilistic}: the generalized
probabilistic approximate optimization algorithm \citep{abdelrahman2025generalized} learns transferable
inverse-temperature schedules and layer-wise transition couplings, and on its logic-gate examples legal assignments are
not preserved as ground states of a single energy, whereas IsingFormer \citep{bunaiyan2025isingformer} trains nonlocal
parallel-tempering proposals on fixed pairwise multipliers, an orthogonal advance that could be paired with learned
energies. In-place hardware Boltzmann learning and exact Markov-flow checks in these studies stay on single gates and
full adders (Supplementary Note~\snote{sn:discussion}{S7}).

A modular route composes small optimized blocks: \citet{patel2022logically} optimize small restricted Boltzmann
machines by contrastive divergence, merge them along shared visible nodes, fine-tune the merged machine and its weights
for low-precision fixed point, and factor 16-bit products on an FPGA. Our approach differs in three ways. First,
correctness: their accuracy is the fraction of 300 random instances whose most frequently sampled state is correct after
a given number of samples, with no audit over all cases reported, whereas here every legal state lies at least $g$ below
every illegal state on every clamp case by exact audit and one forward pass verifies each restart. Second, size:
representing higher-order correlations via hidden units uses 80, 164 and 412 nodes at 8-, 10- and 12-bit products (the
last merged) and 544 or 654 at 16 bits, versus 25, 35, 45 and 55 circuit nodes from 8- to 14-bit products here using
explicit interactions up to order four under an incident-weight budget. Third, fitting: fixed-pool likelihood is convex,
whereas training a model with hidden units is not; they report that direct training did not reach a good joint-density
model of their 80-node 8-bit multiplier even after long training, and that their 32-bit adder fell behind most of their
other units after a week of training, which they attribute to sampling a data set too large to enumerate (hitting-time
and platform-speed notes in Supplementary Note~\snote{sn:discussion}{S7}).

Earlier designs also shape the excited spectrum beyond ground states and the minimum gap \citep{onizawa2021high}.
\citet{he2023many} merge four-body linear-programming gates that hold wrong gate states at a single level, raising
correct-factor shares on individual products of wider factorizers, while \citet{he2024direct} expand a truth table
directly into a two-level energy without auxiliary nodes at interaction orders growing with width; both check Boltzmann
laws on small circuits without reporting time to solution or tuned samplers (Supplementary
Note~\snote{sn:discussion}{S7}). Counting the unique multilinear expansion of \citet{he2024direct} via a descriptive
Walsh transform gives 56{,}106, 994{,}300, 16{,}298{,}019 and 264{,}469{,}813 nonzero terms (86 to 99\% of all
monomials) on the 16, 20, 24 and 28 nodes of 4$\times$4 to 7$\times$7 multipliers, versus 571, 1{,}072, 1{,}692 and
2{,}318 terms up to order four on 25, 35, 45 and 55 nodes here. Although its node count is lower and its scaled gap is larger (values in Supplementary
Note~\snote{sn:discussion}{S7}), a $p$-bit network computing local fields from incident terms averages roughly 28{,}000,
500{,}000, 8.2 million and 130 million terms per node against 61, 83, 102 and 115 here, so we did not run it on the
panels' term-by-term kernels. A kernel testing legality directly would avoid that cost by departing from term-by-term
local-field evaluation; we ran none, so these counts say nothing of its speed in updates.

Holding an entire spectrum at two levels thus outgrows bounded fan-in, motivating constrained spectral fitting under an
explicit term budget. On this study's pool, term budget and floor at 5$\times$5, maximizing the minimum gap already
reduces updates relative to the hand-written encoding at higher work, and replacing gap maximization by the likelihood
objective cuts updates roughly in half again (Section~\ref{sec:res-width} and Supplementary
Note~\snote{sn:maxgap}{S9}). Prior gate-level invertible circuits have run in hardware emulation, FPGA or software
simulation, and CMOS or magnetic tunnel-junction silicon across 4- to 32-bit adders and 2$\times$2 to 10$\times$10
multipliers
\citep{pervaiz2017hardware,nishino2018study,camsari2017stochastic,bunaiyan2025isingformer,smithson2019efficient,duffee2025integrated},
alongside non-gate cost functions written directly on factor or clause variables
\citep{borders2019integer,xu2025probabilistic,he2026solving} (detailed node counts in Supplementary
Note~\snote{sn:discussion}{S7}). The Hamiltonians fitted here reach a 55-node 7$\times$7 multiplier (41 free spins when
factoring) and a 129-node 32-bit adder, combining one static energy fitted by likelihood across clamp cases on the
circuit's own nodes, a hard gap floor certified over all clamp cases, and equal sampler tuning, while retaining
multi-spin couplings up to four bodies for their convergence benefits
\citep{onizawa2021high,bybee2023efficient,nikhar2024all} and tracking every term's evaluation and routing cost
explicitly.

Problems in NP such as factoring and subset sum \citep{camsari2019pbits,pervaiz2019weighted,borders2019integer} motivate
invertible logic because any proposed spin configuration can be checked by one deterministic forward pass of the circuit
at negligible cost relative to sampling, so practical solvers restart independent trajectories until verification
succeeds \citep{patel2022logically,duffee2025integrated,xu2025probabilistic}. Following standard Ising-benchmarking
convention
\citep{ronnow2014defining,hamerly2019experimental,isakov2015optimised,aramon2019physics,kowalsky2022regular,mohseni2022ising},
we score each Hamiltonian by $\TTS$, the update budget for 99\% verified success at its optimal restart length, across
nine sampler families with identical hyperparameter grids and all replica updates counted. Whereas invertible-logic
designs have typically been tested at a single schedule \citep{onizawa2021high,smithson2019efficient}, inexpensive
parameter sweeps accelerate hand-written Hamiltonians markedly, making equal tuning across families essential to isolate
the effect of the learned energy.

Three contributions anchor this study: (i) a constrained Gibbs-likelihood formulation for many-body invertible-logic
Hamiltonians, convex on a fixed pool, certified by exact ground-state audits, and scaled from 4$\times$4 to 7$\times$7
multipliers via a sampled-likelihood engine; (ii) an equal-tuning evaluation across a multiplier width series and an
adder ladder showing that learned Hamiltonians cut update counts in underdetermined modes across every sampler family,
except for a few unresolved cells at 7$\times$7 and on the adders up to 12 bits, whereas hand-written encodings take less work in determined modes; and (iii) an
ablation separating coefficient values from hypergraph support at 4$\times$4 alongside end-to-end hardware costs from
graph coloring and integer quantization to FPGA synthesis and board execution.

\section{Model and method}
\label{sec:method}

\paragraph{Circuits, modes and energies.} An $n$-bit ripple adder ADD$n$ sums two $n$-bit inputs across four clamp
modes: \emph{ab} (both inputs clamped, the forward mode), \emph{asum} and \emph{bsum} (one input and the sum clamped),
and \emph{reverse} (only the sum clamped). A $W{\times}W$ multiplier MUL$W$ operates in \emph{factor} mode (product
clamped) and \emph{multiply} mode (both factors clamped). Reverse and factor modes are \emph{underdetermined}, admitting
multiple legal completions per clamp case; the remaining modes are \emph{determined}, admitting exactly one. Over spins
$s_i\in\{-1,+1\}$, each Hamiltonian takes the polynomial form
\begin{equation}
\begin{aligned}
&E_\theta(s) = -\sum_{S\in\mathcal P}\theta_S\,\phi_S(s), \\
&\phi_S(s) = \prod_{i\in S}s_i,\qquad |S|\le 4,
\end{aligned}
\label{eq:energy}
\end{equation}
on a support or candidate pool $\mathcal P$. Each clamp case $c$ fixes the clamped spins, restricting configurations to
the compatible set $X_c$ with legal subset $L_c\subseteq X_c$. At inverse temperature $\beta$, case $c$ induces the
Gibbs law
\begin{equation}
\begin{aligned}
&p_{\theta,\beta}(s\mid c) = \frac{e^{-\beta E_\theta(s)}}{Z_c(\theta)}, \\
&Z_c(\theta) = \sum_{s\in X_c}e^{-\beta E_\theta(s)},\qquad s\in X_c.
\end{aligned}
\label{eq:gibbs}
\end{equation}
Spin $i$ experiences the local field $h_i(s)=\sum_{S\ni i}\theta_S\prod_{j\in S\setminus\{i\}}s_j$. Learned and
hand-written Hamiltonians alike satisfy the per-node incident-weight budget
\begin{equation}
\ell_i(\theta)=\sum_{S\ni i}|\theta_S|\le 1\qquad\text{for every node } i,
\label{eq:budget}
\end{equation}
bounding $|h_i(s)|\le 1$ and placing noise and temperature schedules on a common scale. Hand-written baselines, termed
\emph{canonical} encodings and normalized to $\max_i\ell_i=1$, sum squared half- and full-adder residuals plus a copy
penalty on the adder, and sum squared column equations on the multiplier
\citep{lucas2014ising,jiang2018quantum,luo2026general} using direct three- and four-body spin products
\citep{borders2019integer} (Table~\ref{tab:structures}).

\paragraph{Objective.} With $p^\ast_c$ uniform on $L_c$ for each mode $m$ and case $c\in\mathcal C_m$, we minimize
\begin{equation}
\begin{aligned}
J(\theta) &= \sum_{m}\frac{\omega_m}{|\mathcal C_m|}\sum_{c\in\mathcal C_m}
  \mathrm{KL}\bigl(p^\ast_c\,\big\|\,p_{\theta,\beta_{\rm fit}}(\cdot\mid c)\bigr) \\
&= \sum_{m}\frac{\omega_m}{|\mathcal C_m|}\sum_{c\in\mathcal C_m}
  \Bigl[\log Z_c(\theta) \\
&\qquad\quad +\beta_{\rm fit}\,\mathbb E_{p^\ast_c}[E_\theta]-H(p^\ast_c)\Bigr],
\end{aligned}
\label{eq:objective}
\end{equation}
where $\omega_m\ge 0$ weights mode $m$ and $H$ denotes Shannon entropy. The multiplier fits use the factor and multiply
modes with $\omega_m=1$ and every case of each (90 and 256 cases at 4$\times$4, 340 and 1{,}024 at 5$\times$5, 1{,}238
and 4{,}096 at 6$\times$6, 4{,}647 and 16{,}384 at 7$\times$7); the adder fits use the four modes with $\omega_m=1/4$,
at $\beta_{\rm fit}=128$ with the legal energies of each sum held equal, on every case of each mode at 6 and 8 bits and,
above 8 bits, on a uniform seeded sample that estimates the mean over a mode's cases (Appendix~\ref{app:engines}).
Differentiating
Eq.~\ref{eq:objective} yields the feature-moment residual and positive-semidefinite covariance matrix,
\begin{equation}
\begin{aligned}
&\nabla J = \beta_{\rm fit}\sum_{m}\frac{\omega_m}{|\mathcal C_m|}\sum_{c\in\mathcal C_m}
  \bigl(\mathbb E_{p_{\theta}(\cdot\mid c)}[\phi]-\mathbb E_{p^\ast_c}[\phi]\bigr), \\
&\nabla^2 J = \beta_{\rm fit}^2\sum_{m}\frac{\omega_m}{|\mathcal C_m|}\sum_{c\in\mathcal C_m}
  \mathrm{Cov}_{p_{\theta}(\cdot\mid c)}[\phi]\succeq 0,
\end{aligned}
\label{eq:grad}
\end{equation}
establishing convexity of $J$: an unconstrained optimum equates model and target moments in weighted average, whereas
under active gap and budget constraints normal-cone multipliers balance the remaining moment residual (Appendix~\ref{sec:proof-convexity}). Rewriting a single case's divergence $D_c$ clarifies how likelihood goes beyond
extremal gap design:
\begin{equation}
\begin{aligned}
&D_c = \log\left[\frac1{|L_c|}\sum_{s\in X_c}e^{-\beta_{\rm fit}(E_\theta(s)-\bar E_{L_c})}\right], \\
&P_{\theta,\beta_{\rm fit}}(L_c) \ge e^{-D_c},
\end{aligned}
\label{eq:mechanism-spectrum}
\end{equation}
where $\bar E_{L_c}$ is the mean legal energy. Whereas the minimum gap inspects only one worst legal--illegal pair,
$D_c$ sums the Boltzmann weight of every competing state (Appendix~\ref{sec:proof-likelihood}). Once
$\beta_{\rm fit}$ sets the fitted Hamiltonian, each sampler tunes its own operating noise or temperature schedule
independently.

\begin{table*}[!tbp]
\centering
\footnotesize
\caption{\textbf{Structures.} Terms: nonzero support, significant support (above $10^{-3}$ of the largest coefficient) in
parentheses. Work: incident terms per update over the mode's free nodes, in the same two counts. Fallback: the canonical
terms added where the pruned start violated the floor. Minimum case gap: the exact audit on every case of both modes;
the canonical encodings are exact by construction.}
\label{tab:structures}
\setlength{\tabcolsep}{4pt}
\begin{tabular}{@{}ll>{\raggedright\arraybackslash}p{3.3cm}rrr@{}}
\toprule
Circuit & Structure & Pool / rule & Terms & Work per update & Min.\ case gap \\
\midrule
\multicolumn{6}{@{}l}{\textit{Multiplier series, factor mode (series engine: fit, prune to $K$, refit)}} \\
\multirow{3}{*}{MUL4} & canonical & hand-written & 308 & 38.1 & exact \\
 & \mbox{MUL4-P3} & P3; $K=462$, fallback +109 & 571 (452) & 76.4 (62.5) & $\ge$0.02 \\
 & \mbox{MUL4-P4} & P4; $K=462$ & 462 (457) & 77.5 (76.7) & $\ge$0.02 \\
\multirow{3}{*}{MUL5} & canonical & hand-written & 565 & 50.0 & exact \\
 & \mbox{MUL5-P3} & P3; $K=848$, fallback +224 & 1{,}072 (847) & 101.0 (83.5) & $\ge$0.005 \\
 & \mbox{MUL5-P4} & P4; $K=848$, fallback +396 & 1{,}244 (922) & 139.9 (111.1) & $\ge$0.005 \\
\multirow{2}{*}{MUL6} & canonical & hand-written & 906 & 62.9 & exact \\
 & \mbox{MUL6-P3-seed1} & P3; $K=1{,}359$, fallback +333 & 1{,}692 (1{,}350) & 125.2 (104.7) & $\ge$0.0048485 \\
\multirow{2}{*}{MUL7} & canonical & hand-written & 1{,}312 & 75.4 & exact \\
 & \mbox{MUL7-P3-seed2} & P3; $K=1{,}968$, fallback +350 & 2{,}318 (1{,}764) & 140.3 (111.2) & $\ge$0.0047059 \\
\midrule
\multicolumn{6}{@{}l}{\textit{Adder ladder, reverse mode (column-chain engine, equal legal energies; prune at $10^{-3}$, refit)}} \\
\multirow{2}{*}{ADD8} & canonical & hand-written & 81 & 5.2 & exact \\
 & \mbox{ADD8-learned} & column-local pool & 169 & 14.4 (13.9) & 0.503 \\
\multirow{2}{*}{ADD16} & canonical & hand-written & 161 & 5.3 & exact \\
 & \mbox{ADD16-learned-seed1} & column-local pool & 287 & 12.4 (12.2) & 0.504 \\
\multirow{2}{*}{ADD24} & canonical & hand-written & 241 & 5.3 & exact \\
 & \mbox{ADD24-learned-seed2} & column-local pool & 462 & 13.1 (12.8) & 0.511 \\
\multirow{2}{*}{ADD32} & canonical & hand-written & 321 & 5.3 & exact \\
 & \mbox{ADD32-learned-seed2} & column-local pool & 557 & 12.1 (11.8) & 0.512 \\
\bottomrule
\end{tabular}
\end{table*}

\paragraph{The gap floor.} Because unconstrained likelihood fitting introduces spurious ground states (breaking 21 of
90 factor products on the 4$\times$4 column-local pool P4 at $\beta_{\rm fit}=8$), we enforce per-case gap inequalities,
\begin{equation}
\begin{aligned}
&E_\theta(y)-E_\theta(x) = \theta^{\top}\bigl(\phi(x)-\phi(y)\bigr)\ge g \\
&\qquad \forall\,c,\;x\in L_c,\;y\in X_c\setminus L_c,
\end{aligned}
\label{eq:percase}
\end{equation}
all linear in $\theta$; on the adders all legal states of a sum are instead constrained to one common energy at least
$g$ below every illegal state. Exactness is certified via the minimum case gap
\begin{equation}
\gamma(\theta)=\min_{c}\Bigl[\min_{y\in X_c\setminus L_c}E_\theta(y)-\max_{x\in L_c}E_\theta(x)\Bigr],
\label{eq:gap}
\end{equation}
where $\gamma(\theta)\ge g>0$ guarantees legal-only ground states in every case (Appendix~\ref{sec:proof-gap}). Under hard-threshold SSA the same floor also provides a conditional retention bound:
whenever noise increments stay below $g/2$, no single-spin update can exit $L_c$ (Appendix~\ref{sec:proof-ssa}). Introducing slack variables $u_S\ge|\theta_S|$ renders Eq.~\ref{eq:budget} linear,
preserving convexity over any fixed pool.

\paragraph{Solving, pruning and refitting.} Violated gap inequalities are separated dynamically by cutting planes inside
a logarithmic barrier method \citep{boyd2004convex}. Given an active cut set $K$ of pairs $(x,y)$, stage $t$ minimizes
\begin{equation}
\begin{split}
F_t(\theta,u) &= t\,J(\theta)-\sum_{i=1}^{n}\log\Bigl(1-\sum_{S\ni i}u_S\Bigr) \\
&\quad -\sum_{S\in\mathcal P}\log\bigl(u_S^2-\theta_S^2\bigr) \\
&\quad -\sum_{(x,y)\in K}\log\bigl(\theta^{\top}(\phi(x)-\phi(y))-g\bigr)
\end{split}
\label{eq:barrier}
\end{equation}
via damped Newton iterations; exact separation identifies any illegal state below the floor, appends the worst violators
to $K$, and advances $t$ tenfold once feasible. On 25-node circuits, moments and gap audits are computed over all
$2^{25}$ configurations; on column-local pools of arbitrary width, a column-chain dynamic program evaluates them
exactly on each case by sweeping column by column with carry states as boundary variables. Each converged fit is pruned by
coefficient magnitude, refitted on the surviving support warm-started with its accumulated cuts, and re-audited
exactly. On one problem, the 25-node multiplier on pool P4 at $\beta_{\rm fit}=16$ with gap 0.015, starting from the
canonical encoding, from zero or from random weights within the budget reaches one solution numerically, objectives within
$2.4\times10^{-12}$ and coefficients within $5\times10^{-10}$, with identical sampling panels. In the series the hand-written
encoding sets the floor and supplies both the start and the fallback's terms.

\paragraph{The width series.} Above 5$\times$5, exact partition sums over all clamp cases become intractable, and one
sampled-likelihood engine fits the multiplier series: each partition function combines exact enumeration of $L_c$ with
illegal states collected from color-sweep tempering chains and violator searches, weighted against the empirical legal
fraction along the chains while keeping cutting planes, barrier stages and exact ground-state audits unchanged. One
recipe at $\beta_{\rm fit}=40$ governs MUL4-P3, MUL5-P3, MUL6-P3-seed1 and MUL7-P3-seed2 on pool P3 (column-local terms
up to order three plus the canonical terms): fit from 0.8 times the canonical encoding under a gap floor equal to 0.64
of the normalized canonical minimum gap, prune to $K=\lceil 1.5\,n_{\rm can}\rceil$ terms with missing canonical terms
restored whenever pruning drops below the floor (tracked for MUL4-P3 in Fig.~\ref{fig:concept}b--f), refit carrying the
cuts, and normalize to $\max_i\ell_i=1$. Two seeds run per width and every delivered structure is checked against five
acceptance gates, including full frozen CPU exact audits on MUL4-P3 through MUL7-P3-seed2 (21{,}031 cases at
7$\times$7, with companion seed MUL7-P3-seed1 audited on GPU across all cases and on CPU across an 80-case subset;
candidate counts, floors, seed rules and gate exceptions in Appendix~\ref{app:engines}, Supplementary
Note~\snote{sn:engines}{S1} and Table~\sref{tab:si-audit}{S2}).

At 4$\times$4 and 5$\times$5 the same pipeline also runs on pool P4 (MUL4-P4 and MUL5-P4), which adds fourth-order
column-local terms to P3. Along the adder ladder (ADD8-learned to ADD32-learned-seed2), the column-chain engine
evaluates each case exactly without within-case state sampling on column-local pools of one to four nodes per column
(244, 476, 708 and 940 terms at 8, 16, 24 and 32 bits) with equal legal energies per sum, pruning at $10^{-3}$ of the
largest coefficient and refitting on the kept support (Appendix~\ref{app:engines}).

\section{Readout protocol}
\label{sec:baselines}

\paragraph{Samplers.} Hardware deployment targets hard-threshold SSA \citep{onizawa2023fast,onizawa2023local}, evaluated
in software under random scan: each update picks a free spin $i$, integrates its local field with scaled binary noise
$r_t=\pm 1$, clips the internal state at an annealed bound, and outputs its sign,
\begin{equation}
\begin{aligned}
&I_i \leftarrow \operatorname{clip}\bigl(I_i+h_i(s)+\eta a_i(t)r_t,\,-I_{0,i}(t),I_{0,i}(t)\bigr), \\
&s_i \leftarrow \operatorname{sgn}(I_i),
\end{aligned}
\label{eq:ssa}
\end{equation}
where $\eta=1$ defines the standard setting under schedules N0 and N1. Single-spin heat-bath updates instead set
$s_i=+1$ with probability $(1+e^{-2\beta h_i})^{-1}$, sampling Eq.~\ref{eq:gibbs} at fixed $\beta$
\citep{geman1984stochastic}. Across every width of both ladders, nine sampler families run on both Hamiltonians
(Supplementary Table~\sref{tab:si-families}{S1}): SSA, constant-$\beta$ (HB) and annealed (HBA) heat bath, Metropolis
annealing (SA-M) \citep{kirkpatrick1983optimization,isakov2015optimised}, parallel-trial annealing (PT-SA)
\citep{aramon2019physics}, parallel tempering (PT) \citep{hukushima1996exchange,earl2005parallel} with two- and
four-copy isoenergetic cluster variants (PT+ICM, PT+ICM-4)
\citep{houdayer2001cluster,zhu2015efficient,chowdhury2025pushing}, and adaptive color-sweep tempering (APT-C)
\citep{nikhar2024all}. Reference checks for each kernel and the uniform three-family supplement from 6$\times$6
upward and on the adder ladder appear in Appendix~\ref{app:protocol}.

\paragraph{Metrics.} A trajectory succeeds when its terminal state satisfies the clamped circuit relation on every
node, evaluated on the coldest replica in tempering or on any copy in cluster variants. Because candidate solutions are verified in
a single forward pass, finding any valid completion counts as success, whereas uniform coverage across multiple valid
completions is not reached at finite run length (Supplementary Note~\snote{sn:discussion}{S7}). Given empirical success
fraction $v_c(T)$ on clamp case $c$ at run length $T$,
\begin{equation}
\begin{aligned}
&\TTS = \min_{T}\;\operatorname*{median}_{c}\;\tau_c(T), \\
&\tau_c(T) = \begin{cases}T\,\dfrac{\ln 0.01}{\ln\bigl(1-v_c(T)\bigr)}, & v_c(T)<0.99,\\[4pt] T, & v_c(T)\ge 0.99,\end{cases}
\end{aligned}
\label{eq:tts}
\end{equation}
measures the single-node updates across all replicas needed to reach 99\% verified success under independent restarts.
To account for interaction density, \emph{work} weights updates by the mean incident degree over the mode's free spins
$F$,
\begin{equation}
\begin{aligned}
&W = \TTS\cdot\bar w, \\
&\bar w = \frac{1}{|F|}\sum_{i\in F}\bigl|\{S\in\operatorname{supp}\theta:\ i\in S\}\bigr|,
\end{aligned}
\label{eq:work}
\end{equation}
evaluated on the significant support above $10^{-3}\max_S|\theta_S|$, with full nonzero counts reported in
Table~\ref{tab:structures} and Supplementary Table~\sref{tab:si-series}{S6} and hardware lookup-table cost measured directly in Section~\ref{sec:hw}. Relative to the
canonical encoding, the within-family ratio
\begin{equation}
\rhofam^{(f)}=\frac{\TTS^{(f)}(\theta_{\rm learned})}{\TTS^{(f)}(\theta_{\rm canonical})}
\label{eq:rhofam}
\end{equation}
compares both Hamiltonians at their own tuning-selected settings, whereas the best-solver ratio
\begin{equation}
\rhostar=\frac{\min_{f}\TTS^{(f)}(\theta_{\rm learned})}{\min_{f}\TTS^{(f)}(\theta_{\rm canonical})}
\label{eq:rho}
\end{equation}
re-selects each side's winning family and restart length on evaluation medians inside every bootstrap replicate, paired
with a tuning-frozen check $\rhopre$ (Appendix~\ref{app:protocol}).

\paragraph{Selection and statistics.} Hyperparameters and run lengths are chosen on a dedicated tuning seed
namespace, with pre-specified grid extensions whenever an optimum hits a boundary, and are then scored once on a
disjoint evaluation namespace. At every width, canonical evaluation outputs are frozen before any learned structure is
evaluated, using matched clamp panels (every case at 4$\times$4, 5$\times$5 and 6 bits; 64 products at 6$\times$6 and
7$\times$7; 129 to 153 sums and 256 input pairs from 8 to 32 bits; stratification rules in Appendix~\ref{app:protocol}),
tempering pilots and pseudorandom streams. Nested 95\% bootstrap intervals \citep{efron1979bootstrap} over clamp cases and
trajectories classify each ratio as \emph{surviving} (S, strictly below 1), \emph{reversed} (R, strictly above 1) or
\emph{unresolved} (U). Giving hand-written encodings this equal tuning budget accelerates them 2.8- to 6.5-fold over
standard SSA, compared with 1.4- to 1.5-fold on learned Hamiltonians, and often selects adaptive color-sweep
tempering \citep{nikhar2024all}. Every replica sweep is charged in full, whereas exchange tests and cluster flips carry
zero update charge (Appendix~\ref{app:protocol}).

\section{Results}
\label{sec:results}

Throughout this section, primary comparisons are read in $\TTS$ at each Hamiltonian's optimal restart length, measuring
verified solution cost across all nine tuned sampler families. Fixed-length single-run failure is examined descriptively
alongside $\TTS$, primarily when one sampler family runs on both Hamiltonians: within the canonical encoding's best
family on the multiplier and under SA-M on the adder, the learned energy lowers failure at every common evaluated run
length and width. Comparing each Hamiltonian's $\TTS$-optimal family directly shows the same ordering across the adder
ladder and the P3 multipliers, apart from one short-restart selection at 5$\times$5 examined below (Supplementary
Fig.~\sref{fig:si-failure-best}{S5}).

\subsection{The multiplier series}
\label{sec:res-mult}

\begin{figure*}[!tbp]
\centering
\includegraphics{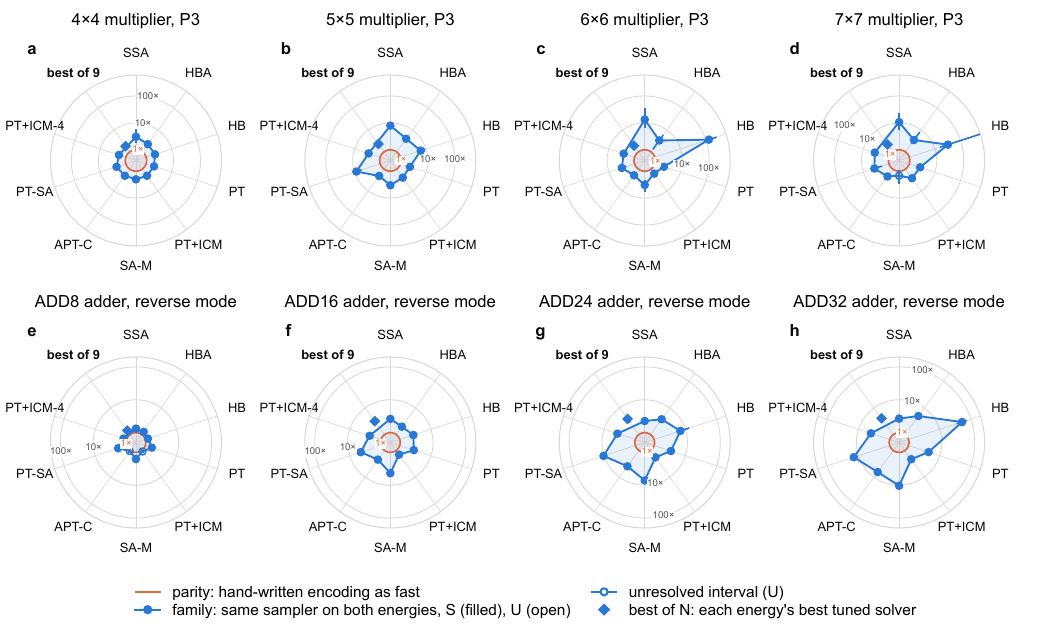}
\caption{\textbf{The learned structures need fewer updates under every tuned sampler family, except three unresolved cells: SA-M at 7$\times$7, and PT+ICM and APT-C at 8 bits.}
Speed-up, the canonical encoding's median $\TTS$ in updates over the learned structure's (log radius; orange circle,
parity): family spokes $1/\rhofam$ (Eq.~\ref{eq:rhofam}), both energies in that family at their own tuned
configuration and length; diamond, $1/\rhostar$ (Eq.~\ref{eq:rho}), each energy's best tuned solver. Bars, 95\%
intervals; filled S, open U; HB at 7$\times$7 has a lower bound of 0, read as direction only. (a--d) Multiplier, factor
mode, pool P3; (e--h) adder, reverse mode; nine families and the best of nine on every panel.}
\label{fig:perfamily}
\end{figure*}

Across the pool-P3 multiplier series (MUL4-P3, MUL5-P3, MUL6-P3-seed1 and MUL7-P3-seed2), learned Hamiltonians require
fewer updates than canonical encodings within every tested sampler family from 4$\times$4 through 6$\times$6
(Fig.~\ref{fig:perfamily}a--d, Tables~\ref{tab:series} and \ref{tab:families}): all nine families survive at
4$\times$4 ($\rhofam$ from 0.33 to 0.54), 5$\times$5 (0.12 to 0.49) and 6$\times$6 (0.01 to 0.63). At 7$\times$7, eight
of nine families survive (0.03 to 0.46) while SA-M is unresolved at its tuning-selected length (0.68 [0.33, 1.16]),
where the learned tuning median rests on one success in 32 trials; at each side's best evaluation length, SA-M survives
there as well (Supplementary Note~\snote{sn:series}{S3}). No sampler family reverses in updates or in either work
metric on any multiplier panel, and in significant work five, nine, six and eight of the nine families survive from
4$\times$4 to 7$\times$7. Companion runs evaluated at the chosen settings match the headline update budgets:
MUL6-P3-seed2 and MUL7-P3-seed1 (the latter also differing in GPU model and engine revision; Supplementary
Note~\snote{sn:engines}{S1}) take 1.00 and 1.05 times the chosen structures' updates, both unresolved. Factor-mode
color counts match the canonical encodings at every width, leaving parallel color-sweep step counts unchanged.

\begin{table*}[!tbp]
\centering
\scriptsize
\caption{\textbf{The width series under equal tuning.} Multiplier, factor mode (P3 above, P4 below), and adder, reverse
mode, each against the canonical encoding's best of nine; from 6$\times$6 upward and on the adder the best of six gives
the same value in every cell. Best family: each side's best of nine and its median $\TTS$, canonical $\to$ learned. $\rhostar$: 95\% intervals, readings S
(below 1), U (containing 1), R (above 1); $\rhopre$: family and length fixed by each side's tuning selection, low-count tuning cells on the multiplier re-measured first (descriptive; Appendix~\ref{app:protocol}), and on the adder, with no pick below 8 successes in 32 on its median sum, as first measured.
$\rhofam$: range, counts S/U/R. Work: the significant count; $\rhostar$ in the nonzero count is in Supplementary Table~\sref{tab:si-series}{S6}.}
\label{tab:series}
\setlength{\tabcolsep}{2.4pt}
\begin{tabular}{@{}>{\raggedright\arraybackslash}p{2.60cm}l>{\raggedright\arraybackslash}p{2.30cm}>{\raggedright\arraybackslash}p{1.98cm}>{\raggedright\arraybackslash}p{1.98cm}>{\raggedright\arraybackslash}p{1.98cm}>{\raggedright\arraybackslash}p{1.55cm}@{}}
\toprule
Structure & Panel & Best family, median $\TTS$ (canonical $\to$ learned) & $\rhostar$, updates & $\rhopre$, updates & $\rhostar$, work (significant) & $\rhofam$, updates \\
\midrule
\mbox{MUL4-P3} & 4$\times$4 P3 & SA-M 6{,}317 $\to$ SA-M 3{,}486 & 0.55 [0.45, 0.65] S & 0.47 [0.35, 0.64] S & 0.90 [0.73, 1.06] U & 0.33--0.54; 9/0/0 \\
\mbox{MUL5-P3} & 5$\times$5 P3 & APT-C 45{,}508 $\to$ HBA 19{,}879 & 0.44 [0.37, 0.47] S & 0.49 [0.43, 0.56] S & 0.73 [0.62, 0.79] S & 0.12--0.49; 9/0/0 \\
\mbox{MUL6-P3-seed1} & 6$\times$6 P3 & APT-C 152{,}737 $\to$ APT-C 79{,}926 & 0.52 [0.38, 0.66] S & 0.52 [0.39, 0.70] S & 0.87 [0.64, 1.09] U & 0.01--0.63; 9/0/0 \\
\mbox{MUL7-P3-seed2} & 7$\times$7 P3 & APT-C 708{,}961 $\to$ APT-C 327{,}237 & 0.46 [0.38, 0.56] S & 0.46 [0.36, 0.56] S & 0.68 [0.56, 0.82] S & 0.03--0.68; 8/1/0 \\
\midrule
\mbox{MUL4-P4} & 4$\times$4 P4 & SA-M 6{,}317 $\to$ SA-M 2{,}218 & 0.35 [0.30, 0.45] S & 0.29 [0.23, 0.40] S & 0.71 [0.61, 0.91] S & 0.22--0.40; 9/0/0 \\
\mbox{MUL5-P4} & 5$\times$5 P4 & APT-C 45{,}508 $\to$ SA-M 13{,}553 & 0.30 [0.27, 0.35] S & 0.35 [0.28, 0.38] S & 0.66 [0.61, 0.78] S & 0.08--0.36; 9/0/0 \\
\midrule
\mbox{ADD8-learned} & ADD8 reverse & APT-C 1{,}112 $\to$ SA-M 794 & 0.71 [0.63, 0.80] S & 0.74 [0.65, 0.85] S & 1.91 [1.67, 2.13] R & 0.54--0.96; 7/2/0 \\
\mbox{ADD16-learned-seed1} & ADD16 reverse & APT-C 6{,}622 $\to$ SA-M 2{,}081 & 0.31 [0.28, 0.36] S & 0.32 [0.27, 0.36] S & 0.73 [0.64, 0.82] S & 0.23--0.70; 9/0/0 \\
\mbox{ADD24-learned-seed2} & ADD24 reverse & SSA 13{,}945 $\to$ SA-M 3{,}613 & 0.26 [0.23, 0.35] S & 0.26 [0.21, 0.33] S & 0.62 [0.55, 0.84] S & 0.10--0.55; 9/0/0 \\
\mbox{ADD32-learned-seed2} & ADD32 reverse & SSA 22{,}411 $\to$ SA-M 5{,}537 & 0.25 [0.20, 0.32] S & 0.24 [0.19, 0.31] S & 0.55 [0.45, 0.72] S & 0.02--0.47; 9/0/0 \\
\bottomrule
\end{tabular}
\end{table*}

\begin{table*}[!tbp]
\centering
\footnotesize
\caption{\textbf{Every sampler family against itself.} The same-family ratio $\rhofam$ in updates (Eq.~\ref{eq:rhofam}), each
energy at its own tuned configuration and length, with its reading (S: 95\% case-bootstrap interval below 1; U:
containing 1); medians in Supplementary Table~\sref{tab:si-family-medians}{S7}. $^\dagger$Both medians rest on two
(learned) and three (canonical) successes in 128 trajectories on the median product.}
\label{tab:families}
\setlength{\tabcolsep}{4.5pt}
\begin{tabular}{@{}lrrrr@{\hspace{10pt}}rrrr@{}}
\toprule
 & \multicolumn{4}{c}{Multiplier, factor mode} & \multicolumn{4}{c}{Adder, reverse mode} \\
\cmidrule(lr){2-5}\cmidrule(l){6-9}
Family & 4$\times$4 & 5$\times$5 & 6$\times$6 & 7$\times$7 & 8 bits & 16 bits & 24 bits & 32 bits \\
\midrule
SSA & 0.33\,S & 0.13\,S & 0.075\,S & 0.094\,S$^\dagger$ & 0.76\,S & 0.38\,S & 0.44\,S & 0.37\,S \\
HBA & 0.44\,S & 0.25\,S & 0.30\,S & 0.29\,S & 0.80\,S & 0.51\,S & 0.27\,S & 0.20\,S \\
HB & 0.45\,S & 0.16\,S & 0.008\,S & 0.031\,S & 0.83\,S & 0.37\,S & 0.14\,S & 0.019\,S \\
PT & 0.49\,S & 0.43\,S & 0.44\,S & 0.38\,S & 0.62\,S & 0.35\,S & 0.29\,S & 0.23\,S \\
PT+ICM & 0.50\,S & 0.41\,S & 0.63\,S & 0.38\,S & 0.92\,U & 0.70\,S & 0.55\,S & 0.47\,S \\
SA-M & 0.50\,S & 0.30\,S & 0.31\,S & 0.68\,U & 0.63\,S & 0.23\,S & 0.14\,S & 0.097\,S \\
APT-C & 0.50\,S & 0.49\,S & 0.52\,S & 0.46\,S & 0.96\,U & 0.45\,S & 0.25\,S & 0.15\,S \\
PT-SA & 0.44\,S & 0.12\,S & 0.32\,S & 0.27\,S & 0.54\,S & 0.23\,S & 0.096\,S & 0.071\,S \\
PT+ICM-4 & 0.54\,S & 0.35\,S & 0.38\,S & 0.28\,S & 0.80\,S & 0.43\,S & 0.27\,S & 0.25\,S \\
\bottomrule
\end{tabular}
\end{table*}

\begin{figure*}[!tbp]
\centering
\includegraphics{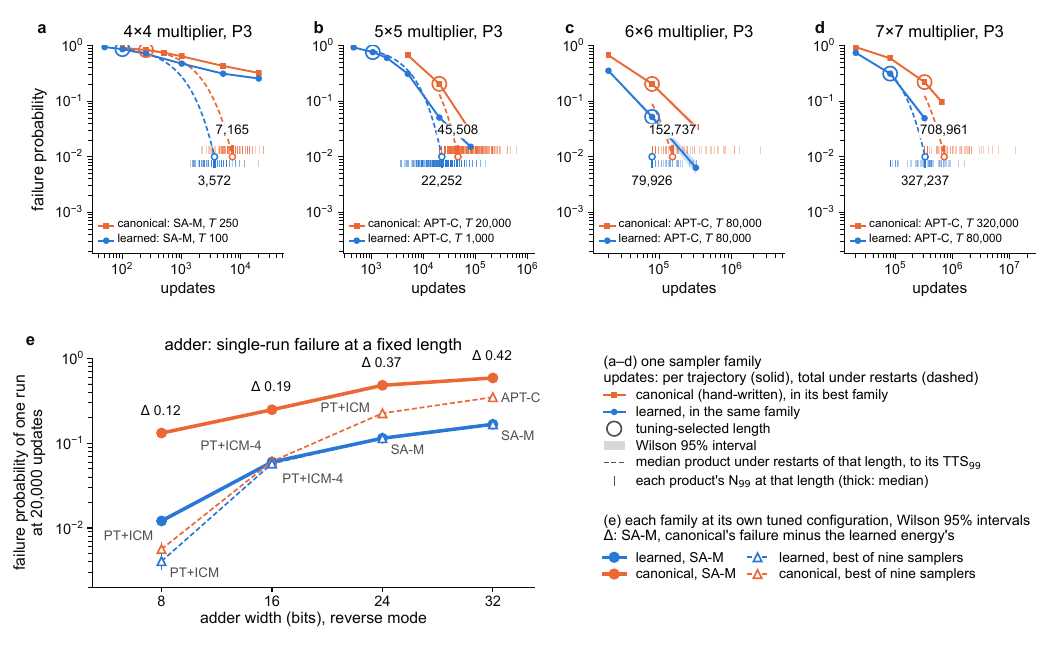}
\caption{\textbf{Failure against run length within one sampler family, and the adder's single-run failure.} (a--d) Multiplier,
both energies in the canonical encoding's best family, each at its own tuned configuration: failure pooled over the panel's products against run length (log--log), Wilson 95\% bands \citep{wilson1927probable}; ring, the tuned length $T^\ast$; dashed, the median product
under restarts of length $T^\ast$ to its $\TTS$, giving $\rhofam$ as the end points' ratio (0.50, 0.49, 0.52 and 0.46).
(e) Adder, reverse mode: single-run failure probability near 20{,}000 updates against width (log axis); thick with solid
circles, SA-M on both energies; thin dashed with hollow triangles, each energy's best of the nine families at that length; bars, Wilson
95\% intervals.}
\label{fig:failure}
\end{figure*}

Comparing best tuned solver with best tuned solver, $\rhostar$ in updates reads 0.55 [0.45, 0.65], 0.44 [0.37, 0.47],
0.52 [0.38, 0.66] and 0.46 [0.38, 0.56] from 4$\times$4 to 7$\times$7, surviving at every width. Significant-work ratios
are unresolved at 4$\times$4 (0.90 [0.73, 1.06]) and 6$\times$6 (0.87 [0.64, 1.09]) and survive at 5$\times$5 (0.73
[0.62, 0.79]) and 7$\times$7 (0.68 [0.56, 0.82]), with no work reversal in $\rhostar$. Per single-node update, the
learned structures evaluate 62.5, 83.5, 104.7 and 111.2 significant incident terms against 38.1, 50.0, 62.9 and 75.4
for the canonical encodings, offsetting the update reduction at 4$\times$4 and 6$\times$6. Winning solver families
shift with width (Table~\ref{tab:series}), and $\rhofam$ corresponds to the end-point ratio of the median-product
restart lines on the failure curves (Fig.~\ref{fig:failure}). At 5$\times$5, the learned structure's two lowest medians are a
near tie between HBA at 19{,}879 updates (selected for 500-update restarts, falling behind canonical APT-C on single
long runs) and APT-C at 20{,}023, which takes 0.44 [0.38, 0.48] of canonical APT-C in $\TTS$ at its best length (0.49
at its tuned length) and leads at every common length (Supplementary Note~\snote{sn:series}{S3}).

We checked the $10^{-3}$ significant-work threshold against the integer word lengths certified on the board
(Section~\ref{sec:hw}) to test whether terms below that cutoff survive hardware quantization. Across the four multiplier
networks built (MUL4-P3 at 7 bits; MUL5-P3, MUL6-P3-seed1 and MUL7-P3-seed1 at 13 bits), 100, 93, 97 and 68\% of
non-significant terms round directly to zero, and not a single significant term is lost at 13 bits (43 round to zero at
7 bits). Consequently, the quantized networks retain 409, 862, 1{,}359 and 1{,}957 nonzero terms compared with 452, 847,
1{,}350 and 1{,}791 significant terms in floating point, while all non-significant terms combined account for at most
0.15\% of summed coefficient magnitude. Every rounded network remains exact on every clamp case of both modes, with its
post-rounding minimum case gap sitting below the floating-point design floor $g$, so the floor guarantee stays with the
floating-point structure.

\subsection{The adder ladder}
\label{sec:res-adder}

In the reverse mode, learned adders (ADD8-learned, ADD16-learned-seed1, ADD24-learned-seed2 and ADD32-learned-seed2)
cut best-solver updates to 0.71 [0.63, 0.80], 0.31 [0.28, 0.36], 0.26 [0.23, 0.35] and 0.25 [0.20, 0.32] of the
canonical encoding's best at 8, 16, 24 and 32 bits (Fig.~\ref{fig:perfamily}e--h, Table~\ref{tab:series}). All nine
families survive in updates from 16 bits upward, and seven of nine survive at 8 bits with PT+ICM and APT-C unresolved;
canonical winners move from APT-C at 8 and 16 bits to SSA at 24 and 32 bits, while SA-M wins on every learned adder.
Significant work starts reversed at 8 bits (1.91 [1.67, 2.13], 13.9 versus 5.2 incident terms per update), crosses
below parity at 16 bits (0.73 [0.64, 0.82]), and reaches 0.62 [0.55, 0.84] and 0.55 [0.45, 0.72] at 24 and 32 bits as
learned degree drops to 11.8 against 5.3. Freezing selections on tuning data ($\rhopre$) or splitting evaluation
trajectories in half ($\rhosplit$) reproduces every update reading down to 0.24 S at 32 bits (Supplementary
Table~\sref{tab:si-series}{S6}), and companion seed ADD32-learned-seed1 reads 1.00 [0.90, 1.20] U against
ADD32-learned-seed2.

In single trajectories without restarts, read descriptively (Fig.~\ref{fig:failure}e), the learned adder fails less often
under SA-M at every common run length and width, lowering the empirical failure fraction at 20{,}000 updates by 0.120,
0.188, 0.368 and 0.422 from 8 to 32 bits. Comparing each Hamiltonian's best of the nine families near 20{,}000 updates
yields failure differences of 0.002, 0.003, 0.111 and 0.182 (0.012, 0.039, 0.121 and 0.182 across the six families
evaluated first), essentially tied at 8 and 16 bits where PT+ICM and PT+ICM-4 lead on both sides, corresponding to
failure ratios (canonical over learned) of 1.39, 1.06, 1.97 and 2.09. Near 20{,}000 updates the learned adder leads or
ties within every tuned family, whereas at 80{,}000 updates the canonical encoding's best family pulls ahead at 16 and
24 bits (Supplementary Note~\snote{sn:adder}{S4}).

\subsection{Width, pool and objective}
\label{sec:res-width}

\begin{figure}[!tbp]
\centering
\includegraphics{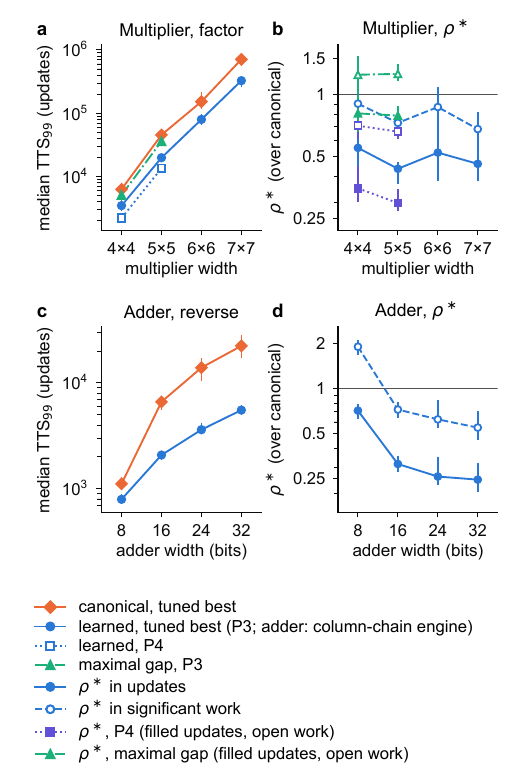}
\caption{\textbf{The two ladders under equal tuning.} (a) Best-of-nine median $\TTS$ in updates against multiplier width, factor
mode, 95\% intervals: the canonical encoding's tuned best (diamonds), the learned structure's in P3 (circles) and P4 (open squares, dotted), and the
maximal-gap fit's in P3 (triangles, dash-dotted), the last two at 4$\times$4 and 5$\times$5. (b) Each over the canonical encoding's best of nine: $\rhostar$ in updates
(filled) and in significant work (open), with the same markers. (c) Adder, reverse
mode, best of nine: both energies' tuned best against width. (d) The adder's $\rhostar$ in updates (filled) and in
significant work (open).}
\label{fig:width}
\end{figure}

Setting both series side by side in Fig.~\ref{fig:width} contrasts how their update advantages scale with circuit width.
Along the multiplier series, $\rhostar$ in updates reads 0.55, 0.44, 0.52 and 0.46 from 4$\times$4 to 7$\times$7;
canonical and learned best medians grow 3.4- versus 4.0-fold from 5$\times$5 to 6$\times$6, and 4.6- versus 4.1-fold from
6$\times$6 to 7$\times$7. Tested against pre-registered Expectation~3 (Section~\ref{sec:methods}), the
4$\times$4-to-5$\times$5 step resolves below 1 (0.79 [0.62, 0.96]), whereas the two wider steps remain unresolved (1.20
[0.91, 1.62] and 0.88 [0.67, 1.26]). Along the adder ladder, the reverse-mode update ratio deepens sharply between 8 and
16 bits before leveling off near one quarter at 24 and 32 bits, where the 24-to-32-bit step reads 0.95 [0.66, 1.22] as
best medians grow 1.61- and 1.53-fold. Relative to the 6-bit adder baseline (0.77 [0.69, 0.85]), the 32-bit ratio falls
to 0.32 [0.26, 0.43] of that value, confirming pre-registered expectation~(c). Significant work improves in its point
estimates from its 8-bit reversal (1.91) through 0.73 and 0.62 to 0.55 [0.45, 0.72] at 32 bits.

Running the series pipeline on the fourth-order pool P4 (MUL4-P4 and MUL5-P4) yields lower multiplier update counts
at both tested widths: $\rhostar$ reads 0.35 [0.30, 0.45] at 4$\times$4 and 0.30 [0.27, 0.35] at 5$\times$5, every
sampler family survives in updates (0.22 to 0.40 and 0.08 to 0.36), and $\rhostar$ survives in significant work as well
(0.71 [0.61, 0.91] and 0.66 [0.61, 0.78]). Descriptively, P4 structures need roughly two thirds of P3 update budgets.
Because MUL4-P4 retains 462 terms directly without the canonical fallback invoked by MUL4-P3 (571 terms), whereas MUL5-P4
selects seed~2 with 396 fallback terms restored (1{,}244 total versus 1{,}072 in MUL5-P3), the P3--P4 pair is not a
single-factor perturbation; P4's own 4$\times$4-to-5$\times$5 step reads 0.85 [0.66, 1.06] U.

Fig.~\ref{fig:width}a, b also carries the maximal-gap fit on pool P3, which against the canonical encoding's
best of nine reads 0.81 and 0.79 in updates (unresolved at 4$\times$4, surviving at 5$\times$5) and 1.25 and 1.26 in
significant work (both reversed). With the pool, node budget, floor and rules for the term budget and the fallback the
same, and the objective replaced by the linear program that maximizes the minimum gap with all legal energies equal,
the likelihood fit needs 0.55 [0.46, 0.58] of the maximal-gap fit's updates and 0.58 of its significant work at
5$\times$5, the pre-registered comparison, and 0.68 [0.56, 0.79] and 0.72 of them at 4$\times$4, whose direction was known
before the comparison was fixed; read descriptively on fresh streams, the same ratios in updates are 0.56 and 0.75
(Supplementary
Note~\snote{sn:maxgap}{S9} and Table~\sref{tab:si-maxgap}{S13}), and we claim no trend from two widths.

\subsection{The standard setting at 25 nodes}
\label{sec:res-standard}

\begin{figure}[!tbp]
\centering
\includegraphics{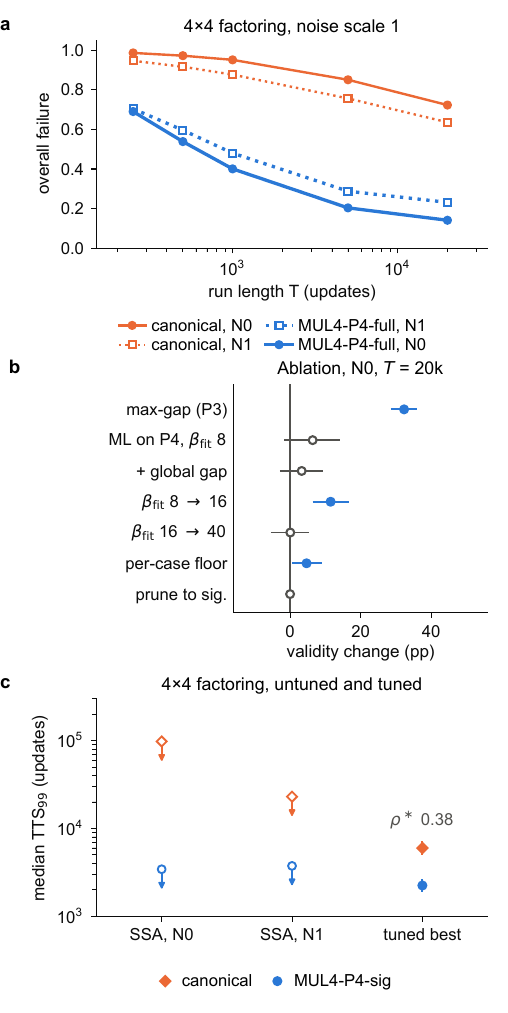}
\caption{\textbf{The standard setting at 25 nodes.} (a) 4$\times$4 factoring at noise
scale 1: failure pooled over the 90 products (512 trajectories each) against run length, canonical and
MUL4-P4-full under noise rules N0 and N1 (Confirmation~B, fresh seeds). (b) The Confirmation~B ablation
ladder at N0 and $T$ 20{,}000: change in factoring validity from the previous rung, 95\% intervals (descriptive);
filled markers exclude zero. (c) Median $\TTS$ in updates over the same products, canonical (diamonds) and
MUL4-P4-sig (circles): standard SSA under N0 and N1 (open; upper bounds, the minimum at the shortest run length,
250) and each energy's best of nine tuned samplers (filled, 95\% intervals; $\rhostar$ 0.38 [0.31, 0.45]).}
\label{fig:standard}
\end{figure}

Under untuned standard SSA, learned Hamiltonians substantially outperform canonical encodings in inverse modes
(Fig.~\ref{fig:standard}a, b). On the 4$\times$4 multiplier, MUL4-P4-full is fitted by the exact engine at
$\beta_{\rm fit}=40$ across all 1{,}984 terms of pool P4 (588 significant) under the 0.02 per-case floor without
pruning, unlike the pruned series point MUL4-P4. In Confirmation~B at noise rule N0 and $T=20{,}000$, it reaches 85.9\%
[81.2, 90.3] factoring validity versus 27.6\% [25.2, 30.3] for the canonical encoding, a +58.3 percentage-point (pp)
gain [+53.7, +62.5] across 87 of 90 products. An equal-gap P3 linear-programming construction following
\citet{onizawa2021high} (MUL4-P3-maxgap, factor-mode gap 0.0672 versus 0.0667) trails MUL4-P4-full by 26.0 pp [+20.1,
+31.4]; there candidate pool (P3 versus P4) and objective change together, whereas Section~\ref{sec:res-width} isolates
the objective on pool P3 and Confirmation~C compares the two on a shared 88-term 6-bit adder support, where they
perform alike.

Across the Confirmation~B ablation ladder (Fig.~\ref{fig:standard}b), gains come from fitted weights, higher fitting
inverse temperature (+11.5 pp from $\beta_{\rm fit}=8$ to 16, flat to 40), and per-case rather than global equal-energy
gap constraints (+4.7 pp [+0.6, +9.0]), while keeping only the 588 significant terms in MUL4-P4-sig preserves validity
within $\pm$0.2 pp at 27\% of the per-update work. On the 6-bit adder, both ADD6-compact and ADD6-compact-maxgap beat
standard-noise canonical SSA in work across all four modes (Confirmation~C; Supplementary Note~\snote{sn:25node}{S2}).
Equal sampler tuning closes most of those untuned gaps (Fig.~\ref{fig:standard}c; Supplementary
Table~\sref{tab:si-25node}{S5}), leaving MUL4-P4-sig at 0.38 [0.31, 0.45] of canonical best updates and matching work
(0.99 [0.81, 1.20]).

\subsection{Coefficients and support at 4\texorpdfstring{$\times$}{x}4}
\label{sec:res-e5}

\begin{figure}[!tbp]
\centering
\includegraphics{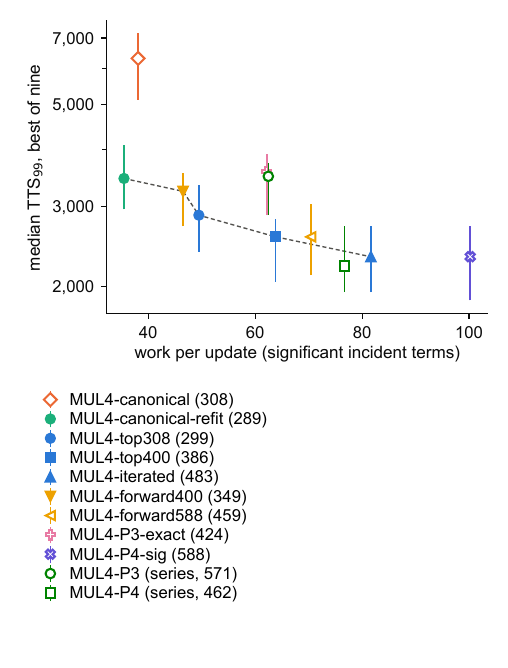}
\caption{\textbf{Coefficients and support at 4$\times$4.} Best-of-nine median $\TTS$ in updates against work per update
(significant count) in the equal-workload comparison, factor mode, 90 products, 95\% intervals: the canonical encoding,
refits of chosen supports with the exact engine's recipe (MUL4-canonical-refit, MUL4-top308, MUL4-top400,
MUL4-iterated, MUL4-forward400, MUL4-forward588 and MUL4-P3-exact) and MUL4-P4-sig; dashed, the frontier on point
estimates. Green open markers: the series points MUL4-P3 and MUL4-P4 on the same frozen canonical runs (descriptive).}
\label{fig:e5}
\end{figure}

Learned 4$\times$4 multipliers evaluate more terms per update than the canonical encoding, raising the question of how
much convergence comes from new interactions versus better weights on existing ones. Our equal-workload comparison
(Fig.~\ref{fig:e5}) refits fixed supports on 4$\times$4 with the exact engine's recipe, audits every fit over all cases,
and scores best-of-nine $\TTS$ on all 90 products against the frozen canonical baseline. Refitting the canonical
308-term support alone in MUL4-canonical-refit, pruned to 289 terms at 35.5 versus 38.1 incident terms per update, cuts
best-solver updates to 0.55 [0.47, 0.69] and work to 0.51 [0.44, 0.64] despite a $\beta_{\rm fit}=40$ objective well
above MUL4-P4-full's (1.47 versus 0.064), showing that weight optimization delivers the primary gain and extra terms
supply the rest. Adding terms ranked by MUL4-P4-sig magnitude trades work for fewer updates up to 400 terms
(MUL4-top308, MUL4-top400) before leveling off (MUL4-P4-sig takes 0.90 [0.80, 1.09] of MUL4-top400's updates at 1.42
[1.26, 1.72] times its work), whereas iterated pruning and refitting to 483 terms (MUL4-iterated) matches MUL4-P4-sig's
updates (1.00 [0.87, 1.17]) at 0.81 [0.71, 0.95] of its work. Forward selection by zero-weight gradient score
(MUL4-forward400) trails magnitude truncation at 400 terms by 1.26 [1.06, 1.51] in updates (Supplementary
Note~\snote{sn:e5}{S5}).

\subsection{Hardware-facing cost}
\label{sec:hw}

Among the benchmarked sampler families, color-sweep hard-threshold SSA fits digital FPGA fabric most directly: whereas
heat-bath, Metropolis and replica-exchange kernels require nonlinear $\tanh$ or exponential lookup tables, multi-bit
uniform comparators per $p$-bit, or replicated state arrays and swap logic across temperatures
\citep{pervaiz2019weighted,chowdhury2023full,nikhar2024all}, SSA reduces each spin update to saturating integer
addition, sign readout and a single random bit \citep{onizawa2023fast,shin2023memory}, while conflict-free graph
coloring updates an entire independent class per clock \citep{gonzalez2011parallel,nikhar2024all,onizawa2026unified}.
Hardware measurements therefore implement single-replica color-sweep SSA on the practical deployment structures of the
main text: the pool-P3 multiplier series (MUL4-P3, MUL5-P3, MUL6-P3-seed1 and MUL7-P3-seed1) alongside the 8-to-32-bit
adder ladder (ADD8-learned through ADD32-learned-seed2), each paired with its canonical encoding. MUL7-P3-seed1 is the
companion 7$\times$7 seed fixed before the headline seed rule resolved; its all-case audit is the GPU one, accompanied
by an 80-case frozen CPU audit.

\begin{figure*}[!tbp]
\centering
\includegraphics{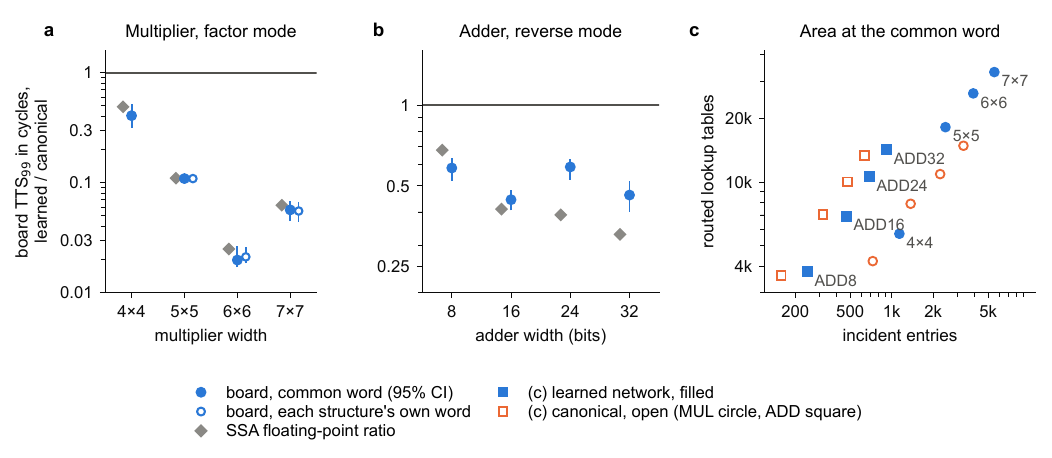}
\caption{\textbf{Hardware-facing cost of the main-text structures.} Certified $W$-bit integer coefficients on the KR260 board at
99.999\,MHz, each pair at its common word and its panel's color-sweep SSA setting. (a) Multiplier, factor mode, and
(b) adder, reverse mode: learned over canonical median $\TTS$ in clock cycles, paired case-bootstrap 95\% intervals
(filled, common word, every reading S; open, own words where they differ); gray diamonds, the
floating-point ratio within SSA, in other units. At 7$\times$7 the learned network is MUL7-P3-seed1, the other seed's structure. (c) Routed lookup tables against incident entries,
learned filled, canonical open.}
\label{fig:hw}
\end{figure*}

Real coefficients are quantized to $W$-bit integers under a single scale per structure maximizing the normalized minimum
case gap and certified at the narrowest tested ladder word keeping every legal state below every illegal one across all
cases and modes at that word and all wider ladder words: 3 bits on all adders, 7 bits on both 4$\times$4 multipliers,
and 13 bits learned versus 9 bits canonical from 5$\times$5 to 7$\times$7, with each pair benchmarked at the larger of
its two words. Executed on a Kria KR260 FPGA at 99.999\,MHz under each panel's frozen SSA setting without re-tuning, all
38 replay and main board runs match bit-exact software references. At the common word, learned multipliers take 0.41
[0.32, 0.52], 0.109 [0.102, 0.121], 0.020 [0.017, 0.027] and 0.057 [0.045, 0.068] of canonical median clock cycles to
99\% success from 4$\times$4 to 7$\times$7 (with the 7$\times$7 operating point set by two successes in 128
trajectories), and learned adders take 0.58, 0.44, 0.59 and 0.46 from 8 to 32 bits in reverse mode, every interval below
1 and aligned in direction with the floating-point SSA ratios (Fig.~\ref{fig:hw}a, b; Supplementary
Note~\snote{sn:hw}{S6}).

Synthesized as standalone term-table cores updating one color class per clock, both Hamiltonians run on the same
single-replica color-sweep SSA architecture; learned networks preserve canonical color counts while requiring more
routed lookup tables except at 16 bits because they evaluate more incident entries (Fig.~\ref{fig:hw}c). A
multi-replica tempering core, which would require replicated state arrays and temperature-swap logic, was not built. In
software, where Metropolis annealing leads on the canonical 4$\times$4 multiplier and multi-replica tempering leads from
5$\times$5 upward, single-replica SSA trails tempering at 6$\times$6 and 7$\times$7 by about fourfold on the learned
energy and about twentyfold on the canonical encoding (Supplementary Note~\snote{sn:hw}{S6}), whereas tempering against
tempering gives the 0.52 and 0.46 update ratios of Table~\ref{tab:series}. Across all board pairs, wall-clock time
tracks cycle counts within 0.5\%, and the idle board with either network loaded draws equal power within one sensor
step, from which we infer that energy per solution should scale with clock cycles (Supplementary
Note~\snote{sn:hw}{S6} and Table~\sref{tab:si-hw}{S12}).

\section{Discussion}
\label{sec:discussion}

\paragraph{Landscape shaping and finite-time mixing.} Giving hand-written encodings equal sampler tuning closes much of
the untuned gap through low-noise SSA, short heat-bath restarts, Metropolis schedules and color-sweep tempering, yet a
consistent update reduction remains across sampler families in underdetermined modes.
Refitting canonical supports at 4$\times$4 shows that tuned weights on existing interactions already cut update and work
budgets roughly in half, while running on pool P4 lowers update counts further and incident degree
sets the per-node fan-in cost. Physically, the constrained likelihood objective suppresses excited illegal states in bulk
and balances weight across legal basins while the hard gap floor preserves exact ground-state ordering and bounds
low-noise SSA escape. Because finite-length trajectories sample out of equilibrium, terminal readout couples stationary
concentration $e^{-D_c}$ to total-variation mixing distance for terminal law $\nu_{c,T}$,
\begin{equation}
\nu_{c,T}(L_c)\ge \max\{0,e^{-D_c}-\|\nu_{c,T}-p_c\|_{\rm TV}\}
\label{eq:discussion-transfer}
\end{equation}
(Appendix~\ref{sec:proof-finite}), motivating direct $\TTS$ evaluation across diverse update rules rather than relying
on equilibrium mass alone.

\paragraph{Scope and outlook.} Because the fitting pipeline requires only a circuit's legal input--output assignments
and no hand-designed penalty structure, it applies unchanged to any gate-level circuit whose clamp cases can be listed
and audited, the representation into which integer factoring, Boolean satisfiability and subset-sum instances compile
\citep{camsari2017stochastic,grimaldi2022spintronics,pervaiz2019weighted}. Extending the approach past block sizes
that admit exhaustive case listing points to two directions for future work: composing certified subcircuit blocks, where
naive weight tiling breaks ground-state legality (tiling 5$\times$5 weights onto 6$\times$6 yields a minimum case gap
of $-2.7$; Supplementary Note~\snote{sn:discussion}{S7}), or fitting on sampled clamp cases and testing generalization on
held-out cases. Within audited blocks, we hypothesize that update reductions carry over to other underdetermined clamp
modes, with work ratios set by incident fan-in.

\paragraph{Implications for $p$-bit hardware.} Evaluated on an SSA engine's native update rule (SSA versus SSA), learned
Hamiltonians reduce update counts on every panel of both ladders (Supplementary Fig.~\sref{fig:si-forest}{S4} and
Table~\sref{tab:si-family-medians}{S7}; at 7$\times$7 at the evaluation's counting resolution) while matching canonical
chromatic numbers in the inverse modes run on the board, reaching 0.020 of canonical clock cycles at 6$\times$6 in
exchange for larger lookup-table area. Because canonical and learned Hamiltonians share circuit spins and inverse-mode
color counts, whereas adder forward mode takes 4 colors against 3, a reconfigurable $p$-bit array could load canonical
weights for determined forward passes and learned weights for underdetermined inverse queries.

\paragraph{Limitations.} Determined modes admit a unique output that conventional digital logic computes directly in one
forward pass, so speedups here are claimed only for underdetermined modes; earlier clamped-circuit studies similarly
evaluate speed primarily on inverse queries and show forward operation as a functional check or omit it, though
\citet{camsari2017stochastic}, \citet{nishino2018study} and \citet{patel2022logically} also report forward-mode speed. In determined modes
hand-written encodings use less work, and fewer updates in forward addition at every width and half-clamped addition
from 12 bits upward: at 32 bits (Supplementary Fig.~\sref{fig:si-add32-modes}{S7}), learned adders take 2.12 [1.94,
2.37] times canonical best updates in \emph{ab}, 2.22 in \emph{asum} and \emph{bsum}, and 4.82 to 4.92 times canonical
significant work (forward addition reads 1.07, 1.40 and 2.04 at 6, 8 and 16 bits, whereas half-clamped modes favor
ADD6-learned at 0.77 and 0.81 at 6 bits and are unresolved at 8 bits; Supplementary Table~\sref{tab:si-adder}{S9}).
Forward 4$\times$4 multiplication reverses for MUL4-P4-sig under multi-family tuning (1.81 [1.52, 2.04]), although
MUL4-P4-full holds 0.91 exact Gibbs legal mass there at $\beta=32$ versus 0.05 for the canonical encoding (Supplementary
Note~\snote{sn:25node}{S2}); board color-sweep SSA at an 8-bit word still favors MUL4-P4-sig (0.86 [0.68, 0.99]), and
exact SA-M propagation puts MUL4-P3 at 0.88 in updates and 1.95 in work versus 1.10 and 3.35 on MUL4-P4 (Supplementary
Note~\snote{sn:series}{S3}).

We hypothesize that canonical local fields propagate determined inputs directly along carry chains, whereas
single-temperature likelihood rewards equilibrium mass rather than directed kinetic pathways
(Appendix~\ref{sec:proof-ssa}, Eq.~\ref{eq:proof-trap}); incorporating finite-time hitting cost into the constrained
fit renders the objective non-convex and is left to future work. Methodologically, re-selecting winning families and
lengths on evaluation data in $\rhostar$ introduces a finite-sample winner's curse, though freezing tuning choices and
reading descriptively on fresh streams (Appendix~\ref{app:protocol}) still yields surviving multiplier update ratios of
0.40, 0.48, 0.51 and 0.44 from 4$\times$4 to 7$\times$7 in P3 and 0.29 and 0.31 in P4, with significant work surviving
except at 6$\times$6 (0.86 [0.68, 1.17]). Finally, the 4$\times$4 support ablation uses one seed per support with a
point-estimate frontier, board runs share a fixed 100\,MHz clock rather than per-network $F_{\max}$ while board-level
sensors do not separate dynamic core draw from static platform power, and the multiplier series ends at 7$\times$7
(8$\times$8 planned, not run).

\section{Methods}
\label{sec:methods}

\paragraph{Protocol records and statistics.} Confirmations~A, B and C at 25 nodes were pre-registered with fixed
hypotheses, testing sequences and equivalence margins on fresh random streams or held-out clamp cases. All subsequent
campaigns recorded dated pre-run protocols approved prior to execution, froze canonical tuning and evaluation outputs
before reading any learned counterpart at the same width, and fixed each registered analysis before running it, reporting any rerun. Confidence intervals
report 95\% nested-bootstrap percentiles over clamp cases and trajectories, re-selecting winning families and lengths
within each replicate; analyses outside the registered protocols are marked descriptive or development evidence where
they appear (records in Supplementary Note~\snote{sn:records}{S8}).

\paragraph{Registered expectations of the multiplier series.} Prior to fitting any structure of the multiplier series,
three expectations were recorded: Expectation~1, that every point except MUL4-P3 retains an update advantage ($\rhostar$
surviving) over the canonical encoding's best of $N$, read with $N$ nine at every width; Expectation~2, that MUL4-P3 does not; and Expectation~3, that
every pairwise cross-width ratio of $\rhostar$ in updates across MUL4-P3 through MUL7-P3-seed2 has a 95\% interval
containing 1. Empirically, Expectation~1 holds across MUL4-P4, MUL5-P4, MUL5-P3, MUL6-P3-seed1 and MUL7-P3-seed2;
Expectation~2 is refuted because MUL4-P3 survives as well (0.55 [0.45, 0.65]); and Expectation~3 is refuted on the
4$\times$4-to-5$\times$5 step (0.79 [0.62, 0.96], resolving in the favorable direction) while holding from 5$\times$5
to 6$\times$6 (1.20 [0.91, 1.62]) and from 6$\times$6 to 7$\times$7 (0.88 [0.67, 1.26]), with P4's descriptive
4$\times$4-to-5$\times$5 step reading 0.85 [0.66, 1.06]. Pre-registered adder-ladder expectations and outcomes appear in
Supplementary Table~\sref{tab:si-expect}{S8}.

\section*{Data and code availability}
The data and code supporting this study are available from the corresponding author on reasonable request and will be
released as an open-source repository once the paper has been submitted to a journal. The Supplemental Material, which
includes Ref.~\cite{newcombe1998interval}, gives fitting-engine and structure provenance, full benchmark tables,
hardware details and protocol records.

\appendix

\section{Fitting algorithms and exact audits}
\label{app:engines}

\paragraph{Exact solver at 25 nodes and support pruning.}
On the 25-node circuits every partition sum and feature moment in Eqs.~\ref{eq:objective}--\ref{eq:grad} is evaluated
exactly over all $2^{25}$ spin configurations via Walsh--Hadamard transforms. Cutting-plane separation in
Eq.~\ref{eq:barrier} initializes $K$ with every illegal one-flip neighbor of a legal state together with the
lowest-gap illegal states at the starting point; after each damped Newton round, an exact audit lists every illegal
state violating Eq.~\ref{eq:percase}, appends the worst violators to $K$, and pulls the iterate back along the segment
toward the strictly feasible start until every logarithmic barrier argument is strictly positive again. Once no
violation remains across any clamp case, the barrier parameter advances tenfold from $t_0=10$ until the standard bound
$(n+2|\mathcal P|+|K|)/t$ on the barrier duality gap \citep{boyd2004convex} drops below $5\times 10^{-6}$; each inner
Newton loop terminates when half the squared Newton decrement falls below $10^{-10}$. Starting from the canonical
encoding, from zero or from random weights within the budget reaches the same optimum on one problem (coefficients within
$5\times 10^{-10}$ on pool P4 at $\beta_{\rm fit}=16$ with gap 0.015; at $\beta_{\rm fit}=40$ the optimal objective and sampled behavior remain unique even where flat
directions leave individual coefficients non-unique; Appendix~\ref{sec:proof-convexity}). Relative magnitude pruning
retains the significant support
\begin{equation}
\begin{aligned}
&\operatorname{supp}_\epsilon(\theta) = \bigl\{S\in\mathcal P:\ |\theta_S|>\epsilon\max_{S'}|\theta_{S'}|\bigr\},\\
&\epsilon = 10^{-3}.
\end{aligned}
\label{eq:prune}
\end{equation}
Inventory details for the individual 25-node structures are given in Supplementary Note~\snote{sn:engines}{S1}.

\paragraph{Sampled-likelihood estimator.}
In the series on the multiplier, each partition function $Z_c(\theta)$ in Eq.~\ref{eq:objective} is
replaced by a rescaled state-set sum,
\begin{equation}
\hat Z_c(\theta) = \sum_{x\in L_c}e^{-\beta_{\rm fit}E_\theta(x)}
  +\kappa_c\sum_{y\in S_c}e^{-\beta_{\rm fit}E_\theta(y)},
\label{eq:sampled}
\end{equation}
where all legal completions $L_c$ are enumerated explicitly, $S_c$ collects distinct illegal states visited by the cold
replicas of color-sweep tempering chains together with states returned by violator searches, and $\kappa_c\ge 1$ is
recomputed at each refresh of $S_c$ and held fixed across inner Newton steps; states returned by the violator search that
follows a refresh join $S_c$ with $\kappa_c$ unchanged. At a refresh at the current point $\theta^{(r)}$,
\begin{equation}
\begin{aligned}
&\tilde\kappa_c = \frac{1-\hat v_c}{\hat v_c}\,
  \frac{\sum_{x\in L_c}e^{-\beta_{\rm fit}E_{\theta^{(r)}}(x)}}{\sum_{y\in S_c}e^{-\beta_{\rm fit}E_{\theta^{(r)}}(y)}},\\
&\kappa_c = \min\bigl\{10^3,\ \max\bigl[1,\ \tilde\kappa_c\,(1-2\sigma_c)_+\bigr]\bigr\},
\end{aligned}
\label{eq:kappa}
\end{equation}
with $\hat v_c$ the legal fraction over the $n_c$ chain snapshots of the case, limited to $[1/(n_c+1),\,n_c/(n_c+1)]$, and
$\sigma_c=[n_c\hat v_c(1-\hat v_c)]^{-1/2}$. The value $\tilde\kappa_c$ would match the model's legal fraction on
$L_c\cup S_c$ to $\hat v_c$ at $\theta^{(r)}$; the factor shrinks it by two standard errors and the limits keep it at or
above one. At fixed $S_c$ and $\kappa_c$ the objective is convex. Refreshing them changes the objective, so a run is a
sequence of such problems and not one convex problem, and one constant per case does not calibrate the law within the
illegal states, so $\hat Z_c$ is a surrogate with no claim of being unbiased. Gradients, Hessians, cuts
and barrier updates follow Eqs.~\ref{eq:grad}--\ref{eq:barrier} with $\hat Z_c$ in place of $Z_c$, while the exact
ground-state audit of Eq.~\ref{eq:gap} remains exact over the full state space. Because every delivered structure clears
that exact audit, it lies inside the exact feasible polytope under the gap floor, so its exact $J$ is at or above the
constrained minimum on its support. Proposition~2 (Appendix~\ref{sec:proof-bound}) bounds how far above from the
run's saved state, and at MUL4-P3 and MUL5-P3 a barrier run with the exact objective places the delivered refits 0.02
to 0.03\% and 0.06 to 0.09\% above that minimum. Finite-time $\TTS$ is evaluated directly on independent readout runs.

\paragraph{Multiplier-series recipe and acceptance gates.}
Across MUL4-P3, MUL5-P3, MUL6-P3-seed1 and MUL7-P3-seed2, pool P3 unites column-local terms up to interaction order
three with the canonical terms (748, 1{,}659, 2{,}922 and 4{,}528 candidates from 4$\times$4 to 7$\times$7); at
4$\times$4 and 5$\times$5 the same pipeline also runs on pool P4, which adds fourth-order column-local terms (1{,}984
versus 748 candidates at 4$\times$4). Each fit starts from 0.8 times the canonical encoding at $\beta_{\rm fit}=40$,
advances the barrier parameter from $t_0=10$ to $t=10^7$, prunes by magnitude to $K=\lceil 1.5\,n_{\rm can}\rceil$ terms
(462, 848, 1{,}359 and 1{,}968), restores missing canonical terms as the sole fallback whenever the pruned start violates
the floor, refits to $t=10^7$ carrying the accumulated cuts (from the pruned point, or from 0.8 times the canonical
encoding under the fallback), and normalizes delivery to $\max_i\ell_i=1$. The gap floor $g$ equals 0.8 times the
starting gap, or 0.64 times the canonical encoding's normalized minimum case gap (1/32, 1/128, 1/132 and 1/136), giving
0.02, 0.005, 0.0048485 and 0.0047059. Two fitting seeds run per width, and a pre-registered rule chooses the larger
exact minimum case gap or, within 1\%, the lower exact objective. Five acceptance gates govern delivery: converged fit
and refit schedules, kept-budget or fallback accounting, feasible GPU exact audits, agreement with the frozen CPU exact
audit within $10^{-6}$ relative across all cases, and an exact objective at most $10^{-2}$ relative above an earlier
registered fit on the same width and pool. At 7$\times$7 the full frozen CPU audit ran on the chosen seed MUL7-P3-seed2
(zero of 21{,}031 cases at or below the floor), whereas the companion seed MUL7-P3-seed1 (deployed on the board) rests
on the full GPU exact audit plus an 80-case frozen CPU cross-check; the objective gate has no prior reference on MUL4-P3
or MUL5-P4, is descriptive at 7$\times$7, and at MUL4-P4 passes the $10^{-2}$ margin after missing the initial
$3\times 10^{-3}$ mark (arithmetic staging, gate outcomes and seed records in Supplementary
Note~\snote{sn:engines}{S1} and Table~\sref{tab:si-audit}{S2}).

\paragraph{Column-chain engine on the adder ladder.}
Each learned adder is fitted on the column-local pool containing all nonempty subsets of up to four nodes within a bit
column, with the most significant column also including carry-out (186, 244, 360, 476, 708 and 940 candidate terms at 6,
8, 12, 16, 24 and 32 bits). Every pool term lies in one column $\{a_k,b_k,c_k,s_k,c_{k+1}\}$ (the top column also holds
$s_n$), so the carry bit $c_{k+1}$ is the only variable passed between columns. With $F_k(c)$ the log partition sum over
columns up to $k$ with $c_{k+1}=c$ and $c_0=0$,
\begin{equation}
F_k(c)=\log\sum_{c'\in\{0,1\}}\sum_{x_k}e^{-\beta_{\rm fit}E_k(x_k,c',c)+F_{k-1}(c')},
\label{eq:adder-chain}
\end{equation}
where $x_k$ runs over the column's free bits and $E_k$ collects the terms assigned to that column, a term on a carry alone
going to one of its two columns, so that no term counts twice. One clamp case, in any of the four modes, is a two-state
sweep of $n$ steps with no Monte Carlo sampling, and a backward sweep gives its means. Its covariances across columns
carry each column's moments along the chain through the carry's transition probabilities, so one Hessian visits every
pair of columns: one evaluation on the full pool and the case sets below took 46, 75 and 164~s at 8, 12 and 16 bits on four threads. Evaluations repeat this case by case, so the objective's cost scales with the number of
cases it holds, $2^{2n}$ per determined mode in full, which is why it is sampled above 8 bits. At 6 and 8 bits the objective runs
over every case of every mode. Above 8 bits it is a sampled estimate of the uniform mean over a mode's cases: 65{,}536
cases per determined mode, every sum at 12 and 16 bits (8{,}191 and 131{,}071) and 131{,}072 sums at 24 and 32 bits, each
sample drawn uniformly without replacement, independently per mode, from a seeded generator. The two fitting seeds at 12
bits and above differ only in this sample. All legal completions of a clamped sum are constrained to share one
energy level, the barrier schedule runs until $(N+2m)/t<5\times 10^{-6}$, and the converged vector is pruned by
Eq.~\ref{eq:prune} and refitted on the surviving support. Every delivered adder is certified over all cases of all four
modes by a min-plus recursion along the same column chain, achieving minimum case gaps of 0.503 to 0.512 from 8 to 32
bits while retaining 1.6 to 2.1 times the canonical term count (seed provenance in Supplementary
Note~\snote{sn:engines}{S1}). Every fit of every seed was correct on all cases at its first audit.

\section{Sampler protocol, ratio estimators and baseline strength}
\label{app:protocol}

\paragraph{SSA noise schedules and idealized Gibbs limit.}
In the hard-threshold SSA update of Eq.~\ref{eq:ssa}, the noise amplitude $a_i(t)$ is piecewise constant across four
equal run stages and $\operatorname{sgn}(0)=+1$. Schedule N0 applies stationary statistical noise
$a_i=0.6745\,\sigma_i$, where $\sigma_i$ is the standard deviation of spin $i$'s local field
\citep{onizawa2023local}; schedule N1 tapers linearly by stage from $a_i^{(0)}=0.6745\,\sigma_i$ down to
$\min(a_i^{(0)},0.9\,m_i)$, with $m_i$ denoting node $i$'s minimum local-field magnitude across legal assignments.
Memory-efficient FPGA realizations of SSA are given in \citet{shin2023memory}.

An idealized local-relaxation limit connects clipped integration to Gibbs sampling. Holding $s_{-i}$ fixed while site
$i$ takes microsteps of size $\varepsilon>0$ on $[-B_i,B_i]$,
\[
I_i\leftarrow\operatorname{clip}\bigl(I_i+\varepsilon h_i+\sqrt{\varepsilon}\,\varsigma_i r;\;-B_i,B_i\bigr)
\]
with independent equiprobable increments $r=\pm 1$, sends $I_i$ as $\varepsilon\to 0$ to a reflected Brownian motion
with drift $h_i$ and variance $\varsigma_i^2$. Its stationary density on $[-B_i,B_i]$ is
$\rho_i(I_i)\propto\exp(2h_iI_i/\varsigma_i^2)$, whose integral over $[0,B_i]$ versus $[-B_i,0]$ yields
$\Pr(s_i=+1\mid s_{-i})=(1+e^{-2B_ih_i/\varsigma_i^2})^{-1}$. If $B_i/\varsigma_i^2=\beta$ at every free spin and each
integrator equilibrates between spin reads, single-site updates form a heat-bath chain with invariant law
Eq.~\ref{eq:gibbs}. Discrete SSA instead takes one binary-noise step per visit under an annealed bound rather than equilibrating each
integrator between spin reads, so this limit motivates Eq.~\ref{eq:gibbs} as a continuous-relaxation guide rather than
establishing SSA's stationary law; finite-time SSA trajectories are measured directly.

\paragraph{Sampler grids and tempering ladders.}
Supplementary Table~\sref{tab:si-families}{S1} specifies the hyperparameter grids and literature sources of all nine
sampler families. Every family runs on both Hamiltonians at every width of both series: PT-SA, PT+ICM and PT+ICM-4
joined the panels from 6$\times$6 upward and on the adder ladder under one uniform amendment on the registered random
streams after the initial six families had run, and none of the three wins $\TTS$ for any structure or mode there,
leaving every best-of-nine $\TTS$ cell identical to best of six while allowing leads at fixed run lengths
(Section~\ref{sec:res-adder}). Within any circuit and clamp mode, every tested family shares the same length grid $\mathcal T$ and
runs every grid point at each length in $\mathcal T$. Parallel tempering (PT and PT+ICM) sets its replica count on the
tuning namespace from a neighbor swap-acceptance floor, whereas adaptive color-sweep tempering (APT-C) spaces replicas
at constant intervals $\lambda$ of thermodynamic length computed from pilot-case energy variances. Those variances are
evaluated exactly by column-chain recursion above 5$\times$5 where feasible, or by a sampled estimate validated against
exact enumeration at 5$\times$5.

\paragraph{Hyperparameter selection, evaluation panels and grid extensions.}
For each structure, mode and sampler family $f$ with hyperparameter grid $G_f$, the operating configuration and restart
length are chosen exclusively on the tuning seed namespace,
\begin{equation}
(q_f^\ast,T_f^\ast)=\operatorname*{arg\,min}_{q\in G_f,\;T\in\mathcal T}\;
  \operatorname*{median}_{c\in\mathcal C_{\rm tune}}\;\tau_c(T;q),
\label{eq:select}
\end{equation}
breaking ties toward fewer updates. Fresh evaluation runs then score $q_f^\ast$ at $T_f^\ast$, its two adjacent grid
lengths and shared diagnostic lengths. Evaluation covers every case at 4$\times$4 (90 products), 5$\times$5 (340) and
6 bits; 64 products at 6$\times$6 and at 7$\times$7, stratified by the number of ordered factor pairs; and, from 8 to
32 bits on the adder, 129 to 153 sums stratified by the logarithm of the number of legal completions together with 256
input pairs shared across the determined modes, stratified by the longest carry-propagate run. Whenever $T_f^\ast$ lands on the longest base length, pre-registered extensions
double run lengths on fresh streams; from 5$\times$5 upward, selecting an edge value of a tempering-ladder spacing
parameter likewise triggers a wider ladder grid.

\paragraph{Ratio estimators and sensitivity variants.}
Beyond the primary definitions in Eqs.~\ref{eq:rhofam}--\ref{eq:rho}, the bootstrap and sensitivity estimators treat
selection variance explicitly. The per-family ratio $\rhofam$ pairs learned and canonical cells at each side's fixed
tuning-selected $(q_f^\ast,T_f^\ast)$ across case-bootstrap replicates \citep{efron1979bootstrap}. The best-solver
ratio $\rhostar$ re-evaluates $\min_f\TTS^{(f)}$ across the evaluated families and lengths inside every bootstrap
replicate identically on both sides, folding family-selection noise into the confidence interval while leaving a
finite-sample winner's curse, which the pre-selected counterpart below avoids. That counterpart, $\rhopre$,
freezes both the winning family (lowest tuning median, ties broken by registered family order) and $T_f^\ast$ on the
tuning namespace and holds that choice fixed in
every evaluation bootstrap replicate, thereby omitting tuning-selection variance. On the multiplier panels two descriptive steps go with it. On both sides, a tuning cell within a factor of 2 of the lowest median, with at most 5 successes in 32 on the products its median interpolates, is re-measured at 1{,}024 trials per product on a fresh namespace before the choice is frozen, and the test repeats against the new lowest; $\rho_{\rm fresh}$ reads the frozen choices on a second fresh namespace, at 128 trials per product or 1{,}024 where the cell was re-measured. No adder pick has fewer than 8 successes in 32 on its median sum, so the adder rows of $\rhopre$ are the tuning selection as first measured. On the adder ladder the split-half estimator $\rhosplit$ selects each structure's winning family and length on the
even-indexed (or odd-indexed) half of the 128 evaluation trajectories per sum and scores them on the complementary half,
averaging the two directions. Confidence bounds rounding to 1.00 at two decimals are printed to three decimals to
preserve S/U/R classification.

\paragraph{Baseline tuning and kernel checks.}
Five protocol choices define the baseline comparison:
(1) \emph{Grid range and extensions.} At 25 nodes every Hamiltonian receives the identical 172 configurations per mode
(Supplementary Table~\sref{tab:si-families}{S1}). Inverse temperatures reach $\beta=1{,}024$, well past the exact
enumeration threshold at which the median clamp case already places 99\% of its Gibbs mass on $L_c$ (roughly $\beta=16$
on learned adders, 64 on the canonical adder and MUL4-P4-full, and 256 on the canonical 4$\times$4 multiplier). Boundary
selections triggered extensions: a labeled post-hoc extension of SSA noise scale down to $\eta=0.05$ brackets the
canonical adder's optimum at $\eta=0.5$, and widening APT-C's spacing grid to $\lambda=4$ at 5$\times$5 left every
optimum at $\lambda=2$.
(2) \emph{Update accounting.} Every replica sweep is charged in full and PT-SA is charged all $n_{\rm free}$
candidate evaluations per step, whereas replica-exchange tests and isoenergetic cluster moves carry zero update charge,
favoring multi-replica competitors.
(3) \emph{Kernel verification.} Every kernel is checked before its panel: against an independent bit-exact reference
where one exists, by bit-exact reduction to a verified kernel otherwise, and for replica families by exact Gibbs
statistics on enumerable cases. Specifically, SSA reproduces earlier panel trajectories bit for bit; Metropolis
and PT-SA match reference implementations recomputing $\Delta E_i$ from the full interaction list; APT-C matches
the generic tempering kernel driven by the same color-sweep order; cluster moves conserve energy identically across all
interaction orders up to four; each newly added sampler family reproduced 234 to 318 stored benchmark values of
preceding families; and cross-host reruns were bit-identical.
(4) \emph{Effect of tuning on canonical encodings.} Tuning accelerates canonical Hamiltonians 2.8- to 6.5-fold over
standard SSA: across the four 6-bit adder modes canonical $\TTS$ drops to 701, 525, 624 and 714 updates from 2{,}102,
1{,}748, 1{,}748 and 3{,}401, and on the 4$\times$4 multiplier to 5{,}972 (factor) and 2{,}092 (multiply) from 22{,}998
and 13{,}553, whereas learned structures speed up by 1.4- to 1.5-fold (ADD6-compact: 832, 506, 519 and 517 versus
1{,}189, 748, 765 and 724; MUL4-P4-sig: 2{,}244 versus roughly 3{,}430).
(5) \emph{Higher-order tempering baseline.} Adaptive color-sweep tempering \citep{nikhar2024all} wins for the
canonical encoding in 4$\times$4 multiply mode, in 5$\times$5 through 7$\times$7 factoring, and on 8- and 16-bit
reverse adders, cutting the canonical 5$\times$5 factoring baseline to 0.56 of the best four-family result.

\section{Mathematical derivations}
\label{app:proofs}

These seven propositions establish the optimization, gap-certification and probability statements cited in the main
text, give an a posteriori bound on the fitted objective, and separate two dynamical questions: retaining a legal state once reached and entering one from an illegal start. The
two-spin constructions in Appendices~\ref{sec:proof-toy} and \ref{sec:proof-ssa} are exact analytical examples involving
no fitted or empirical parameters. Natural logarithms are written $\log$ throughout this appendix.

\subsection*{Assumptions and notation}
Index a mode--case pair by $c$, with finite compatible configuration space $X_c$, nonempty proper legal subset
$L_c\subset X_c$, illegal complement $I_c=X_c\setminus L_c$, and cardinalities $\ell_c=|L_c|$ and $r_c=|I_c|$. Let
$\phi(s)$ stack a fixed finite pool of nonempty spin monomials so that $E_\theta(s)=-\theta^\top\phi(s)$ with one vector
$\theta$ shared across all cases. At fixed inverse temperature $\beta>0$, define
\begin{equation}
\begin{aligned}
&p_c(s) = \frac{e^{-\beta E_\theta(s)}}{Z_c},\quad
u_c(s) = \begin{cases}\ell_c^{-1}, & s\in L_c,\\ 0, & s\in I_c,\end{cases}\\
&D_c = \mathrm{KL}(u_c\|p_c),\qquad M_c = p_c(L_c).
\end{aligned}
\label{eq:proof-notation}
\end{equation}
Finite parameters give $p_c(s)>0$ on $X_c$. Assume that the mode weights in Eq.~\ref{eq:objective} satisfy $\omega_m\ge 0$
and are not all zero, and that every included mode set $\mathcal C_m$ is nonempty. Set $W=\sum_m\omega_m>0$ and normalized case
weights $a_c=\omega_m/(W|\mathcal C_m|)$ for $c\in\mathcal C_m$ (omitting zero-weight cases whenever dividing by $a_c$),
so $\sum_c a_c=1$ and $\bar J(\theta)=\sum_c a_cD_c=J(\theta)/W$, leaving minimizers unchanged. Unless noted otherwise,
all statements refer to exact finite state-space sums at fixed pool, weights and $\beta$.

\subsection{Convexity, existence and the moment condition}
\label{sec:proof-convexity}

\paragraph{Proposition 1.}
Let $\mathcal F$ be the feasible set defined by the per-node incident-weight budgets (Eq.~\ref{eq:budget}), the linear
gap inequalities (Eq.~\ref{eq:percase}) and any equal-legal-energy linear equalities. If $\mathcal F\neq\emptyset$,
$\bar J$ attains a global minimum on $\mathcal F$. Writing $\mu_c=\mathbb E_{p_c}\phi$ and
$\mu_c^\ast=\mathbb E_{u_c}\phi$, its gradient and Hessian are
\begin{align}
&\nabla\bar J = \beta\sum_c a_c(\mu_c-\mu_c^\ast),
\label{eq:proof-gradient}\\
&\nabla^2\bar J = \beta^2\sum_c a_c\operatorname{Cov}_{p_c}(\phi)\succeq 0.
\label{eq:proof-hessian}
\end{align}
At any finite unconstrained minimizer, the case-weighted aggregate moments match,
\begin{equation}
\sum_c a_c\mu_c = \sum_c a_c\mu_c^\ast,
\label{eq:proof-moments}
\end{equation}
without requiring moment matching in each clamp case individually.

\emph{Proof.} Expanding $D_c=\log Z_c-\beta\theta^\top\mu_c^\ast-\log\ell_c$ and differentiating the finite log
partition sum \citep{wainwright2008graphical} yields $\nabla\log Z_c=\beta\mu_c$ and
$\nabla^2\log Z_c=\beta^2\operatorname{Cov}_{p_c}(\phi)$, giving Eqs.~\ref{eq:proof-gradient}--\ref{eq:proof-hessian}.
Along any vector $v$,
\begin{equation}
v^\top\nabla^2\bar J\,v = \beta^2\sum_c a_c\operatorname{Var}_{p_c}(v^\top\phi)\ge 0,
\label{eq:proof-null}
\end{equation}
so $\bar J$ is convex. Because every monomial in $\mathcal P$ contains at least one spin, Eq.~\ref{eq:budget} implies
$|\theta_S|\le 1$ for every $S\in\mathcal P$; intersected with finitely many closed linear gap and equal-energy
constraints, $\mathcal F$ is compact in $\mathbb R^{|\mathcal P|}$, so continuous $\bar J$ attains its minimum on
$\mathcal F$. A vanishing gradient at a finite unconstrained minimum gives Eq.~\ref{eq:proof-moments} because $\theta$ is
shared across the sum over $c$. $\square$

The word ``finite'' is essential: because $u_c$ assigns zero mass to $I_c$, an unconstrained infimum of $\bar J$ can
lie at infinity along diverging rays. The incident budgets ensure existence on $\mathcal F$ directly. Moreover,
Eq.~\ref{eq:proof-null} characterizes every flat direction: $v^\top\phi$ must be constant on $X_c$ for every case with
$a_c>0$, shifting case energy offsets without altering $p_c$, so optimal probabilities can be unique even when $\theta$
is not.

\paragraph{Constrained optimality.}
A feasible vector $\theta^\ast\in\mathcal F$ minimizes $\bar J$ on $\mathcal F$ if and only if
\begin{equation}
\langle\nabla\bar J(\theta^\ast),\,\theta-\theta^\ast\rangle\ge 0
\qquad\forall\,\theta\in\mathcal F,
\label{eq:proof-firstorder}
\end{equation}
by directional differentiation along feasible segments and the supporting-hyperplane inequality for differentiable
convex functions \citep{boyd2004convex}. Writing the polytope $\mathcal F$ via linear sign, budget and gap inequalities
$A\theta\ge b$ and equalities $H\theta=0$, normal-cone generation yields the Karush--Kuhn--Tucker system
\begin{equation}
\begin{aligned}
&\beta\sum_c a_c(\mu_c-\mu_c^\ast)-A^\top\lambda+H^\top\nu = 0,\\
&\lambda\ge 0,\qquad \lambda_j(A_j\theta^\ast-b_j) = 0,
\end{aligned}
\label{eq:proof-kkt}
\end{equation}
showing how active gap and budget constraints balance a nonzero aggregate moment residual. Whenever a reference encoding
$\theta_0$ lies in $\mathcal F$, global optimality guarantees $\bar J(\theta^\ast)\le\bar J(\theta_0)$ for the exact
problem, distinct from any numerical solver tolerance.

\subsection{How far a fitted point lies above the constrained minimum}
\label{sec:proof-bound}

The series on the multiplier minimizes the surrogate of Eq.~\ref{eq:sampled}, and refreshing its state sets makes a run
a sequence of convex problems rather than one (Appendix~\ref{app:engines}). A delivered point that clears the exact
audit lies in $\mathcal F$, so its exact $\bar J$ is at or above $\min_{\mathcal F}\bar J$. The next proposition bounds
how far above from quantities a run saves, with no assumption on how the state sets were sampled and none on the
barrier path.

\paragraph{Proposition 2.}
Fix a point $\vartheta$ and, for each case $c$, a set $S_c\subseteq I_c$ and a constant $\kappa_c\ge 1$. Let $w_c(s)$
be $1$ on $L_c$, $\kappa_c$ on $S_c$ and $0$ on $I_c\setminus S_c$, so that $\hat Z_c(\theta)=\sum_s
w_c(s)e^{-\beta E_\theta(s)}$ is Eq.~\ref{eq:sampled} at $\beta$, and define
\begin{equation}
\begin{aligned}
&\hat J(\theta)=\sum_c a_c\bigl[\log\hat Z_c(\theta)-\beta\theta^\top\mu_c^\ast-\log\ell_c\bigr],\\
&\hat p_c(s)=w_c(s)\,e^{-\beta E_\vartheta(s)}/\hat Z_c(\vartheta),\\
&\psi=\nabla\hat J(\vartheta)=\beta\sum_c a_c\bigl(\mathbb E_{\hat p_c}\phi-\mu_c^\ast\bigr),\\
&\underline J=\hat J(\vartheta)-\sum_c a_c\,\hat p_c(S_c)\log\kappa_c.
\end{aligned}
\label{eq:proof-surrogate}
\end{equation}
Then (i) for every $\theta$,
\begin{equation}
\bar J(\theta)\ge\underline J+\psi^\top(\theta-\vartheta);
\label{eq:proof-minorant}
\end{equation}
(ii) for every set $\mathcal G\supseteq\mathcal F$,
\begin{equation}
\min_{\mathcal F}\bar J\ge\underline J-\Delta_{\mathcal G},\qquad
\Delta_{\mathcal G}=\sup_{\theta\in\mathcal G}\psi^\top(\vartheta-\theta);
\label{eq:proof-lower}
\end{equation}
(iii) if $\mathcal G$ is cut out by the budgets of Eq.~\ref{eq:budget} and the inequalities of Eq.~\ref{eq:percase} for
any set $\mathcal K$ of pairs $(x,y)$, $x\in L_c$ and $y\in I_c$ of a common case, then for every
$\alpha\in\mathbb R^n_{\ge 0}$ and $\zeta\in\mathbb R^{\mathcal K}_{\ge 0}$ with
\begin{equation}
\Bigl|\psi_S-\sum_{\mathcal K}\zeta_{xy}\bigl(\phi_S(x)-\phi_S(y)\bigr)\Bigr|\le\sum_{i\in S}\alpha_i
\quad\forall\,S\in\mathcal P,
\label{eq:proof-dual-feasible}
\end{equation}
\begin{equation}
\Delta_{\mathcal G}\le\psi^\top\vartheta+\sum_i\alpha_i-g\sum_{\mathcal K}\zeta_{xy};
\label{eq:proof-dual}
\end{equation}
(iv) with $p_c$ the Gibbs law at $\vartheta$ and $\mathrm{kl}(a\|b)=a\log\frac ab+(1-a)\log\frac{1-a}{1-b}$,
\begin{equation}
\begin{aligned}
&\bar J(\vartheta)-\underline J=\sum_c a_c\,\mathrm{KL}(\hat p_c\|p_c),\\
&\mathrm{KL}(\hat p_c\|p_c)=-\log\bigl[1-p_c(I_c\setminus S_c)\bigr]\\
&\qquad+\mathrm{kl}\bigl(\hat p_c(S_c)\,\big\|\,p_c(S_c\mid L_c\cup S_c)\bigr).
\end{aligned}
\label{eq:proof-split}
\end{equation}
With $S_c=I_c$ and $\kappa_c=1$, $\hat J=\bar J$ and $\underline J=\bar J(\vartheta)$, and Eq.~\ref{eq:proof-minorant} is
the first-order inequality of the convex $\bar J$ at its exact gradient; this exact form needs no state set.

\emph{Proof.} For any law $q$ on $X_c$, Gibbs' variational inequality gives $\log Z_c(\theta)\ge\beta\theta^\top\mathbb
E_q\phi+H(q)$ \citep{wainwright2008graphical}. Take $q=\hat p_c$, supported on $L_c\cup S_c$, where $\log\hat
p_c(s)=\log w_c(s)+\beta\vartheta^\top\phi(s)-\log\hat Z_c(\vartheta)$, so that
\begin{equation}
H(\hat p_c)=\log\hat Z_c(\vartheta)-\beta\vartheta^\top\mathbb E_{\hat p_c}\phi-\hat p_c(S_c)\log\kappa_c.
\label{eq:proof-entropy}
\end{equation}
Substituting in $D_c=\log Z_c-\beta\theta^\top\mu_c^\ast-\log\ell_c$, summing with the weights $a_c$ and using
$\nabla\log\hat Z_c(\vartheta)=\beta\mathbb E_{\hat p_c}\phi$ gives Eq.~\ref{eq:proof-minorant}. Every
$\theta\in\mathcal F\subseteq\mathcal G$ has $\psi^\top(\theta-\vartheta)\ge-\Delta_{\mathcal G}$, and the minimum exists
by Proposition~1, which gives Eq.~\ref{eq:proof-lower}. Every $\theta\in\mathcal G$ has
$\theta^\top(\phi(x)-\phi(y))\ge g$ on $\mathcal K$ and $\sum_{S\ni i}|\theta_S|\le 1$, so by
Eq.~\ref{eq:proof-dual-feasible}
\begin{equation}
\begin{aligned}
-\psi^\top\theta&=-\Bigl[\psi-\sum_{\mathcal K}\zeta_{xy}\bigl(\phi(x)-\phi(y)\bigr)\Bigr]^\top\theta\\
&\quad-\sum_{\mathcal K}\zeta_{xy}\,\theta^\top\bigl(\phi(x)-\phi(y)\bigr)\\
&\le\sum_i\alpha_i\sum_{S\ni i}|\theta_S|-g\sum_{\mathcal K}\zeta_{xy}\\
&\le\sum_i\alpha_i-g\sum_{\mathcal K}\zeta_{xy},
\end{aligned}
\label{eq:proof-weak-duality}
\end{equation}
and adding $\psi^\top\vartheta$ before taking the supremum over $\mathcal G$ gives Eq.~\ref{eq:proof-dual}
\citep{boyd2004convex}. Expanding $\mathrm{KL}(\hat p_c\|p_c)$ with Eq.~\ref{eq:proof-entropy} leaves $\log
Z_c(\vartheta)-\log\hat Z_c(\vartheta)+\hat p_c(S_c)\log\kappa_c$, which summed with the weights $a_c$ is $\bar
J(\vartheta)-\underline J$. On $L_c\cup S_c$, $p_c$ is $1-p_c(I_c\setminus S_c)$ times its conditional law there, and
$\hat p_c$ has the shape of that conditional law within $L_c$ and within $S_c$, since $w_c$ is constant on each; the
chain rule of relative entropy then leaves only the two-point term $\mathrm{kl}$. $\square$

By Eqs.~\ref{eq:proof-lower} and \ref{eq:proof-split}, $\bar J(\vartheta)$ exceeds the bound by $\sum_c
a_c\mathrm{KL}(\hat p_c\|p_c)+\Delta_{\mathcal G}$, and $\Delta_{\mathcal G}\ge 0$ whenever $\vartheta\in\mathcal G$. No
choice of $\mathcal G$ therefore brings the bound closer to $\bar J(\vartheta)$ than the divergence of the surrogate's
law from the Gibbs law, whose term ${-\log[1-p_c(I_c\setminus S_c)]}$ is at least the Gibbs mass
${p_c(I_c\setminus S_c)}$ that the state sets miss.

\paragraph{Computation and values.}
A run saves $\vartheta$, its sets $S_c$, its constants $\kappa_c$ and its cut set, so $\underline J$ and $\psi$ follow by
enumerating $L_c\cup S_c$, and $\Delta_{\mathcal G}$ is read from the linear program $\max_{\mathcal
G}\psi^\top(\vartheta-\theta)$ over the budgets and the saved gap inequalities. Its dual solution is made strictly
feasible by raising $\alpha$, and Eq.~\ref{eq:proof-dual} is recomputed from those multipliers, so the value used rests
on no solver tolerance. Exact separation at the program's optimum adds violated gap inequalities; each is an inequality
of Eq.~\ref{eq:percase}, so each round's $\mathcal G$ still contains $\mathcal F$ and each round's value is a valid
bound, with no convergence claimed. Where exact separation and a budget check place $\vartheta$ in $\mathcal F$, the exact
form brackets the minimum, $\bar J(\vartheta)\ge\min_{\mathcal F}\bar J\ge\bar J(\vartheta)-\Delta_{\mathcal G}$.

Computed at the 4$\times$4 and 5$\times$5 points (descriptive), with $\mathcal P$ the refit's kept support and the
largest bound among the forms computed, the delivered refit's exact objective exceeds $\min_{\mathcal F}\bar J$ by at most 0.94, 3.49, 1.14
and 1.71\% of its value at MUL4-P3, MUL4-P4, MUL5-P3 and MUL5-P4. A barrier run with the exact objective on the same
support ends inside $\mathcal F$ and brackets the true distance at 0.02 to 0.03\% at MUL4-P3 and 0.06 to 0.09\% at
MUL5-P3, so the bound overstates it by factors of 30 to 44 and 13 to 21. At MUL5-P3 the divergence term of
Eq.~\ref{eq:proof-split} is 0.46\% of $\bar J$, 97\% of it the missed-mass term, and about two fifths of the bound's
distance.

\subsection{What the audited gap proves}
\label{sec:proof-gap}

\paragraph{Proposition 3.}
For a clamp case $c$, define the worst-pair and best-legal gaps
\begin{equation}
\begin{aligned}
\gamma_c &= \min_{y\in I_c}E(y)-\max_{x\in L_c}E(x),\\
\delta_c &= \min_{y\in I_c}E(y)-\min_{x\in L_c}E(x).
\end{aligned}
\label{eq:proof-two-gaps}
\end{equation}
Then:
\begin{align}
&\gamma_c\ge g \Longleftrightarrow E(y)-E(x)\ge g\;\;\forall x\in L_c,\,y\in I_c,
\label{eq:proof-allpairs}\\
&\operatorname*{arg\,min}_{s\in X_c}E(s)\subseteq L_c \Longleftrightarrow \delta_c>0,
\label{eq:proof-groundsubset}\\
&\operatorname*{arg\,min}_{s\in X_c}E(s)=L_c \Longleftrightarrow \text{$E$ const.\ on $L_c$, }\delta_c>0.
\label{eq:proof-groundequal}
\end{align}
Hence $\gamma_c>0$ implies legal-only ground states, whereas the converse need not hold.

\emph{Proof.} Minimizing $E(y)-E(x)$ independently over $y\in I_c$ and $x\in L_c$ gives $\gamma_c$, proving
Eq.~\ref{eq:proof-allpairs}. Every global minimizer on $X_c$ belongs to $L_c$ if and only if $\min_{I_c}E>\min_{L_c}E$,
giving Eq.~\ref{eq:proof-groundsubset}, and $L_c$ equals the full ground-state set if and only if all $x\in L_c$ share
that minimum while $\min_{I_c}E$ lies strictly higher, giving Eq.~\ref{eq:proof-groundequal}. Clearly
$\gamma_c\le\delta_c$; legal energies $\{0,2\}$ with an illegal state at $1$ give $\delta_c=1>0$ yet $\gamma_c=-1$.
$\square$

Thus $\gamma(\theta)\ge g>0$ in Eq.~\ref{eq:gap} certifies more than ground-state membership: every legal completion
sits strictly below every illegal configuration of its case, even when per-case constraints let legal energies split;
imposing equal legal energies recovers Eq.~\ref{eq:proof-groundequal}.

\subsection{Why likelihood adds information beyond the minimum gap}
\label{sec:proof-likelihood}

\paragraph{Proposition 4.}
Let $q_c=p_c(\cdot\mid L_c)$ denote the Gibbs law conditioned on $L_c$. Then
\begin{equation}
\begin{aligned}
&D_c = -\log M_c+\mathrm{KL}(u_c\|q_c),\\
&M_c \ge e^{-D_c},\qquad \sum_c a_cM_c\ge e^{-\bar J}.
\end{aligned}
\label{eq:proof-mass}
\end{equation}
Hence the case-weighted Gibbs failure probability satisfies $1-\sum_c a_cM_c\le 1-e^{-\bar J}$, and for any $a_c>0$,
$M_c\ge e^{-\bar J/a_c}$.

\emph{Proof.} Writing $p_c(s)=M_cq_c(s)$ on $L_c$ and using $\sum_{s\in L_c}u_c(s)=1$ yields the decomposition in
Eq.~\ref{eq:proof-mass}. Nonnegativity of $\mathrm{KL}(u_c\|q_c)$ implies $M_c\ge e^{-D_c}$; convexity of $e^{-x}$
gives $\sum_c a_cM_c\ge\sum_c a_ce^{-D_c}\ge e^{-\sum_c a_cD_c}=e^{-\bar J}$; and $a_cD_c\le\bar J$ since $D_{c'}\ge 0$.
$\square$

Because $u_c$ is uniform on $L_c$, the conditional term equals
\begin{equation}
\mathrm{KL}(u_c\|q_c)
= -\log\left[\ell_c\left(\prod_{x\in L_c}q_c(x)\right)^{1/\ell_c}\right],
\label{eq:proof-coverage}
\end{equation}
which vanishes if and only if $q_c=u_c$: lowering $D_c$ simultaneously suppresses illegal Boltzmann mass and penalizes
unevenness across valid answers (with $D_c=-\log M_c$ identically whenever legal energies are constrained equal).

\paragraph{Sensitivity to the full excited spectrum.}
With mean legal energy $\bar E_{L_c}=\ell_c^{-1}\sum_{x\in L_c}E(x)$, define
\begin{equation}
\begin{aligned}
A_c &= \frac1{\ell_c}\sum_{x\in L_c}e^{-\beta(E(x)-\bar E_{L_c})},\\
B_c &= \frac1{\ell_c}\sum_{y\in I_c}e^{-\beta(E(y)-\bar E_{L_c})}.
\end{aligned}
\label{eq:proof-spectrum}
\end{equation}
Direct substitution gives
\begin{equation}
\begin{aligned}
&D_c = \log(A_c+B_c),\qquad M_c = \frac{A_c}{A_c+B_c},\\
&\mathrm{KL}(u_c\|q_c) = \log A_c,\qquad A_c\ge 1,
\end{aligned}
\label{eq:proof-spectrum-identity}
\end{equation}
where $A_c\ge 1$ follows from Jensen's inequality on centered legal energies. When legal energies are equal ($A_c=1$),
minimizing $D_c$ minimizes $B_c$, the summed Boltzmann weight of \emph{every} illegal state above the legal level,
whereas a minimum gap inspects only one extremal pair. By contrast, $\gamma_c\ge g$ alone guarantees only
\begin{equation}
\gamma_c\ge g\quad\Longrightarrow\quad
M_c\ge\left[1+\frac{r_c}{\ell_c}e^{-\beta g}\right]^{-1},
\label{eq:proof-gapmass}
\end{equation}
obtained from $Z_I\le r_ce^{-\beta(\max_{L_c}E+g)}$ and $Z_L\ge\ell_ce^{-\beta\max_{L_c}E}$: whenever the illegal
multiplicity ratio $r_c/\ell_c$ is exponential in circuit size, a positive gap can still leave $M_c\ll 1$ at finite
$\beta$. Treating state energies formally as independent coordinates likewise gives
\begin{equation}
\frac{\partial D_c}{\partial E(s)}=\beta\bigl(u_c(s)-p_c(s)\bigr),
\label{eq:proof-energy-gradient}
\end{equation}
pushing occupied illegal states upward and underrepresented legal states downward, coupled across states through
$E=-\theta^\top\phi$ and constrained by Eq.~\ref{eq:proof-kkt}.

\subsection{Same support, same maximal gap, different probability and hitting time}
\label{sec:proof-toy}

\paragraph{Proposition 5.}
On two free spins with unique legal state $(+1,+1)$, consider for $\tfrac12\le d<1$ the family
\begin{equation}
E_d(s_1,s_2)=-d(s_1+s_2)-(1-d)s_1s_2.
\label{eq:proof-toy-energy}
\end{equation}
Every member shares the same three-term support, saturates both node budgets ($\ell_1=\ell_2=1$), and achieves the
largest gap $\gamma=2$ attainable under Eq.~\ref{eq:budget}. Increasing $d$ strictly raises the Gibbs legal mass $M_d$
and, under random-scan heat-bath dynamics at fixed $\beta>0$, strictly reduces the expected first-passage time to
$(+1,+1)$ from every illegal state.

\emph{Proof.} Each spin has incident $L_1$ weight $d+(1-d)=1$, and relative energies above $(+,+)$ are
\begin{equation}
E_d(s)-E_d(+,+)=
\begin{cases}
0, & s=(+,+),\\
2, & s=(+,-)\text{ or }(-,+),\\
4d, & s=(-,-),
\end{cases}
\label{eq:proof-toy-levels}
\end{equation}
fixing $\gamma=2$ across $d\in[1/2,1)$. Because flipping spin $i$ changes any budget-feasible energy by
$|2s_ih_i(s)|\le 2\ell_i(\theta)\le 2$ and both one-flip neighbors of $(+,+)$ are illegal, no Hamiltonian obeying
Eq.~\ref{eq:budget} can exceed $\gamma=2$. Yet
\begin{equation}
\begin{aligned}
&M_d = \frac1{1+2e^{-2\beta}+e^{-4\beta d}},\\
&D_d = \log\bigl(1+2e^{-2\beta}+e^{-4\beta d}\bigr),\\
&\frac{dD_d}{dd} = -\frac{4\beta e^{-4\beta d}}{1+2e^{-2\beta}+e^{-4\beta d}}<0;
\end{aligned}
\label{eq:proof-toy-mass}
\end{equation}
for instance, $d=1/2$ versus $d=3/4$ shifts illegal relative levels from $(2,2,2)$ to $(2,2,3)$, suppressing $(-,-)$.

Under random-scan heat-bath updates \citep{geman1984stochastic}, each step picks spin $1$ or $2$ with equal probability and
draws from its Gibbs conditional. With $\sigma(z)=(1+e^{-z})^{-1}$ and
\begin{equation}
p=\sigma(2\beta),\qquad q=\sigma\bigl(\beta(4d-2)\bigr),
\label{eq:proof-toy-pq}
\end{equation}
symmetry reduces the count $k\in\{0,1,2\}$ of negative spins to a Markov chain with transition probabilities
$P(1,0)=p/2$, $P(1,2)=(1-q)/2$ and $P(2,1)=q$ prior to hitting $k=0$. First-step conditioning on the finite mean
hitting times $H_k$ gives
\begin{equation}
\begin{aligned}
H_0 &= 0,\\
H_2 &= 1+(1-q)H_2+qH_1,\\
H_1 &= 1+\Bigl(1-\frac{p}{2}-\frac{1-q}{2}\Bigr)H_1+\frac{1-q}{2}H_2,
\end{aligned}
\label{eq:proof-toy-system}
\end{equation}
whose solution is
\begin{equation}
H_1=\frac{1+q}{pq},\qquad H_2=\frac{1+q+p}{pq}.
\label{eq:proof-toy-hitting}
\end{equation}
Since $p$ is independent of $d$ while $q$ increases strictly with $d$, $H_1$, $H_2$ and the uniform-start mean
$H_1/2+H_2/4$ all strictly decrease with $d$: raising $d$ from $1/2$ to $3/4$ lifts the outward local field at $(-,-)$
from $0$ to $1/2$ while keeping the inward field at $(+,+)$ equal to $1$. (This analytical first-passage time is
distinct from the terminal-state restart metric $\TTS$ measured in the main text.) $\square$

\subsection{What the gap guarantees for the integrator sampler}
\label{sec:proof-ssa}

Write the SSA update \citep{onizawa2023fast} on free spin $i$ with retained integrator $z_i$, output spin $s_i$ (using a
fixed tie rule at zero), clipping bound $I_i(t)>0$ and total signed noise increment $\xi_i(t)$ (including $\eta$) as
\begin{equation}
\begin{aligned}
z_i^+ &= \operatorname{clip}_{[-I_i(t),I_i(t)]}\bigl(z_i+h_i(s)+\xi_i(t)\bigr),\\
s_i^+ &= \operatorname{sgn}(z_i^+),
\end{aligned}
\label{eq:proof-ssa-update}
\end{equation}
calling $(s,z)$ sign-consistent whenever $s_iz_i\ge 0$ for all $i$.

\paragraph{Proposition 6.}
Suppose $s\in L_c$ and $(s,z)$ is sign-consistent. If $\gamma_c\ge g>0$ and $|\xi_i(t)|<g/2$ at every subsequent update,
all subsequent spin states remain in $L_c$.

\emph{Proof.} Updating spin $i$ from $s\in L_c$ can only leave $L_c$ if the flipped neighbor $s^{(i)}$ lies in $I_c$,
in which case Eq.~\ref{eq:proof-allpairs} gives
\begin{equation}
E(s^{(i)})-E(s)=2s_ih_i(s)\ge g.
\label{eq:proof-boundary-field}
\end{equation}
Sign consistency ($s_iz_i\ge 0$) then yields $s_i(z_i+h_i(s)+\xi_i(t))\ge g/2-|\xi_i(t)|>0$. Symmetric clipping at
$I_i(t)>0$ preserves strict sign, keeping $s_i^+=s_i$ and $s_i^+z_i^+>0$; induction over steps completes the proof
(refinable per node by replacing $g/2$ with that node's minimum inward boundary field against actual scaled noise).
$\square$

\paragraph{Why a separate entry condition is needed.}
Neither $\gamma_c>0$ nor $M_c\to 1$ suffices to guarantee reaching $L_c$ from an illegal start under SSA. For
\begin{equation}
E(s_1,s_2)=-\tfrac34s_1s_2-\tfrac14(s_1+s_2),\quad L=\{(+,+)\},
\label{eq:proof-trap}
\end{equation}
both node budgets equal $1$, illegal relative energies are $(1,2,2)$ with $\gamma=1$, and $M\to 1$ as
$\beta\to\infty$; yet at $(-,-)$ both local fields equal $-1/2$, so from negative sign-consistent integrators with
$|\xi_i(t)|\le 1/2$ every update stays negative indefinitely. Reaching $L_c$ therefore depends on noise amplitude,
schedule and accessible transition paths beyond static ground-state or Gibbs certificates.

\subsection{Analytical bridge to finite-time success}
\label{sec:proof-finite}

\paragraph{Proposition 7.}
Let $\nu_{c,T}$ denote the terminal spin distribution after $T$ updates from the prescribed initialization, and put
\begin{equation}
\epsilon_c(T)=\|\nu_{c,T}-p_c\|_{\rm TV},\qquad
v_c(T)=\nu_{c,T}(L_c).
\label{eq:proof-tv}
\end{equation}
Then
\begin{equation}
\begin{aligned}
&v_c(T) \ge \max\bigl\{0,\,e^{-D_c}-\epsilon_c(T)\bigr\},\\
&\sum_c a_cv_c(T) \ge e^{-\bar J}-\sum_c a_c\epsilon_c(T).
\end{aligned}
\label{eq:proof-finite-bound}
\end{equation}
Whenever $b_c=e^{-D_c}-\epsilon_c(T)\in(0,1)$, $N_{\rm rest}$ independent restarts fail with probability at most
$(1-b_c)^{N_{\rm rest}}$, so 99\% verified success requires at most
\begin{equation}
N_{\rm rest}\ge\left\lceil\frac{\log(0.01)}{\log(1-b_c)}\right\rceil
\label{eq:proof-restarts}
\end{equation}
restarts.

\emph{Proof.} By definition of total variation on the event $L_c$, $v_c(T)\ge M_c-\epsilon_c(T)$; combining with
$v_c(T)\ge 0$ and Eq.~\ref{eq:proof-mass} gives Eq.~\ref{eq:proof-finite-bound}. Independent restarts multiply failure
probabilities, and taking logarithms yields Eq.~\ref{eq:proof-restarts}. $\square$

Proposition~7 is purely an analytical probability bridge relating the static divergence $D_c$ to finite-time readout; no
finite-time term enters the fitting objective of this paper. Under constant-temperature heat bath, $p_c$ is stationary
and $\epsilon_c(T)$ measures mixing error; under annealed SSA, the same inequality holds with $\epsilon_c(T)$ absorbing
integrator memory and schedule nonstationarity, which static likelihood alone does not bound. Thus $D_c$ shapes
equilibrium legal mass and coverage, $\gamma_c$ certifies correctness and low-noise retention (Proposition~6),
Proposition~5 exhibits an exact kernel speed-up from suppressing higher excited states, and the benchmark panels measure
finite-time transfer directly across sampler families.

\input{main.bbl}

\onecolumngrid
\clearpage
\setcounter{section}{0}
\setcounter{equation}{0}
\renewcommand{\theequation}{S\arabic{equation}}
\renewcommand{\theHequation}{S\arabic{equation}}
\setcounter{figure}{0}
\renewcommand{\thefigure}{S\arabic{figure}}
\renewcommand{\theHfigure}{S\arabic{figure}}
\setcounter{table}{0}
\renewcommand{\thetable}{S\arabic{table}}
\renewcommand{\theHtable}{S\arabic{table}}

\begin{center}
\Large\textbf{Supplementary information}
\end{center}

\noindent Notes S1 to S9 give the fitting-engine and structure provenance, every empirical result in full, hardware
details, further discussion, the protocol records and the comparison with maximal-gap fits, with their figures and tables, while the fitting
algorithms, sampler protocol and mathematical derivations appear in Appendices~\ref{app:engines}--\ref{app:proofs}.
Supplementary Table~\ref{tab:si-families} lists the nine sampler families.
Labels: the multiplier series structures are named MUL4-P3, MUL4-P4, MUL5-P3, MUL5-P4, MUL6-P3-seed1 (with second seed
MUL6-P3-seed2) and MUL7-P3-seed2 (with first seed MUL7-P3-seed1, measured on the board), where seed~1 is 20261026 and
seed~2 is 20261027; the first sampled engine's multiplier points are MUL4-P4-v1, MUL5-P3-v1 and MUL6-P3-v1
(Supplementary Note~S1); the adder ladder's structures are ADD6-learned (its own 6-bit fit, 141 terms, not ADD6-compact), ADD8-learned,
ADD12-learned-seed2, ADD16-learned-seed1,
ADD24-learned-seed2 and ADD32-learned-seed2 (with the suffix -seed1 or -seed2 for the other seed from 12 bits on); the 25-node and equal-workload structures carry descriptive circuit--pool--role names; MUL5-P3-maxgap is the 5$\times$5 maximal-gap fit of Supplementary Note~S9; and MUL5-P3-v0 is the
5$\times$5 fit of an earlier CPU sampled pipeline, which preceded both sampled engines.

\begin{table}[!bp]
\centering
\caption{\textbf{The nine sampler families.} Every family runs on every structure and mode with the same grid per circuit,
PT-SA, PT+ICM and PT+ICM-4 joining the panels from 6$\times$6 upward and on the adder ladder by a uniform amendment, on
the panels' registered streams (the first sampled engine's 6$\times$6 point keeps six families,
Table~\ref{tab:si-engine-a}); configurations count grid points at 25 nodes, each run at every length of the circuit's length grid. $h_i$:
Section~\ref{sec:method}; $\Delta E_i=2s_ih_i$. Updates count single-node updates over all replicas. A run succeeds when every node of its terminal state is legal.}
\label{tab:si-families}
\footnotesize
\setlength{\tabcolsep}{3pt}
\begin{tabular}{l>{\raggedright\arraybackslash}p{8.4cm}>{\raggedright\arraybackslash}p{2.6cm}r}
\toprule
Family & Update and acceptance & Source & Configs \\
\midrule
SSA & Eq.~\ref{eq:ssa}, random scan; noise scale $\eta\in\{0.5,0.75,1,1.25,1.5\}$ $\times$ seven noise rules (N0, N1 and
five end levels of a tapered rule) & \citet{onizawa2023fast,onizawa2023local} & 35 \\
HB & heat bath at constant $\beta$, random scan & \citet{geman1984stochastic} & 9 \\
HBA & heat bath with $\beta$ rising geometrically from $\beta_0\in\{0.5,2.5\}$ to $\beta_1\in\{8,\dots,1{,}024\}$ &
\citet{geman1984stochastic} & 16 \\
PT & heat-bath replicas on a geometric ladder from $\beta=1$ to $\beta_{\max}$; after every round, neighbor swaps accepted
with probability $\min\{1,e^{(\beta_r-\beta_{r+1})(E_r-E_{r+1})}\}$; success read from the coldest replica &
\citet{hukushima1996exchange,earl2005parallel} & 8 \\
PT+ICM & PT on two copies with one isoenergetic cluster move between them at every temperature and round; success read from
the coldest replica of any copy &
\citet{houdayer2001cluster,zhu2015efficient,chowdhury2025pushing} & 8 \\
PT+ICM-4 & four copies per temperature, paired by a random perfect matching, on APT-C's ladders, whose $\lambda$ grid widens by APT-C's rule (up to $\lambda$ 16 at 24 and 32 bits); success read from the coldest
replica of any copy &
\citet{chowdhury2025pushing} & 16 \\
SA-M & Metropolis single flips, random scan: accept if $\Delta E_i\le0$ or $u<e^{-\beta\Delta E_i}$, $\beta$ linear from
$\beta_0$ to $\beta_1$ & \citet{kirkpatrick1983optimization,isakov2015optimised} & 16 \\
PT-SA & every free node's flip accepted with probability $\min\{1,e^{-\beta(\Delta E_i-\varepsilon)}\}$; one accepted
flip, chosen uniformly, is applied and $\varepsilon\leftarrow0$; after a step without acceptance
$\varepsilon\leftarrow\varepsilon+\delta$, $\delta\in\{0,0.02,0.1\}$; a step is charged $n_{\rm free}$ updates &
\citet{aramon2019physics} & 48 \\
APT-C & heat-bath replicas updated in color-sweep order; $R=\min\{32,\max\{2,\lceil\Lambda(\beta_{\max})/\lambda\rceil+1\}\}$
replicas at equal steps of the thermodynamic length $\Lambda(\beta)=\int_1^{\beta}\sigma_E(\beta')\,d\beta'$, with
$\sigma_E$ the Gibbs standard deviation of the energy; $\lambda\in\{1,2\}$ at 25 nodes and 4$\times$4,
$\{1,2,4\}$ from 5$\times$5 upward and on the adder ladder, where the next doubling is added while a selection sits at
the top $\lambda$, until none does or the top $\lambda$ already gives $R=2$ at every $\beta_{\max}$ (never needed on
the multiplier; up to $\lambda$ 8 or 16 on the adders) &
\citet{nikhar2024all} & 16 \\
\midrule
All & & & 172 \\
\bottomrule
\end{tabular}
\end{table}

\section*{Supplementary Note S1: fitting-engine and structure provenance}
\phantomsection\label{sn:engines}

\paragraph{The 25-node structures and the first 5$\times$5 fit.}
The exact 25-node solver and pruning threshold (Eq.~\ref{eq:prune}) are specified in Appendix~\ref{app:engines}. The
individual 25-node structures are: on ADD6, the canonical encoding (61 terms of interaction order at most two, work 5.2
to 5.8 per mode), ADD6-compact (an 88-term information-ranked support, exact likelihood at $\beta$ 128; 81 significant
terms), ADD6-compact-maxgap (the same support, a maximum-gap quadratic program; 67 significant) and ADD6-large (a
256-term information-ranked support, likelihood at $\beta$ 8 with a global gap; 130 to 137 significant over its five
initializations, work 29.8 to 31.2, pruned as ADD6-large-sig); on MUL4, the canonical encoding (308 terms, work 38.1, 12 colors for 17 free nodes in
factor mode), MUL4-P4-full (pool P4 of 1{,}984 terms, exact engine likelihood at $\beta$ 40 with the per-case gap; 588
significant terms, the pruned MUL4-P4-sig, work 100.2; global-gap rung MUL4-P4-global; P3 maximum-gap rung
MUL4-P3-maxgap) and MUL4-dense (all 15{,}275 terms of interaction order at most four, $\beta$ 16, zero start; 949 terms,
work 139.3, no parallelism left); and on MUL5, MUL5-P3-v0, fitted by an earlier CPU sampled pipeline (pool P3, sampled likelihood at $\beta$ 40 with
floor 0.005, one seed; 921 terms after the prune, work 91.4), measured in the pre-registered 5$\times$5 confirmation of Note~S2.

\paragraph{The series engine.}
The multiplier series runs on one kernel revision implementing the sampled-likelihood estimator of Eq.~\ref{eq:sampled}
(Appendix~\ref{app:engines}), whose source files are checked byte for byte against a fixed manifest at the start of
every run and before every stage. Its arithmetic is a staged TF32 form: the moment Gram and the cut Gram of the Newton matrix are formed in
TF32, the reduced matrix is factored once per Newton iteration in float64, and at most five retained-factor corrections
bring each step to the float64 residual tolerance; a step still outside it is recomputed in float64, and after five
such misses in a row the rest of the fit or refit runs in float64. The pipeline and
its pools, floors, counts and seeds are those of Section~\ref{sec:method} and Appendix~\ref{app:engines}. The five gates of a delivered point are: (a)
a verified end of the fit and the refit at $t=10^7$, the final phases converged; (b) the kept count $K$, or the fallback
with both counts; (c) the GPU exact audits of the fit, the refit start and the refit feasible; (d) the frozen exact
audit on every case of both modes agreeing with the GPU audit within $10^{-6}$ relative (chunked on CPU at 6$\times$6 and
7$\times$7; at 7$\times$7 it ran on the chosen seed alone, and the other seed rests on the GPU audit, cross-checked by the
frozen audit on an 80-case subset to $10^{-9}$ relative); and (e) the fit's exact objective at most $10^{-2}$ relative
above its registered reference, the first sampled engine's fit at the same width and pool, the registered $3\times10^{-3}$ outcome reported beside. The margin of (e) was widened from the registered
$3\times10^{-3}$ to $10^{-2}$ for every point after the $3\times10^{-3}$ miss at MUL4-P4; before the runs, an amendment had
already relaxed the pre-run note's tighter values ($10^{-9}$ in (d), $10^{-3}$ in (e)). Gate (e) does not apply at MUL4-P3 and MUL5-P4, whose pools have no such
reference, and is descriptive at 7$\times$7, where no registered reference exists (Supplementary Table~\ref{tab:si-audit}); MUL6-P3-seed1 lies 6.42\% below
its reference. The 4$\times$4 P4 point (MUL4-P4) misses (e) at
$3\times10^{-3}$ (its fit's exact objective 0.063962 is 0.52\% above the reference 0.063633, the first sampled engine's
exact 4$\times$4 refit on the same pool) and meets it at $10^{-2}$. The fallback added 109, 224, 333 and 350 canonical
terms to the P3 points at 4$\times$4, 5$\times$5, 6$\times$6 and 7$\times$7 and 396 to the 5$\times$5 P4 point; MUL4-P4
kept 462 terms without it. Each delivered point sits above its floor on every case, by the audits of gate (d) (for example 0.0200053 against 0.02
at 4$\times$4 in P3). The seed rule chose seed~1 (20261026) at 4$\times$4 in both pools (MUL4-P3 and MUL4-P4) and at
5$\times$5 and 6$\times$6 in P3 (MUL5-P3 and MUL6-P3-seed1), and seed~2 (20261027) at 5$\times$5 in P4 (MUL5-P4) and at
7$\times$7 in P3 (MUL7-P3-seed2). At 7$\times$7 the two minimum case gaps lie within one percent (0.004706318 for
seed~1, 0.004706270 for seed~2, against the floor 0.0047059), and the lower exact objective at $\beta$ 40 of the
delivered refits decided (8.874514 for seed~2 against 9.109005 for seed~1, 2.6\% lower). The fits were in the other
order (8.844497 for seed~1 against 8.860444 for seed~2): the refit raised seed~1's objective by 3.0\% and seed~2's by
0.16\%. The chosen structure, MUL7-P3-seed2, keeps 2{,}318 terms (1{,}764
significant), and the other, MUL7-P3-seed1, keeps 2{,}309 (1{,}791), both after the fallback. Evaluated at the chosen
seed's selections, the other seed's structure (MUL6-P3-seed2 at 6$\times$6, MUL7-P3-seed1 at 7$\times$7) needs 1.00 of
the chosen structure's updates at 6$\times$6 and 1.05 [0.83, 1.16] at 7$\times$7 (both unresolved), and 0.52 and 0.48
[0.37, 0.55] of the canonical encoding's, surviving. Fits on a GPU are reproducible to rounding, not bit for bit: on one
card a repeated fit can follow either of two deterministic rounding paths, which end with one term exchanged at the $K$
boundary and the exact objective about $1.4\times10^{-5}$ relative apart; such differences, between cards and between
runs, are reported as a property of the engine, and the gates and exact audits that judge a point do not depend on the
path. At 7$\times$7 both seeds run one route of two settings: the TF32 steps' residual tolerance is held at its cap
of $10^{-3}$ from the first barrier stage, where the profile reaches it only at $t=10^4$, and the exact search for
cases below the floor runs every 20 accepted Newton updates instead of every 10. The route was adopted after a test at
6$\times$6 on seed~1, fixed before the run, delivered a structure whose exact objective lay $3.6\times10^{-4}$ relative
above MUL6-P3-seed1's (criterion $10^{-3}$), with every case above the floor. The two seeds ran on different cards and
engine versions: seed~1 (20261026) with engine version~2.1 on an A800 card, and seed~2 (20261027) with engine
version~2.3 on A100 SXM4 cloud instances, a version that adds a chunked search for violated cuts, a float32 stage
before the float64 fallback and a resume from snapshots at the ends of barrier stages, across which a run moves between
instances of one card model. Both seeds pass gates (a) to (c) and the subset cross-check of (d). MUL7-P3-seed2
also passes the frozen audit of every case: over its 4{,}647 factor and 16{,}384 multiply cases the exact objective and the
minimum case gap agree with the GPU audit to $4.8\times10^{-15}$ and $1.9\times10^{-13}$ relative, no case is broken or at
or below the floor, and the minimum, 0.00470627 in the multiply mode, lies 79 parts per million above it. Its chunks ran on
two CPU hosts; on four chunks run on both, the gap fields are bit for bit equal and the log partition functions differ by
at most two units in the last place.

\begin{table}[!tbp]
\centering
\scriptsize
\caption{\textbf{Acceptance gates and exact-audit scope of the delivered structures.} Multiplier series (top, including
the companion 7$\times$7 seed) and adder ladder (bottom): engine version and hardware card; candidate-pool size, pruning
budget $K$ (or relative threshold) and canonical-union fallback count ($+\Delta \to \text{actual}$); outcomes on gates
(a)--(e) (Section~\ref{sec:method} and Supplementary Note~S1) or column-chain barrier and exact audit checks; exact-audit
scope; and exceptions. Slices: the parts into which the GPU exact audit divides the enumeration of every case's
states. Round 0: the adder fit's first audit, which found no incorrect case. End offset:
the amount by which a fit stopped at $t=10^7$ lies above the fully converged objective on its pool. DP: dynamic
program. $^\dagger$Companion 7$\times$7 seed, evaluated as the other-seed arm and deployed on the FPGA board, its frozen audit on the 80-case subset only.}
\label{tab:si-audit}
\setlength{\tabcolsep}{2.5pt}
\begin{tabularx}{\textwidth}{@{}>{\raggedright\arraybackslash}p{2.45cm}>{\raggedright\arraybackslash}p{2.15cm}>{\raggedright\arraybackslash}p{2.65cm}>{\raggedright\arraybackslash}p{2.70cm}>{\raggedright\arraybackslash}X>{\raggedright\arraybackslash}X@{}}
\toprule
Structure & Engine and card & Pool; budget and fallback & Gates (a)--(e) & Exact-audit scope & Exceptions and disclosures \\
\midrule
\multicolumn{6}{@{}l}{\textit{Multiplier series (series engine: fit to $t=10^7$, prune to $K$, refit to $t=10^7$)}} \\
\mbox{MUL4-P3} & series v2.1, A100 PCIe & P3 (748); $K=462$, $+109 \to 571$ & (a)--(d) pass; (e) n/a (no P3 ref.) & GPU exact + full frozen CPU (90 factor, 256 multiply; $<10^{-11}$ rel.) & none \\
\mbox{MUL4-P4} & series v2.1, A100 PCIe & P4 (1{,}984); $K=462$, $+0 \to 462$ & (a)--(d) pass; (e) $+0.52\%$ (pass $10^{-2}$, miss $3\times10^{-3}$) & GPU exact + full frozen CPU (90 factor, 256 multiply; $<10^{-11}$ rel.) & misses $3\times10^{-3}$ on (e) ($t=10^7$ end offset on 1{,}984-term pool); meets $10^{-2}$ \\
\mbox{MUL5-P3} & series v2.1, A100 PCIe & P3 (1{,}659); $K=848$, $+224 \to 1{,}072$ & (a)--(d) pass; (e) $+0.04\%$ (pass $10^{-2}$, $3\times10^{-3}$) & GPU exact + full frozen CPU (340 factor, 1{,}024 multiply; $<10^{-11}$ rel.) & none \\
\mbox{MUL5-P4} & series v2.1, A100 PCIe & P4 (5{,}633); $K=848$, $+396 \to 1{,}244$ & (a)--(d) pass; (e) n/a (no P4 ref.) & GPU exact + full frozen CPU (340 factor, 1{,}024 multiply; $<10^{-11}$ rel.) & none \\
\mbox{MUL6-P3-seed1} & series v2.1, A100 PCIe (card~2) & P3 (2{,}922); $K=1{,}359$, $+333 \to 1{,}692$ & (a)--(d) pass; (e) $-6.42\%$ (pass $10^{-2}$, $3\times10^{-3}$) & GPU exact + chunked full frozen CPU (1{,}238 factor, 4{,}096 multiply; $<10^{-11}$ rel.) & fit switched to float64 at $t=10$; second A100 PCIe card \\
\mbox{MUL7-P3-seed2} & series v2.3, A100 SXM4 & P3 (4{,}528); $K=1{,}968$, $+350 \to 2{,}318$ & (a)--(d) pass; (e) descr.\ ($+0.16\%$ refit) & GPU exact + chunked full frozen CPU ($2^{28}$ slices: 4{,}647 factor, 16{,}384 multiply; $<10^{-12}$ rel.) & (e) descriptive; two-segment resume; seed pair crosses v2.1/A800 and v2.3/A100 \\
\mbox{MUL7-P3-seed1}$^\dagger$ & series v2.1, A800 & P3 (4{,}528); $K=1{,}968$, $+341 \to 2{,}309$ & (a)--(c) pass; (d) GPU + subset pass; (e) descr.\ ($+3.0\%$ refit) & GPU exact ($2^{28}$ slices: 4{,}647 factor, 16{,}384 multiply) + frozen CPU 80-case subset ($<10^{-9}$ rel.) & companion seed (board structure); full frozen CPU audit run on chosen seed only \\
\midrule
\multicolumn{6}{@{}l}{\textit{Adder ladder (column-chain engine on CPU: equal legal energies, prune at $10^{-3}$, refit to $t=10^9$)}} \\
\mbox{ADD8-learned} & column chain, CPU & col.-local (244); $>10^{-3}$, $+0 \to 169$ & barrier $<5\times10^{-6}$; round-0 audit pass (min.\ gap 0.503) & exact column-chain min-plus DP + compiled all-case check across all 4 modes & separate column-chain engine (no state sampling; equal legal energies) \\
\mbox{ADD12-learned-seed2} & column chain, CPU & col.-local (360); $>10^{-3}$, $+0 \to 228$ & barrier $<5\times10^{-6}$; round-0 audit pass (min.\ gap 0.507) & exact column-chain min-plus DP + compiled all-case check across all 4 modes & separate column-chain engine (no state sampling; equal legal energies) \\
\mbox{ADD16-learned-seed1} & column chain, CPU & col.-local (476); $>10^{-3}$, $+0 \to 287$ & barrier $<5\times10^{-6}$; round-0 audit pass (min.\ gap 0.504) & exact column-chain min-plus DP + compiled all-case check across all 4 modes & separate column-chain engine (no state sampling; equal legal energies) \\
\mbox{ADD24-learned-seed2} & column chain, CPU & col.-local (708); $>10^{-3}$, $+0 \to 462$ & barrier $<5\times10^{-6}$; round-0 audit pass (min.\ gap 0.511) & exact column-chain min-plus DP (all 4 modes) + all $3.36\times10^{7}$ sums + $10^{6}$ cases/det.\ mode & separate column-chain engine (no state sampling; equal legal energies) \\
\mbox{ADD32-learned-seed2} & column chain, CPU & col.-local (940); $>10^{-3}$, $+0 \to 557$ & barrier $<5\times10^{-6}$; round-0 audit pass (min.\ gap 0.512) & exact column-chain min-plus DP (all 4 modes) + all $8.59\times10^{9}$ sums + $10^{6}$ cases/det.\ mode & separate column-chain engine (no state sampling; equal legal energies) \\
\bottomrule
\end{tabularx}
\end{table}

\paragraph{The first sampled engine's multiplier points.}
Before the series engine, the first sampled engine fitted 4$\times$4 (MUL4-P4-v1, pool P4), 5$\times$5 (MUL5-P3-v1, pool
P3) and 6$\times$6 (MUL6-P3-v1, pool P3) with a state-set cap of 300 per case, the floor at 0.8 times the start's
minimum gap, a prune at $10^{-3}$ of the largest coefficient, one refit, a schedule ending at $t=10^{10}$ and the Newton
steps solved densely or, from 6$\times$6, by preconditioned conjugate gradients after a registered check against the
dense form at 5$\times$5 (coefficient cosine 0.99891, objective difference $4.4\times10^{-5}$). Its points remain
registered evidence (Supplementary Table~\ref{tab:si-engine-a}): against the canonical encoding's best of nine or six
they need 0.35, 0.41 and 0.52 of its updates, all surviving, at 0.93, 0.74 and 1.04 of its significant work. At
4$\times$4 MUL4-P4-v1 has coefficient cosine 0.99914 to the exact refit and 0.996 times its minimum case gap; at
6$\times$6 the cap binds in 99\% of the multiply cases. Refitted with an adaptive cap, the fit's missing mass (the
Gibbs mass outside the capped state sets) at the end of the schedule fell about fourfold, from 0.21 to 0.047, while its
exact objective rose by 0.047 (4.077 against 4.029), and the best tuned $\TTS$ was unchanged to the digit (one width
and one seed, development evidence).

\begin{table}[!tbp]
\centering
\scriptsize
\caption{\textbf{The first sampled engine's multiplier points} (Supplementary Note~S1), registered evidence beside the multiplier
series, on the same frozen canonical runs: terms nonzero (significant), the learned structure's best of $N$ with its
median $\TTS$, and $\rhostar$ against the canonical encoding's best of nine (4$\times$4, 5$\times$5) or six
(6$\times$6; that panel was not extended).}
\label{tab:si-engine-a}
\setlength{\tabcolsep}{4pt}
\begin{tabular}{@{}lllllll@{}}
\toprule
Point & Terms & Learned best, median $\TTS$ & $\rhostar$, updates & work (nonzero) & work (significant) \\
\midrule
\mbox{MUL4-P4-v1} & 596 (591) & SA-M 2{,}218 & 0.35 [0.29, 0.44] S & 0.94 [0.77, 1.18] U & 0.93 [0.77, 1.17] U \\
\mbox{MUL5-P3-v1} & 1{,}150 (920) & SA-M 18{,}474 & 0.41 [0.37, 0.48] S & 0.88 [0.80, 1.04] U & 0.74 [0.67, 0.87] S \\
\mbox{MUL6-P3-v1} & 1{,}998 (1{,}627) & APT-C 79{,}926 & 0.52 [0.38, 0.67] S & 1.24 [0.91, 1.58] U & 1.04 [0.76, 1.33] U \\
\bottomrule
\end{tabular}
\end{table}

\paragraph{Adder-ladder seed provenance.}
The column-chain solver, pool sizes and equal-legal-energy formulation for the adder ladder are detailed in
Appendix~\ref{app:engines}. One seed runs at 8 bits (ADD8-learned), and two seeds run at 12 bits and above, differing only in the
objective's sample of cases; at 12 bits the seed rule takes seed~2 (ADD12-learned-seed2, with ADD12-learned-seed1 as the
other-seed arm), at 16 bits seed~1 (ADD16-learned-seed1, with ADD16-learned-seed2), and at 24 and 32 bits it takes seed~2 (ADD24-learned-seed2 and
ADD32-learned-seed2, against ADD24-learned-seed1 and ADD32-learned-seed1; at 32 bits minimum case gap 0.5120 against
0.5032), and the other seed's arm, at the chosen seed's selections, is unresolved against it in every mode (reverse mode
1.00 [0.90, 1.20] at 32 bits). Supplementary Table~\ref{tab:si-audit} summarizes the engine version, card, pool,
fallback count, gate outcomes, exact-audit scope and exceptions for every delivered structure, and Supplementary
Table~\ref{tab:si-cost} the wall times of the fits and of their exact audits.

\begin{table}[!tbp]
\centering
\scriptsize
\caption{\textbf{Fitting and exact-audit cost.} The multiplier series' delivered structures with the companion seeds at
6$\times$6 and 7$\times$7, and the adder ladder from 16 bits on. Wall time of the fit and the refit from each run's own
timers; the total adds the prune and the delivery check. Cards and CPUs differ between rows, so the rows are not a
scaling study. MUL7-P3-seed2 ran as two cloud runs, its fit spanning both. Exact audit: the frozen CPU audit of gate (d)
on the multiplier, at 7$\times$7 the sum of its 4{,}096 single-threaded chunk times, the join apart (2.6 h); the exact audit
of every case on the adder.}
\label{tab:si-cost}
\setlength{\tabcolsep}{4pt}
\begin{tabular}{@{}llrrrl@{}}
\toprule
Structure & Engine and card & Fit (s) & Refit (s) & Total (s) & Exact audit \\
\midrule
\mbox{MUL4-P3} & v2.1, A100 PCIe & 23.0 & 15.5 & 40.1 & frozen, 2 threads, 13.5 s \\
\mbox{MUL4-P4} & v2.1, A100 PCIe & 45.3 & 11.1 & 58.1 & frozen, 2 threads, 14.5 s \\
\mbox{MUL5-P3} & v2.1, A100 PCIe & 141.1 & 86.5 & 231.0 & frozen, 2 threads, 865.5 s \\
\mbox{MUL5-P4} & v2.1, A100 PCIe & 715.4 & 88.5 & 808.4 & frozen, 2 threads, 1{,}097.6 s \\
\mbox{MUL6-P3-seed1} & v2.1, A100 PCIe & 4{,}855.3 & 1{,}081.1 & 5{,}972.2 & frozen, 16 workers, 39.2 CPU-h \\
\mbox{MUL6-P3-seed2} & v2.1, A100 PCIe & 4{,}434.9 & 1{,}054.7 & 5{,}524.3 & frozen, 16 workers, 28.3 CPU-h \\
\mbox{MUL7-P3-seed1} & v2.1, A800 & 43{,}563.7 & 18{,}760.3 & 62{,}794.3 & frozen, 80-case subset only (not the chosen seed) \\
\mbox{MUL7-P3-seed2} & v2.3, A100 SXM4, two runs & 42{,}581.7 & 18{,}516.3 & 61{,}640.7 & frozen, two CPU hosts, 677.7 CPU-h \\
\midrule
\mbox{ADD16-learned}, both seeds & column chain, CPU & 13{,}367; 13{,}542 & 9{,}447; 10{,}002 & & 6{,}600 s \\
\mbox{ADD24-learned-seed2} & column chain, CPU & 10{,}195 & 8{,}889 & & 206 s \\
\mbox{ADD32-learned-seed2} & column chain, CPU & 15{,}940 & 11{,}951 & & 1{,}573 s \\
\bottomrule
\end{tabular}
\end{table}

\section*{Supplementary Note S2: the 25-node confirmations in full}
\phantomsection\label{sn:25node}

\paragraph{Adder.} In Confirmation~A, on held-out cases, the large learned support ADD6-large with noise rule N1 solves the three determined modes of
ADD6 with 98.01\% strict validity at $T$ 1{,}000 against 89.80\% for the canonical encoding with the same rule (+8.21 pp,
[+7.91, +8.48], an interval adjusted for the confirmation's seven tests; pre-registered), and the reverse mode with 82.4\% against 49.5\%; the canonical encoding
does not improve with length (50.5\% at $T$ 5{,}000). ADD6-large's gain is a gain per update, not per unit of work: in the
determined modes it needs 2.4--2.7$\times$ fewer updates than the better canonical rule (per mode, the better of the
canonical encoding under N0 and under N1) but evaluates about 30 incident terms per update against 5.2--5.8, requiring
2.0--2.4$\times$ more work (1.6$\times$ in the reverse mode; descriptive: the updates from the 6-bit adder development panel on the held-out cases, the
work from Confirmation~C's descriptive rows). It is also a gain per update rather than
per parallel step: ADD6-large needs more colors than the canonical encoding (5 against 3 in the determined modes), and in a
development panel (256 cases per determined mode, none of them held out; N1, $T$ 1{,}000) its three-mode advantage of
+7.84 pp at equal updates falls to +0.26 pp at equal parallel steps, while its reverse-mode advantage holds (+37 pp).
The compact learned energies carry the adder result: ADD6-compact needs 0.74, 0.56, 0.57 and 0.29 of the better canonical
rule's work in the four modes, every interval below 1 (Confirmation~C, pre-registered, fresh streams, all 4{,}096 cases). The maximal-gap
fit ADD6-compact-maxgap on the same support, the gap-maximizing criterion of \citet{onizawa2021high} with equal legal energies, is
indistinguishable from it (work ratios 1.00--1.02 between the two): rounded at the designer's scale both keep the same
67 terms at every word, and at every ladder word but 8 and 9 their integers coincide once common factors are divided
out (magnitudes at most 4), the two scales differing by a factor between 0.97 and 1.13 depending on the word, which a
sampler's inverse temperature (for SSA, its noise scale) absorbs. Both keep the canonical
encoding's coloring and their gains per parallel step (+2.3 and +2.2 pp three-mode, +43.5 and +43.7 pp reverse mode at N1,
$T$ 1{,}000). On the canonical encoding's own 61-term support the learned coefficients ADD6-canonical-refit need 0.48 of the canonical
updates in asum and bsum and 0.24 in the reverse mode at the same work (the 6-bit adder development panel, on the 1{,}024
held-out cases per determined mode). The forward mode is the exception: there ADD6-canonical-refit needs 0.85 [0.73,
1.01] of the canonical updates, unresolved, and its validity levels off at 69\%, with 386 of the 1{,}024 cases below a
success fraction of 0.5 at $T$ 5{,}000 (read after the fact).

\paragraph{Multiplier.} At the standard setting MUL4-P4-full needs a median of 3{,}430 updates to 99\% success against 97{,}667
for the canonical encoding with N0 and 22{,}998 with N1, and the canonical encoding needs 10.8 (N0) and 2.5 (N1)
times MUL4-P4-sig's work; all are upper
bounds, because every arm's minimum lies at the shortest run length. Per-case correctness beats the global equal-energy
constraint (MUL4-P4-global) by +4.7 pp [+0.6, +9.0], confirmed under the registered testing sequence; the effect is heterogeneous across
products and is zero at $T\le 1{,}000$. The ablation ladder (Fig.~\ref{fig:standard}b) shows the maximal-gap
coefficients (MUL4-P3-maxgap, +32 pp over canonical) and the fitting temperature (+11.5 pp from $\beta$ 8 to 16). Fitting on all 15{,}275 terms of interaction order at most four from a zero start at
$\beta$ 16 (MUL4-dense) ties MUL4-P4-full in factor mode at N0 ($-0.3$ pp [$-5.1$, $+4.7$]) and beats it at N1 ($+8.2$ pp
[$+2.7$, $+13.9$]) at $T$ 20{,}000, at 1.4$\times$ the work per update in the significant count and with no parallelism left
($k=n_{\rm free}$); refitted on the same pool at $\beta$ 40 with the per-case gap (MUL4-dense-cw, 1{,}082 terms) it reads
$+1.5$ pp [$-3.1$, $+6.4$] and $+9.7$ pp [$+4.5$, $+15.1$] at 1.6$\times$. Both are exploratory evidence, outside the frozen
confirmation's pre-registration. The column-local pool is a hardware choice with a measured price, not a requirement. In multiply mode MUL4-P4-full reaches 62.1\% at N0 (61.3\% at N1) at $T$ 20{,}000 against 87.3\% for the canonical encoding with N1, and its zero-success cases grow with run length,
although its exact Gibbs legal mass at $\beta$ 32 is 0.91 against the canonical encoding's 0.05: the Gibbs masses do
not predict the ordering of the finite-time SSA validities in this mode.

\begin{figure}[!tbp]
\centering
\includegraphics{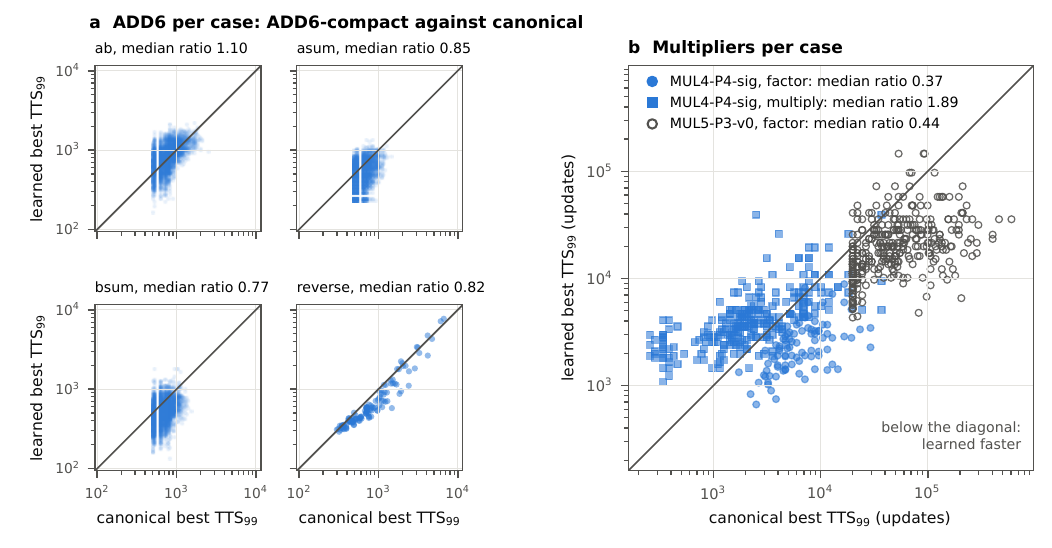}
\caption{\textbf{Per-case $\TTS$ under the best-solver comparator at 25 nodes and 5$\times$5.} Per-case $\TTS$ in single-node
updates (replica updates included), learned best against canonical best, each at its own best family and length;
diagonal: equal. (a) ADD6, ADD6-compact, every case of each mode (4{,}096 per determined mode, 127 sums). (b) MUL4-P4-sig
(90 factor products, 256 multiply cases) and MUL5-P3-v0 (340 products). The median per-case ratio is given per mode;
it differs from the ratio of medians, $\rhostar$. A case solved in at least 99\% of its trajectories has a $\TTS$ equal
to one trajectory's updates, which gives the straight edges in (a).}
\label{fig:si-percase}
\end{figure}

\paragraph{Under equal tuning.} Supplementary Table~\ref{tab:si-25node} gives each structure's best family and
$\rhostar$. The canonical encodings respond strongly to cheap changes: low-noise SSA is their best sampler in the adder's
determined modes, a constant-$\beta$ heat bath in the reverse mode, Metropolis annealing in factor mode at 4$\times$4 and
color-sweep tempering in multiply mode. On the canonical adder, below a noise scale of about 0.3 the SSA trajectories
equal the noise-free ones, and the optimum lies at 0.5. Per update the learned advantage survives in multiplier
factoring (MUL4-P4-sig 0.38), in the adder's reverse mode (ADD6-compact 0.72, ADD6-compact-maxgap 0.68) and in one half-clamped mode (ADD6-compact bsum
0.83); it is unresolved in asum and reversed in the adder's forward mode, in multiply mode and in every cell of the large
support ADD6-large. Per unit of work only ADD6-compact-maxgap in the reverse mode survives (0.80), and the 4$\times$4 multiplier ties (0.99). On the
SSA kernel alone MUL4-P4-sig keeps 0.31 of the canonical encoding's SSA updates. Per case (Supplementary
Fig.~\ref{fig:si-percase}), each structure at its own best family and length, MUL4-P4-sig is faster on 89 of the 90 products
(median per-case ratio 0.37) and MUL5-P3-v0 on 313 of 340 (0.44); in multiply mode the canonical encoding is faster on 203 of
256 cases (1.89); on the adder the learned energy is faster on 111 of the 127 sums (0.82) and on fewer than half of the
forward-mode cases (1.10). Under the width protocol at 4$\times$4 and in a pre-registered confirmation on all 340
products at 5$\times$5, the learned energies need 0.37 [0.30, 0.46] and 0.44 [0.37, 0.48] of the canonical encoding's
best updates and 0.97 [0.78, 1.20] and 0.80 [0.68, 0.87] of its work.
The answer spreads are in Supplementary Figs.~\ref{fig:si-adder-spread} and \ref{fig:si-mult-ladder}
and Supplementary Note~S7.

\begin{table}[!tbp]
\centering
\scriptsize
\caption{\textbf{Best tuned sampler per structure and mode at 25 nodes, and the initial 5$\times$5 fit.} Canonical and learned
best family with its median $\TTS$ in updates, and $\rhostar$ (learned / canonical) in updates and in work
(significant count) with 95\% intervals; all nine families; readings S, U, R as in Table~\ref{tab:series}.
$^\dagger$The width protocol (fewer trajectories per product), whose 4$\times$4 row repeats the comparison under one
protocol with 5$\times$5; the 5$\times$5 row (MUL5-P3-v0) covers all 340 products (pre-registered confirmation).}
\label{tab:si-25node}
\setlength{\tabcolsep}{3pt}
\begin{tabularx}{\textwidth}{@{}>{\raggedright\arraybackslash}p{1.8cm}>{\raggedright\arraybackslash}p{2.1cm}>{\raggedright\arraybackslash}p{2.8cm}>{\raggedright\arraybackslash}p{2.1cm}>{\raggedright\arraybackslash}X>{\raggedright\arraybackslash}X@{}}
\toprule
Circuit, mode & Canonical best (family, TTS99) & Learned energy & Learned best (family, TTS99) & $\rho^\ast$, updates & $\rho^\ast$, work \\
\midrule
\multirow{3}{=}{ADD6 ab} & \multirow{3}{=}{SSA, 701} & ADD6-compact & SA-M, 832 & 1.19 [1.14, 1.19] R & 1.51 [1.45, 1.51] R \\
 &  & ADD6-compact-maxgap & SA-M, 800 & 1.14 [1.11, 1.15] R & 1.32 [1.28, 1.33] R \\
 &  & ADD6-large & SSA, 891 & 1.27 [1.23, 1.27] R & 2.96 [2.87, 2.97] R \\
\multirow{3}{=}{ADD6 asum} & \multirow{3}{=}{SSA, 525} & ADD6-compact & APT-C, 506 & 0.96 [0.90, 1.01] U & 1.21 [1.13, 1.27] R \\
 &  & ADD6-compact-maxgap & SA-M, 538 & 1.02 [0.99, 1.08] U & 1.19 [1.15, 1.25] R \\
 &  & ADD6-large & SA-M, 658 & 1.25 [1.20, 1.32] R & 3.10 [2.97, 3.26] R \\
\multirow{3}{=}{ADD6 bsum} & \multirow{3}{=}{SSA, 624} & ADD6-compact & SA-M, 519 & 0.83 [0.80, 0.96] S & 1.05 [1.01, 1.21] R \\
 &  & ADD6-compact-maxgap & SA-M, 538 & 0.86 [0.83, 1.02] U & 1.00 [0.97, 1.19] U \\
 &  & ADD6-large & SA-M, 658 & 1.05 [1.01, 1.25] R & 2.60 [2.50, 3.09] R \\
\multirow{3}{=}{ADD6 reverse} & \multirow{3}{=}{HB, 714} & ADD6-compact & SA-M, 517 & 0.72 [0.67, 0.78] S & 0.93 [0.86, 1.01] U \\
 &  & ADD6-compact-maxgap & SA-M, 484 & 0.68 [0.64, 0.73] S & 0.80 [0.75, 0.87] S \\
 &  & ADD6-large & SSA, 938 & 1.31 [1.24, 1.48] R & 3.78 [3.57, 4.27] R \\
MUL4 factor & SA-M, 5{,}972 & MUL4-P4-sig & SA-M, 2{,}244 & 0.38 [0.31, 0.45] S & 0.99 [0.81, 1.20] U \\
MUL4 multiply & APT-C, 2{,}092 & MUL4-P4-sig & APT-C, 3{,}793 & 1.81 [1.52, 2.04] R & 6.96 [5.85, 7.84] R \\
\midrule
MUL4 factor$^{\dagger}$ & SA-M, 6{,}317 & MUL4-P4-sig & SA-M, 2{,}325 & 0.37 [0.30, 0.46] S & 0.97 [0.78, 1.20] U \\
MUL5 factor$^{\dagger}$ & APT-C, 45{,}508 & MUL5-P3-v0 & SA-M, 19{,}879 & 0.44 [0.37, 0.48] S & 0.80 [0.68, 0.87] S \\
\bottomrule
\end{tabularx}
\end{table}

\begin{figure}[!tbp]
\centering
\includegraphics{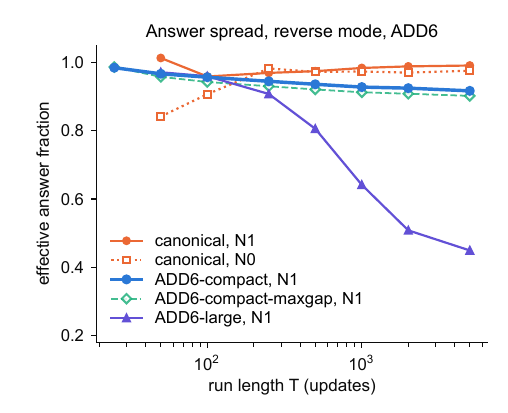}
\caption{\textbf{Answer spread of the 6-bit adder in the reverse mode.} Effective answer fraction over valid input pairs against run
length, in the 6-bit adder development panel (random scan, 127 sums), for the canonical encoding at N1 and N0 and for ADD6-compact, ADD6-compact-maxgap
and ADD6-large at N1.}
\label{fig:si-adder-spread}
\end{figure}

\begin{figure}[!tbp]
\centering
\includegraphics{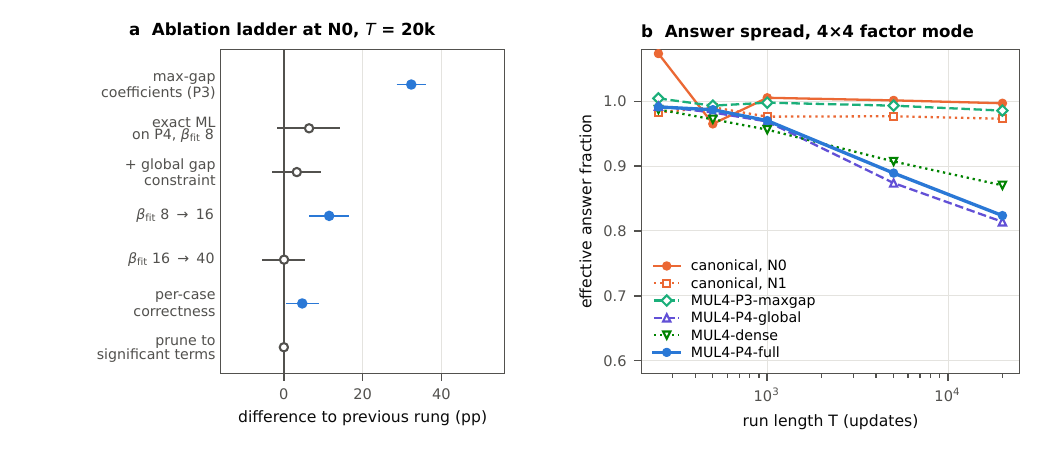}
\caption{\textbf{Ablation ladder and answer spread of the 4$\times$4 multiplier (Confirmation~B).} (a) Descriptive ablation ladder at N0 and
$T$ 20{,}000 (as Fig.~\ref{fig:standard}b): percentage-point differences from the previous rung with 95\% intervals.
(b) Effective answer fraction over factor pairs against run length.}
\label{fig:si-mult-ladder}
\end{figure}

\section*{Supplementary Note S3: the multiplier series in full}
\phantomsection\label{sn:series}

\paragraph{The 4$\times$4 point through the pipeline.} Fig.~\ref{fig:concept}b--f follows MUL4-P3 through
the pipeline. The nodes sit in the multiplier's column layout (a and b above their columns, the carries in the middle,
the product bits below), and a third- or fourth-order term is drawn at its nodes' centroid with thin spokes to them.
The fourth-order terms are the canonical encoding's 14, which P3 holds. The fit moves 252 coefficients by more than
0.01; the prune removes 286 terms, none above $5.3\times10^{-5}$, among them the 109 canonical terms that the fallback
adds back; from the restart the refit returns to within $6.5\times10^{-4}$ of the pruned coefficients (on a fixed
support the objective is convex, with an optimum independent of the start) and leaves the added terms below
$5.1\times10^{-5}$. Every non-significant term, the 109 added back among them, rounds to zero at every word of the hardware ladder, 3 to 17
bits (at its certified word, 7 bits, the network keeps 409 of its 571 terms; Section~\ref{sec:hw} and Supplementary Note~S6). Every stage is read
from the run's stored outputs, each verified by its recorded checksum.

\paragraph{Sensitivity rows.} Supplementary Table~\ref{tab:si-series} gives, for every series point, $\rhostar$ in nonzero work, $\rhopre$, the
descriptive fresh read $\rho_{\rm fresh}$ (Appendix~\ref{app:protocol}), the same-family ratios in both work counts and, on the adder, $\rhosplit$. $\rhopre$ keeps every reading of
$\rhostar$ in updates at all six multiplier points (0.47, 0.49, 0.52 and 0.46 from 4$\times$4 to 7$\times$7 in P3, 0.29 and 0.35 in P4); in the two work
counts it reads 1.04 [0.78, 1.40] and 0.87 [0.65, 1.17], both unresolved, at 6$\times$6 and 0.86 and 0.68 at 7$\times$7. The re-measurement moved two picks, both
learned and both to color-sweep tempering: at MUL5-P3 from SA-M at $T$ 500 to $T$ 1{,}000, and at 6$\times$6 from heat-bath annealing ($\beta_0$ 2.5, $\beta_1$
128) at $T$ 500, which had one success in 32 trials on its median products and whose median at 1{,}024 trials per product is 137{,}542, to $T$ 80{,}000, the
evaluation side's selection as well. The fresh read gives 0.40, 0.48, 0.51 and 0.44 in updates from 4$\times$4 to 7$\times$7 in P3 and 0.29 and 0.31 in P4, all
surviving, and in significant work 0.66, 0.81, 0.86, 0.65, 0.57 and 0.69, all surviving but 6$\times$6's (0.86 [0.68, 1.17]). The fresh learned median at
6$\times$6 sits at the length floor, 79{,}926 updates at $T$ 80{,}000, so that ratio is the floor over the canonical median. In the same-family ratios no family is reversed in either work count on any multiplier panel; in significant
work 5, 9, 6 and 8 families survive at 4$\times$4, 5$\times$5, 6$\times$6 and 7$\times$7 in P3 (of nine each) and
7 and 9 in P4. At 7$\times$7 one family is unresolved in updates, SA-M at the tuning-selected lengths (0.68 [0.33,
1.16]); it was tuned at $T$ 5{,}000 and evaluates best at $T$ 2{,}000, and at each side's evaluation-best length all nine
families survive (0.03 to 0.46). The learned side's tuning medians of HBA, SA-M, SSA, HB and PT-SA there are $\TTS$ at one
success in the 32 tuning trajectories of the median product, so their selections rest on the registered tie-break among
tied cells; none of them is the lowest family or within a factor of 2 of it.

\begin{table}[!tbp]
\centering
\scriptsize
\caption{\textbf{Sensitivity rows of the width series.} Adders in the reverse mode. (a) $\rhostar$ in nonzero work: the comparison of Table~\ref{tab:series} in the nonzero count, every reading as there but 7$\times$7's, which reads 0.68 [0.56, 0.82] S in significant work. $\rhopre$: each side's family and length fixed by its own tuning
selection, low-count tuning cells on the multiplier re-measured first and the adder rows as first measured (Appendix~\ref{app:protocol}), read on the registered evaluation cells with the family held fixed in every replicate. $\rho_{\rm fresh}$: the same choices read on a fresh namespace, descriptive, on the multiplier only. (b) $\rhofam$ in work:
range and counts S/U/R of the same-family ratios in the two work counts. $\rhosplit$, on the adder only: family
and length chosen on one half of the evaluation trajectories and read on the other. Selections run over the nine
families; from 6$\times$6 upward and on the adder the best of six gives the same $\rhopre$ and
$\rhosplit$ in every cell. Readings as in Table~\ref{tab:series}.}
\label{tab:si-series}
\setlength{\tabcolsep}{3pt}
\begin{tabular}{@{}llllll@{}}
\toprule
(a) Structure & \shortstack[l]{$\rhostar$, work\\(nonzero)} & \shortstack[l]{$\rhopre$,\\updates} & \shortstack[l]{$\rhopre$,\\work (sig.)} & \shortstack[l]{$\rho_{\rm fresh}$,\\updates} & \shortstack[l]{$\rho_{\rm fresh}$,\\work (sig.)} \\
\midrule
\mbox{MUL4-P3} & 1.11 [0.90, 1.30] U & 0.47 [0.35, 0.64] S & 0.78 [0.57, 1.05] U & 0.40 [0.33, 0.55] S & 0.66 [0.54, 0.90] S \\
\mbox{MUL5-P3} & 0.88 [0.75, 0.95] S & 0.49 [0.43, 0.56] S & 0.82 [0.72, 0.94] S & 0.48 [0.44, 0.58] S & 0.81 [0.74, 0.97] S \\
\mbox{MUL6-P3-seed1} & 1.04 [0.76, 1.30] U & 0.52 [0.39, 0.70] S & 0.87 [0.65, 1.17] U & 0.51 [0.41, 0.70] S & 0.86 [0.68, 1.17] U \\
\mbox{MUL7-P3-seed2} & 0.86 [0.71, 1.03] U & 0.46 [0.36, 0.56] S & 0.68 [0.53, 0.83] S & 0.44 [0.34, 0.60] S & 0.65 [0.50, 0.88] S \\
\mbox{MUL4-P4} & 0.71 [0.61, 0.92] S & 0.29 [0.23, 0.40] S & 0.59 [0.47, 0.80] S & 0.29 [0.23, 0.38] S & 0.57 [0.47, 0.77] S \\
\mbox{MUL5-P4} & 0.83 [0.77, 0.98] S & 0.35 [0.28, 0.38] S & 0.77 [0.63, 0.85] S & 0.31 [0.28, 0.36] S & 0.69 [0.63, 0.79] S \\
\midrule
\mbox{ADD8-learned} & 1.97 [1.73, 2.20] R & 0.74 [0.65, 0.85] S & 1.97 [1.74, 2.27] R & not read & not read \\
\mbox{ADD16-learned-seed1} & 0.74 [0.65, 0.84] S & 0.32 [0.27, 0.36] S & 0.73 [0.62, 0.84] S & not read & not read \\
\mbox{ADD24-learned-seed2} & 0.64 [0.56, 0.86] S & 0.26 [0.21, 0.33] S & 0.62 [0.50, 0.79] S & not read & not read \\
\mbox{ADD32-learned-seed2} & 0.56 [0.46, 0.73] S & 0.24 [0.19, 0.31] S & 0.53 [0.43, 0.70] S & not read & not read \\
\bottomrule
\end{tabular}

\medskip

\begin{tabular}{@{}llll@{}}
\toprule
(b) Structure & \shortstack[l]{$\rhofam$, work\\(nonzero)} & \shortstack[l]{$\rhofam$, work\\(sig.)} & \shortstack[l]{$\rhosplit$,\\updates} \\
\midrule
\mbox{MUL4-P3} & 0.66--1.08; 0/9/0 & 0.54--0.88; 5/4/0 & --- \\
\mbox{MUL5-P3} & 0.24--0.99; 7/2/0 & 0.20--0.82; 9/0/0 & --- \\
\mbox{MUL6-P3-seed1} & 0.02--1.25; 4/5/0 & 0.01--1.05; 6/3/0 & --- \\
\mbox{MUL7-P3-seed2} & 0.06--1.27; 7/2/0 & 0.05--1.00; 8/1/0 & --- \\
\mbox{MUL4-P4} & 0.45--0.80; 6/3/0 & 0.44--0.80; 7/2/0 & --- \\
\mbox{MUL5-P4} & 0.23--1.00; 5/4/0 & 0.18--0.79; 9/0/0 & --- \\
\midrule
\mbox{ADD8-learned} & 1.48--2.64; 0/0/9 & 1.44--2.56; 0/0/9 & 0.71 [0.62, 0.80] S \\
\mbox{ADD16-learned-seed1} & 0.53--1.66; 4/3/2 & 0.52--1.62; 4/4/1 & 0.30 [0.26, 0.35] S \\
\mbox{ADD24-learned-seed2} & 0.24--1.37; 7/1/1 & 0.23--1.33; 7/1/1 & 0.26 [0.22, 0.33] S \\
\mbox{ADD32-learned-seed2} & 0.04--1.07; 7/2/0 & 0.04--1.04; 7/2/0 & 0.24 [0.19, 0.28] S \\
\bottomrule
\end{tabular}
\end{table}

\begin{figure}[!tbp]
\centering
\includegraphics{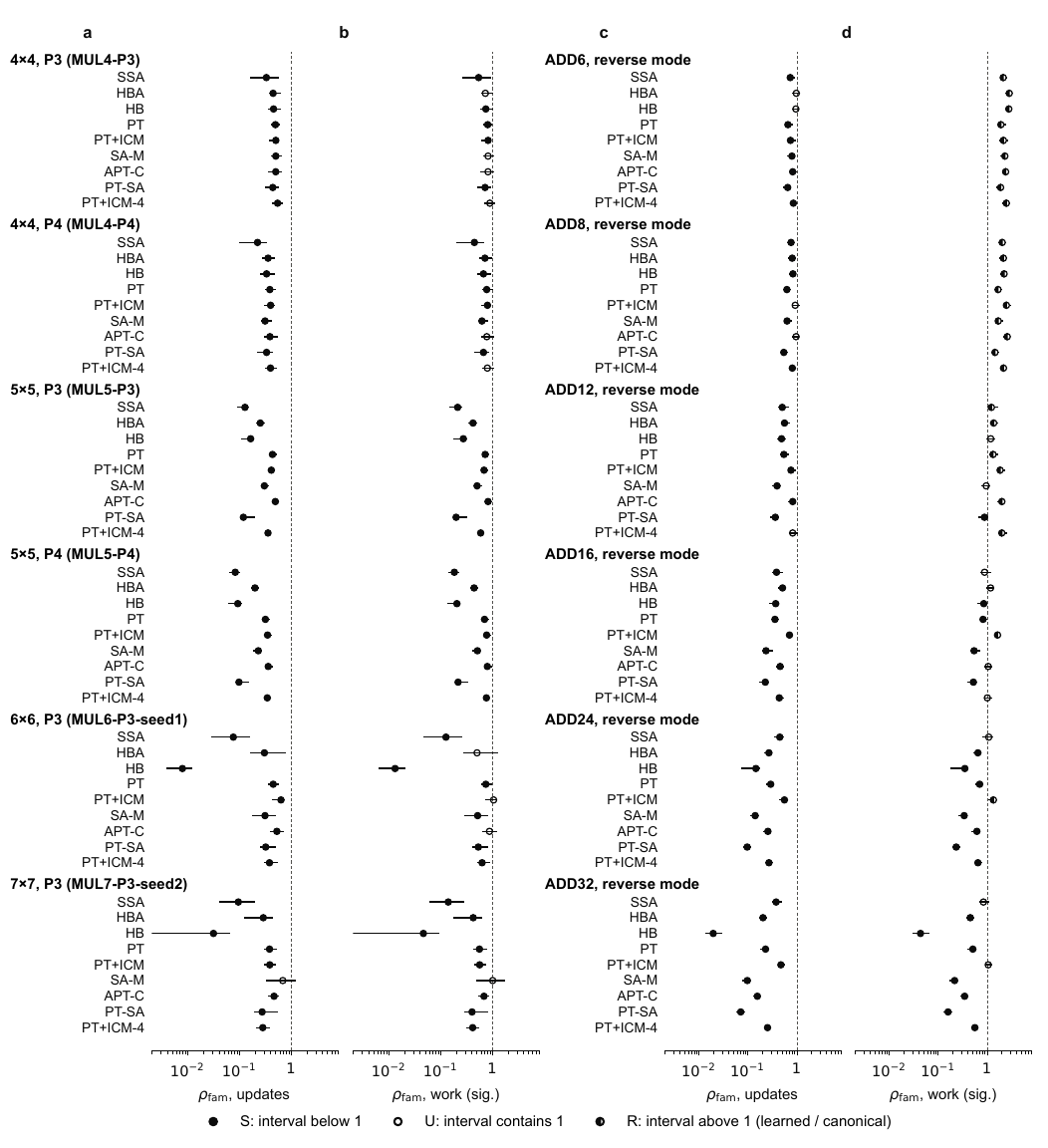}
\caption{\textbf{The per-family comparison as a forest plot.} $\rhofam$ for every comparator family of every series panel (P3
and P4) and of the adder ladder in the reverse mode from 6 to 32 bits (ADD6-learned at 6 bits), learned over canonical, each side at its own
tuning-selected configuration and length, with its 95\% interval on a log axis; dashed, parity. (a, b) Multipliers: updates and work
in the significant count; (c, d) adders: updates and work in the significant count. Filled S, open U, half-filled R. A lower bound of 0 (HB at 7$\times$7) is drawn to the
axis's left edge and read as direction only; the SSA ratios at 7$\times$7 rest on the evaluation's counting resolution
(Supplementary Note~S3).}
\label{fig:si-forest}
\end{figure}

\begin{figure}[!tbp]
\centering
\includegraphics{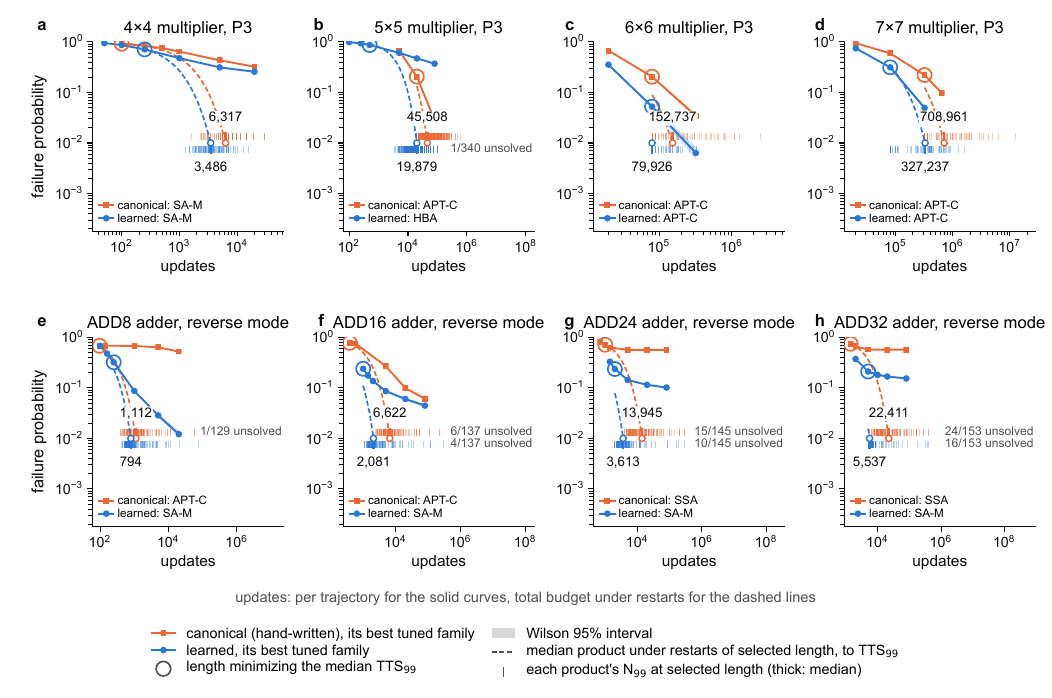}
\caption{\textbf{Failure probability against run length for each energy's best tuned solver.} As Fig.~\ref{fig:failure}a--d, in
the best-of form, for the multiplier (a--d) and for the adder in the reverse mode at 8, 16, 24 and 32 bits (e--h): for each
structure and side the configuration that minimizes the median $\TTS$ (canonical orange, learned blue), with the dashed
restart line ending at the registered median $\TTS$, giving $\rhostar$ as the ratio of the two end points. Each
configuration is selected for restarts at its $T^\ast$, not for success at other lengths; Supplementary Note~S3 lists
them, and Supplementary Note~S4 explains why the adder's best-of curves mislead at long runs.}
\label{fig:si-failure-best}
\end{figure}

\begin{figure}[!tbp]
\centering
\includegraphics{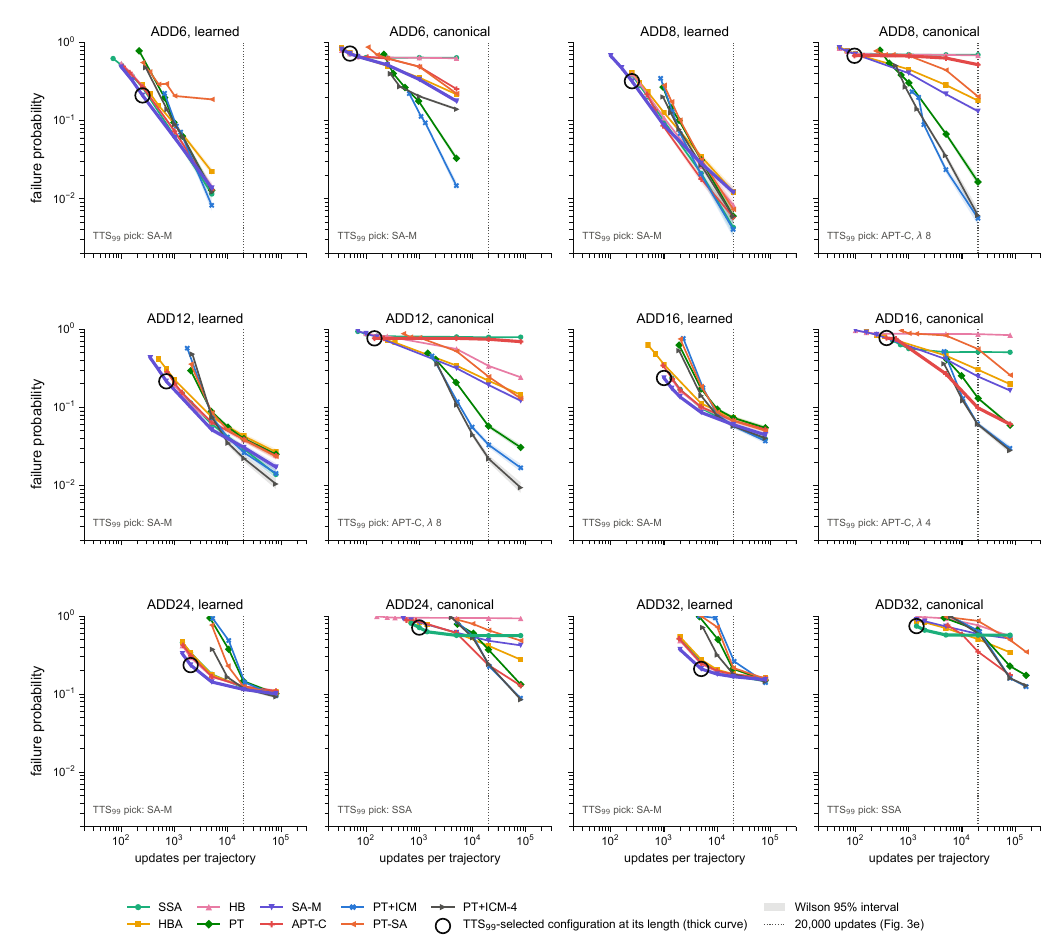}
\caption{\textbf{Single-run failure of the nine sampler families on both adder energies in the reverse mode.} For each width from 6 to 32
bits, the learned energy (left of each pair; ADD6-learned at 6 bits) and the canonical encoding (right): each of the nine at its
tuning-selected configuration, its failure pooled over the panel's sums and trajectories at each evaluated length, on log-log
axes, with Wilson 95\% bands on the pooled counts. Thick and ringed at its length: the configuration that minimizes the
median $\TTS$, which the best-of form (Supplementary Fig.~\ref{fig:si-failure-best}) extends to long runs; its family,
and for APT-C its ladder spacing $\lambda$, is named in each panel. Dotted: 20{,}000 updates, the length of
Fig.~\ref{fig:failure}e.}
\label{fig:si-adder-families}
\end{figure}

\paragraph{Failure curves and families.} Supplementary Fig.~\ref{fig:si-forest} draws every same-family ratio of the
series, P4 included, and of the adder ladder from 6 to 32 bits, in updates and in significant work; Supplementary
Fig.~\ref{fig:si-failure-best} draws the failure curves of each energy's best solver, whose end-point ratio is
$\rhostar$, and Supplementary Table~\ref{tab:si-family-medians} gives the medians behind the per-family comparison. At 6$\times$6 and 7$\times$7 the two seeds' learned structures need 18 and 20 colors in factor mode, as
the canonical encoding does (at 7$\times$7 equal to the clique lower bound), so a color sweep costs them no more
parallel steps. HB's lower bound of 0 at 7$\times$7 (Fig.~\ref{fig:perfamily}d) arises because the canonical encoding's
selection sits at the length cap and its median is infinite in some bootstrap replicates; its bar runs to the panel's
edge. In Fig.~\ref{fig:failure}a--d the family is the one the registered analysis names as the canonical encoding's
best of $N$, whose selections were frozen first, so nothing is chosen on the learned side; the ticks near 1\% mark each
product's $N_{99}=\ln 0.01/\ln(1-v_c)$, the number of runs of length $T^\ast$ that reach 99\% success, and the thick
tick their median. Because $\TTS$ allows restarts, a short run that fails often can cost fewer updates than a long run
that rarely fails; and a replica family can run ladders of different sizes on the two sides. Failure at long runs is
therefore compared between the two energies within one panel and family, not read as $\TTS$. The configurations drawn in
Supplementary Fig.~\ref{fig:si-failure-best} (family, setting, restart length $T^\ast$), canonical then learned: (a)
SA-M, $\beta_0$ 2.5, $\beta_1$ 128, $T^\ast$ 100; SA-M, $\beta_0$ 2.5, $\beta_1$ 32, $T^\ast$ 250. (b) APT-C,
$\beta_{\max}$ 1{,}024, $\lambda$ 2, $T^\ast$ 20{,}000; HBA, $\beta_0$ 2.5, $\beta_1$ 128, $T^\ast$ 500. (c) APT-C,
$\beta_{\max}$ 1{,}024, $\lambda$ 2, $T^\ast$ 80{,}000; APT-C, $\beta_{\max}$ 512, $\lambda$ 2, $T^\ast$ 80{,}000. (d)
APT-C, $\beta_{\max}$ 1{,}024, $\lambda$ 2, $T^\ast$ 320{,}000; APT-C, $\beta_{\max}$ 512, $\lambda$ 2, $T^\ast$
80{,}000. (e) APT-C, $\beta_{\max}$ 64, $\lambda$ 8, $T^\ast$ 100; SA-M, $\beta_0$ 2.5, $\beta_1$ 16, $T^\ast$ 250. (f)
APT-C, $\beta_{\max}$ 64, $\lambda$ 4, $T^\ast$ 350; SA-M, $\beta_0$ 2.5, $\beta_1$ 32, $T^\ast$ 1{,}000. (g) SSA, N1,
$\eta$ 0.5, $T^\ast$ 1{,}000; SA-M, $\beta_0$ 0.5, $\beta_1$ 32, $T^\ast$ 2{,}000. (h) SSA, N1, $\eta$ 0.5, $T^\ast$
1{,}400; SA-M, $\beta_0$ 0.5, $\beta_1$ 32, $T^\ast$ 5{,}000. The text's single-run statements rest on the same-family
form (Fig.~\ref{fig:failure}) and on Supplementary Fig.~\ref{fig:si-adder-families}. In that form the learned structure
also succeeds more often in a single run at every common evaluated length: under Metropolis annealing at 4$\times$4 from
0.134 against 0.079 at 100 updates to 0.744 against 0.677 at 20{,}000, and under color-sweep tempering from 0.688
against 0.328 at 4{,}950 updates to 0.985 against 0.971 at 79{,}950 at 5$\times$5, from 0.646 against 0.328 at 20{,}097
to 0.994 against 0.966 at 319{,}935 at 6$\times$6, and from 0.256 against 0.058 at 20{,}008 to 0.951 against 0.779 at
320{,}128 at 7$\times$7.

\begin{table}[!t]
\centering
\caption{\textbf{The medians behind the per-family comparison.} For every panel of Fig.~\ref{fig:perfamily} and each sampler
family measured on it: the canonical encoding's and the learned structure's evaluation median $\TTS$ in updates, each at
its own tuning-selected configuration and length, and $\rhofam$ with its 95\% case-bootstrap interval and reading; on
each panel's best-of row, each energy's best tuned solver's median and $\rhostar$ (the families named in
Table~\ref{tab:series}), the best of six giving the same value on every best-of row from 6$\times$6 upward and on the adder. A lower bound of 0 (the canonical median infinite in some bootstrap replicates) is read as
direction only. PT+ICM-4's interval at 32 bits collapses to its point because both medians sit at their floor, one run's
length, in every replicate. $^\dagger$Both medians rest on two (learned) and three (canonical) successes in 128 trajectories on the
median product.}
\label{tab:si-family-medians}
\footnotesize
\setlength{\tabcolsep}{3pt}
\begin{tabular}[t]{@{}lrrl@{}}
\toprule
Family & Canonical & Learned & Ratio, updates \\
\midrule
\multicolumn{4}{@{}l}{\textit{Multiplier series in P3, factor mode}} \\
\multicolumn{4}{@{}l}{4$\times$4, P3} \\
SSA & 14{,}621 & 4{,}796 & 0.33 [0.16, 0.54] S \\
HBA & 8{,}351 & 3{,}695 & 0.44 [0.37, 0.61] S \\
HB & 8{,}188 & 3{,}695 & 0.45 [0.36, 0.60] S \\
PT & 13{,}063 & 6{,}409 & 0.49 [0.41, 0.59] S \\
PT+ICM & 13{,}136 & 6{,}550 & 0.50 [0.38, 0.55] S \\
SA-M & 7{,}165 & 3{,}572 & 0.50 [0.41, 0.63] S \\
APT-C & 7{,}547 & 3{,}757 & 0.50 [0.37, 0.65] S \\
PT-SA & 20{,}881 & 9{,}108 & 0.44 [0.32, 0.56] S \\
PT+ICM-4 & 11{,}125 & 6{,}007 & 0.54 [0.43, 0.66] S \\
\textit{best of 9} & 6{,}317 & 3{,}486 & 0.55 [0.45, 0.65] S \\
\addlinespace[2pt]
\multicolumn{4}{@{}l}{5$\times$5, P3} \\
SSA & 292{,}422 & 36{,}947 & 0.13 [0.090, 0.15] S \\
HBA & 81{,}885 & 20{,}471 & 0.25 [0.21, 0.29] S \\
HB & 194{,}176 & 31{,}583 & 0.16 [0.11, 0.18] S \\
PT & 88{,}695 & 38{,}152 & 0.43 [0.39, 0.50] S \\
PT+ICM & 88{,}363 & 36{,}127 & 0.41 [0.37, 0.47] S \\
SA-M & 68{,}975 & 20{,}689 & 0.30 [0.27, 0.36] S \\
APT-C & 45{,}508 & 22{,}252 & 0.49 [0.43, 0.56] S \\
PT-SA & 584{,}845 & 68{,}975 & 0.12 [0.11, 0.19] S \\
PT+ICM-4 & 79{,}800 & 28{,}120 & 0.35 [0.32, 0.37] S \\
\textit{best of 9} & 45{,}508 & 19{,}879 & 0.44 [0.37, 0.47] S \\
\addlinespace[2pt]
\multicolumn{4}{@{}l}{6$\times$6, P3} \\
SSA & 3{,}883{,}511 & 292{,}422 & 0.075 [0.029, 0.15] S \\
HBA & 486{,}598 & 146{,}211 & 0.30 [0.17, 0.76] S \\
HB & 124{,}272{,}368 & 970{,}878 & 0.008 [0.004, 0.012] S \\
PT & 392{,}577 & 174{,}485 & 0.44 [0.37, 0.56] S \\
PT+ICM & 319{,}968 & 201{,}125 & 0.63 [0.44, 0.72] S \\
SA-M & 388{,}351 & 119{,}778 & 0.31 [0.18, 0.48] S \\
APT-C & 152{,}737 & 79{,}926 & 0.52 [0.39, 0.70] S \\
PT-SA & 2{,}168{,}210 & 689{,}682 & 0.32 [0.25, 0.48] S \\
PT+ICM-4 & 319{,}176 & 120{,}182 & 0.38 [0.31, 0.52] S \\
\textit{best of 9} & 152{,}737 & 79{,}926 & 0.52 [0.38, 0.66] S \\
\addlinespace[2pt]
\multicolumn{4}{@{}l}{7$\times$7, P3} \\
SSA & 15{,}534{,}046 & 1{,}462{,}111 & 0.094 [0.042, 0.19] S$^\dagger$ \\
HBA & 3{,}392{,}263 & 970{,}878 & 0.29 [0.12, 0.42] S \\
HB & 751{,}559{,}922 & 23{,}393{,}781 & 0.031 [0, 0.062] S \\
PT & 1{,}587{,}963 & 599{,}915 & 0.38 [0.30, 0.51] S \\
PT+ICM & 1{,}800{,}995 & 686{,}916 & 0.38 [0.30, 0.49] S \\
SA-M & 1{,}427{,}107 & 970{,}878 & 0.68 [0.33, 1.16] U \\
APT-C & 708{,}961 & 327{,}237 & 0.46 [0.36, 0.56] S \\
PT-SA & 10{,}688{,}890 & 2{,}902{,}174 & 0.27 [0.19, 0.53] S \\
PT+ICM-4 & 1{,}724{,}971 & 481{,}736 & 0.28 [0.22, 0.37] S \\
\textit{best of 9} & 708{,}961 & 327{,}237 & 0.46 [0.38, 0.56] S \\
\bottomrule
\end{tabular}
\hfill
\begin{tabular}[t]{@{}lrrl@{}}
\toprule
Family & Canonical & Learned & Ratio, updates \\
\midrule
\multicolumn{4}{@{}l}{\textit{Adder ladder, reverse mode}} \\
\multicolumn{4}{@{}l}{ADD8} \\
SSA & 1{,}306 & 987 & 0.76 [0.65, 0.89] S \\
HBA & 1{,}361 & 1{,}086 & 0.80 [0.67, 0.91] S \\
HB & 1{,}255 & 1{,}036 & 0.83 [0.70, 0.94] S \\
PT & 3{,}073 & 1{,}909 & 0.62 [0.57, 0.73] S \\
PT+ICM & 2{,}356 & 2{,}167 & 0.92 [0.80, 1.12] U \\
SA-M & 1{,}306 & 818 & 0.63 [0.58, 0.77] S \\
APT-C & 1{,}112 & 1{,}065 & 0.96 [0.81, 1.08] U \\
PT-SA & 4{,}859 & 2{,}612 & 0.54 [0.47, 0.63] S \\
PT+ICM-4 & 2{,}377 & 1{,}909 & 0.80 [0.70, 0.90] S \\
\textit{best of 9} & 1{,}112 & 794 & 0.71 [0.63, 0.80] S \\
\addlinespace[2pt]
\multicolumn{4}{@{}l}{ADD16} \\
SSA & 7{,}442 & 2{,}840 & 0.38 [0.32, 0.51] S \\
HBA & 6{,}776 & 3{,}429 & 0.51 [0.43, 0.59] S \\
HB & 8{,}200 & 3{,}010 & 0.37 [0.28, 0.41] S \\
PT & 22{,}323 & 7{,}911 & 0.35 [0.30, 0.41] S \\
PT+ICM & 12{,}721 & 8{,}929 & 0.70 [0.61, 0.78] S \\
SA-M & 9{,}058 & 2{,}107 & 0.23 [0.20, 0.31] S \\
APT-C & 6{,}622 & 2{,}997 & 0.45 [0.39, 0.52] S \\
PT-SA & 40{,}075 & 9{,}017 & 0.23 [0.17, 0.25] S \\
PT+ICM-4 & 12{,}721 & 5{,}528 & 0.43 [0.42, 0.53] S \\
\textit{best of 9} & 6{,}622 & 2{,}081 & 0.31 [0.28, 0.36] S \\
\addlinespace[2pt]
\multicolumn{4}{@{}l}{ADD24} \\
SSA & 13{,}945 & 6{,}203 & 0.44 [0.35, 0.53] S \\
HBA & 22{,}179 & 5{,}919 & 0.27 [0.22, 0.30] S \\
HB & 36{,}263 & 5{,}230 & 0.14 [0.075, 0.17] S \\
PT & 67{,}512 & 19{,}584 & 0.29 [0.24, 0.33] S \\
PT+ICM & 37{,}456 & 20{,}736 & 0.55 [0.45, 0.63] S \\
SA-M & 25{,}863 & 3{,}613 & 0.14 [0.11, 0.16] S \\
APT-C & 21{,}281 & 5{,}409 & 0.25 [0.21, 0.29] S \\
PT-SA & 202{,}982 & 19{,}470 & 0.096 [0.082, 0.11] S \\
PT+ICM-4 & 40{,}593 & 10{,}843 & 0.27 [0.23, 0.32] S \\
\textit{best of 9} & 13{,}945 & 3{,}613 & 0.26 [0.23, 0.35] S \\
\addlinespace[2pt]
\multicolumn{4}{@{}l}{ADD32} \\
SSA & 23{,}160 & 8{,}673 & 0.37 [0.31, 0.48] S \\
HBA & 46{,}500 & 9{,}382 & 0.20 [0.17, 0.24] S \\
HB & 388{,}351 & 7{,}524 & 0.019 [0.014, 0.029] S \\
PT & 105{,}621 & 24{,}028 & 0.23 [0.18, 0.25] S \\
PT+ICM & 79{,}872 & 37{,}456 & 0.47 [0.43, 0.56] S \\
SA-M & 57{,}320 & 5{,}537 & 0.097 [0.077, 0.11] S \\
APT-C & 53{,}571 & 8{,}292 & 0.15 [0.13, 0.17] S \\
PT-SA & 400{,}200 & 28{,}359 & 0.071 [0.060, 0.084] S \\
PT+ICM-4 & 79{,}872 & 19{,}968 & 0.25 [0.25, 0.25] S \\
\textit{best of 9} & 22{,}411 & 5{,}537 & 0.25 [0.20, 0.32] S \\
\bottomrule
\end{tabular}
\end{table}

\paragraph{The best family at 5$\times$5.} At 4$\times$4, 6$\times$6 and 7$\times$7 the learned energy also succeeds
more often at every common length between each energy's best family in $\TTS$; at 5$\times$5 this depends on a
selection. The learned energy's lowest medians, each at its best evaluated length, are a near tie: HBA at 19{,}879
updates (restart length 500), the registered best family, APT-C at 20{,}023 (5{,}000) and SA-M at 20{,}471 (250). At
their tuning-selected lengths they read 20{,}471, 22{,}252 and 20{,}689, so the tie holds at the best evaluated lengths
only. HBA and SA-M, selected for short restarts, level off at 0.529 and 0.344 at 20{,}000 updates and 0.630 and 0.368 at
80{,}000, against 0.795 and 0.971 for the canonical encoding's APT-C. The learned APT-C, at 0.44 [0.38, 0.48] S of the canonical encoding's APT-C in $\TTS$ at its best length (0.49 [0.43, 0.56] S at its tuning-selected length, Table~\ref{tab:si-family-medians}),
reaches 0.949 and 0.985 and leads at all three common lengths; so do PT+ICM-4 and PT+ICM, with medians of 28{,}120 and
36{,}127 updates against the canonical encoding's 45{,}508. The selection is on the learned side only, since the
canonical encoding has no such tie: its next family, SA-M, has a median of 63{,}165 updates, 1.39 times APT-C's. Every
$\rhostar$ in this paper keeps the registered selection.

\paragraph{Against the exact engine's and the first sampled engine's structures.} On the same frozen canonical runs, MUL4-P3 needs 1.50 times
the updates of the exact-engine fit MUL4-P4-sig (reversed), while MUL4-P4's best median (2{,}218) equals MUL4-P4-v1's
(Supplementary Table~\ref{tab:si-engine-a}). From 5$\times$5 to 6$\times$6 the canonical
encoding's best grows 3.36 times and the learned structure's 4.02 times, and from 6$\times$6 to 7$\times$7 4.64 [3.12,
6.53] and 4.09 [3.03, 5.39] times, while the learned structure's significant work per update relative to the canonical
encoding's falls from 1.66 to 1.47.

\paragraph{The multiply mode at 4$\times$4.} A descriptive check reads MUL4-P3 and MUL4-P4 in
the multiply mode under SA-M alone. Each case's SA-M trajectory (a uniform random start, then $T$ random-scan Metropolis
updates at a $\beta$ rising linearly from $\beta_0$ to $\beta_1$) is a Markov chain on the case's $2^{17}$ free-bit
states, and its state distribution is propagated exactly, giving every case's success and $\TTS$ without
sampling. The configuration with the lowest median $\TTS$ is selected over 30 cells of the panel's SA-M grid ($\beta_0$
0.5 or 2.5, $\beta_1$ 8 to 128, $T$ 50, 100 and 250), separately on the 90 products in factor mode and on the 256
multiply cases (both factors clamped, one legal state each), and its work is that median times the mean number of
significant incident terms over the mode's free nodes. Against the canonical encoding's values from an earlier run of
the same computation (6{,}250 updates in factor mode and 3{,}406 in multiply mode, both at $\beta_0$ 2.5, $\beta_1$ 128,
$T$ 100), MUL4-P3 needs 0.53 of the updates at 0.88 of the work in factor mode and 0.88 at 1.95 in multiply mode,
and MUL4-P4 needs 0.38 at 0.76 and 1.10 at 3.35. Both factor-mode selections ($\beta_0$ 2.5, $\beta_1$ 32, $T$ 100)
equal the panels' own SA-M tuning selections (MUL4-P3's evaluation best lies at the neighboring length, $T$ 250); in multiply mode MUL4-P3 selects $\beta_1$ 64 and MUL4-P4 selects $\beta_1$ 32,
both at $\beta_0$ 2.5 and $T$ 100. These are point values without intervals.

\paragraph{Registered expectations.} Supplementary Table~\ref{tab:si-expect} gives the multiplier series' Expectations 1 to 3,
fixed before any fit of the series, and the adder ladder's expectations (c) to (f), with their outcomes.

\begin{table}[!tbp]
\centering
\scriptsize
\caption{\textbf{Registered expectations and their outcomes.} Readings S, U, R as in Table~\ref{tab:series}; $q$, the
learned/canonical work per update. Expectations (a) to (f) form one registered list, fixed before the first sampled
engine's multiplier series and the adder ladder ran: (a) and (b) concern the multiplier series and the prune-and-refit
rule, (c) to (f) the adder ladder. $\exp(H)/|L|$, the exponential of the entropy $H$ of a sum's Gibbs distribution over
its legal completions divided by their number $|L|$, equal to 1 for a uniform distribution.}
\label{tab:si-expect}
\setlength{\tabcolsep}{3pt}
\begin{tabularx}{\textwidth}{@{}>{\raggedright\arraybackslash}p{1.8cm}>{\raggedright\arraybackslash}X>{\raggedright\arraybackslash}p{2.9cm}>{\raggedright\arraybackslash}X@{}}
\toprule
 & Expectation & Registered reading & Outcome \\
\midrule
\multicolumn{4}{@{}l}{\textit{The multiplier series (fixed before any fit; $N$ read as nine at every width)}} \\
Expectation~1 & Every point except 4$\times$4 in P3 keeps the advantage in updates & $\rhostar$ against the canonical
encoding's best of $N$ survives & holds at 4$\times$4 and 5$\times$5 in P4 (0.35, 0.30) and at 5$\times$5 and 6$\times$6
in P3 (0.44, 0.52) and at 7$\times$7 in P3 (0.46 [0.38, 0.56] S) \\
Expectation~2 & 4$\times$4 in P3 does not keep it & $\rhostar$ unresolved or reversed & not met: 0.55 [0.45, 0.65] S \\
Expectation~3 & No resolved width trend & every cross-width ratio of $\rhostar$ among the P3 points contains 1 & not met at
4$\times$4 to 5$\times$5 (0.79 [0.62, 0.96] S, the favorable direction); holds at 5$\times$5 to 6$\times$6 (1.20 [0.91,
1.62] U) and 6$\times$6 to 7$\times$7 (0.88 [0.67, 1.26] U); P4's 4$\times$4 to 5$\times$5, descriptive: 0.85 [0.66, 1.06] U \\
\midrule
\multicolumn{4}{@{}l}{\textit{The adder ladder (registered before its panels; read at 32 bits)}} \\
(c) & reverse mode: the canonical encoding's $\TTS$ grows faster with width than the learned structure's & ratio of ratios
against 6 bits below 1 & holds: 32 against 6 bits 0.32 [0.26, 0.43] S \\
(d) & The forward mode stays with the canonical encoding & $\rhostar$ above 1 at every width & holds: 1.07 to 2.12 \\
(e) & The learned adders keep an even spread of answers in equilibrium & exact $\exp(H)/|L|$ close to 1 & holds by
construction of the equal-legal-energy fit; carries no evidence \\
(f) & The work ratio tracks the updates ratio & $q$ within 0.8 to 1.25 times its 6-bit value at every width & fails at 12,
16, 24 and 32 bits (at 32 bits $q$ is 0.72 to 0.77 times its 6-bit value over the four modes in both work counts) \\
\midrule
\multicolumn{4}{@{}l}{\textit{The first sampled engine's multiplier series, for the record}} \\
(a) & The work and updates ratios come closer with width, with no resolved growth of the advantage & $\rhostar(8\times8)
/\rhostar(5\times5)$; read descriptively at 6$\times$6 & on MUL6-P3-v1 $q$ rises from 1.82 to 2.00
and the 6$\times$6 against 5$\times$5 ratios are unresolved, leaving (a) unmet descriptively; the registered 8$\times$8
endpoint was not run \\
(b) & A fit must be pruned and then refitted & applied as a rule & applied to every point of both series and every adder
fit \\
\bottomrule
\end{tabularx}
\end{table}

\section*{Supplementary Note S4: the adder ladder in full}
\phantomsection\label{sn:adder}

\begin{figure}[!tbp]
\centering
\includegraphics{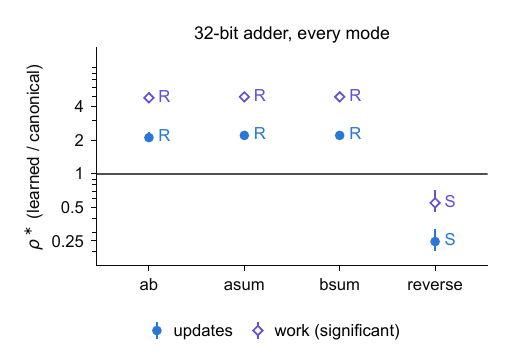}
\caption{\textbf{The 32-bit adder in each mode under equal tuning.} $\rhostar$, learned over canonical with each energy's best of
nine tuned samplers, in updates (filled) and in significant work (open), 95\% intervals, readings S and R.}
\label{fig:si-add32-modes}
\end{figure}

\paragraph{Every mode.} Supplementary Table~\ref{tab:si-adder} gives $\rhostar$ in updates in every mode at every width,
and Supplementary Fig.~\ref{fig:si-add32-modes} draws the 32-bit adder in each mode.
In the reverse mode the advantage in updates holds from 6 to 32 bits; in the determined modes the canonical encoding scales
better. The forward mode ab is reversed at every width, from 1.07 at 6 bits to 2.12 at 32 bits, and the two
half-clamped modes move from favoring the learned energy at 6 bits (0.77 and 0.81) through unresolved at 8 bits to
reversed from 12 bits on (2.22 and 2.22 at 32 bits). From 24 to 32 bits the reversal widens in asum and bsum (steps 1.40
[1.20, 1.59] and 1.33 [1.15, 1.55], reversed), because the canonical encoding's SSA stays near the same run length
while the learned structure's best, Metropolis annealing, grows; ab is unchanged (1.02 [0.90, 1.17]) and the reverse mode's
step is 0.95 [0.66, 1.22]. Under $\rhopre$ and $\rhosplit$ the reverse-mode readings in updates are those of $\rhostar$ at
8, 16, 24 and 32 bits (Supplementary Table~\ref{tab:si-series}); with the sampler fixed, every family survives in updates from 16
bits on, eight of the nine at 12 bits and seven at 6 and 8 bits, the rest unresolved (HBA and HB at 6 bits, PT+ICM and
APT-C at 8, PT+ICM-4 at 12). In nonzero work the same-family ratios are reversed in every family at 8 bits, and from 16
bits on survive in 4, 7 and 7 of the nine families (reversed in HBA and PT+ICM at 16 bits and in PT+ICM at 24). Of the
204 selections of the three added families 23 sit at an edge of their grids (18 PT+ICM-4 at the top $\lambda$, each with
two replicas, 4 at the top $\beta$ and 1 at the top increment), none for a family that is any structure's best.

\begin{table}[!tbp]
\centering
\scriptsize
\caption{\textbf{The adder ladder in every mode.} $\rhostar$ in updates, learned over canonical, best of nine (the best of six gives the same value in every cell), by mode and width,
and the reverse mode's ratio in nonzero work; terms canonical / learned. Readings as in Table~\ref{tab:series}. The 6-bit
learned structure is ADD6-learned, the ladder's own fit, not ADD6-compact of Table~\ref{tab:si-25node}.}
\label{tab:si-adder}
\setlength{\tabcolsep}{2.4pt}
\begin{tabular}{@{}llllllll@{}}
\toprule
Width & Terms (c / l) & ab & asum & bsum & reverse & reverse, work (nonzero) \\
\midrule
ADD6 & 61 / 141 & 1.07 [1.04, 1.07] R & 0.77 [0.75, 0.77] S & 0.81 [0.77, 0.81] S & 0.77 [0.69, 0.85] S & 2.35 [2.10, 2.60] R \\
ADD8 & 81 / 169 & 1.40 [1.27, 1.45] R & 1.11 [1.000, 1.14] U & 1.02 [0.94, 1.11] U & 0.71 [0.63, 0.80] S & 1.97 [1.73, 2.20] R \\
ADD12 & 121 / 228 & 1.67 [1.56, 1.80] R & 1.39 [1.30, 1.53] R & 1.50 [1.35, 1.59] R & 0.57 [0.47, 0.63] S & 1.41 [1.16, 1.58] R \\
ADD16 & 161 / 287 & 2.04 [1.93, 2.30] R & 1.49 [1.34, 1.62] R & 1.83 [1.59, 1.92] R & 0.31 [0.28, 0.36] S & 0.74 [0.65, 0.84] S \\
ADD24 & 241 / 462 & 2.08 [1.91, 2.25] R & 1.58 [1.50, 1.70] R & 1.66 [1.50, 1.77] R & 0.26 [0.23, 0.35] S & 0.64 [0.56, 0.86] S \\
ADD32 & 321 / 557 & 2.12 [1.94, 2.37] R & 2.22 [2.00, 2.41] R & 2.22 [2.00, 2.40] R & 0.25 [0.20, 0.32] S & 0.56 [0.46, 0.73] S \\
\bottomrule
\end{tabular}
\end{table}

\paragraph{The standard setting.} At 8 to 16 bits the canonical adder's failure in the reverse mode at noise scale 1 is
trapping, not a lack of run length: per sum, its success over 32 trajectories at N1 at the longest length of the tuning
runs is spread over the whole range (mean 0.510, 0.484 and 0.498 at 8, 12 and 16 bits, with 5, 10 and 14 sums never and
3, 5 and 7 always solved), while the learned fit solves 0.993, 0.986 and 0.959, with 124 to 127 sums always solved; at
that noise a trajectory's outcome on the canonical energy appears to be decided early, and tempering with enough
replicas escapes it (the best APT-C ladder of the tuning grid reaches 0.977 validity at 12 bits).

\paragraph{A single run at a fixed length.} Fig.~\ref{fig:failure}e gives the pooled failure probability of one run in
the reverse mode near 20{,}000 updates, each family at its own tuned configuration, and Supplementary
Table~\ref{tab:si-adder-single} its success over the nine families and, beside them, over the six evaluated first;
Supplementary Fig.~\ref{fig:si-adder-families} gives each family's failure curve at every evaluated length on both
energies at all six widths. The table names each best family's actual updates, since the two sides' lengths can differ
by a few percent (near 20{,}000 updates, 19{,}968 and 19{,}584 at 16 bits, 20{,}000 and 20{,}736 at 24 bits), and the
Wilson bars pool trajectories and do not separate sums, so differences of a few thousandths are read as near parity.

\begin{table}[!tbp]
\centering
\scriptsize
\caption{\textbf{Single-run success of the adder in the reverse mode.} The ladder's structures (ADD6-learned at 6 bits).
Success of one run pooled over the panel's sums and trajectories, each family at its tuning-selected configuration and at its
evaluated length nearest the stated length and within 10\% of it (4{,}608 to 5{,}376, 19{,}584 to 21{,}504 and 79{,}488
to 80{,}640 updates), at every length evaluated at that width: Metropolis annealing (SA-M) on both energies; and, in two
rows per length, each energy's best sampler at that length over all nine families (set 9) and over the six evaluated
first (set 6), with its family and actual updates, with $\Delta$, the difference in success, learned minus canonical,
and Ratio, the failure ratio, canonical over learned; and the range of success over that set. At 5{,}000 updates the
canonical encoding's PT+ICM-4 at 24 bits and PT+ICM and PT+ICM-4 at 32 bits have no evaluated length within 10\%
(nearest 4{,}032 and 6{,}144 updates, success 0.043, 0.006 and 0.010) and are not counted. A $\Delta$ of a few
thousandths is read as near parity. Fig.~\ref{fig:failure}e draws one minus the SA-M columns and the best of nine at
20{,}000 updates from 8 to 32 bits.}
\label{tab:si-adder-single}
\setlength{\tabcolsep}{2.4pt}
\begin{tabular}{@{}lllllllllllll@{}}
\toprule
 & & \multicolumn{4}{c}{SA-M} & & \multicolumn{4}{c}{Best of that set at that length} & \multicolumn{2}{c}{Range over the set} \\
\cmidrule(lr){3-6}\cmidrule(lr){8-11}\cmidrule(l){12-13}
Width & Length & Learned & Canonical & $\Delta$ & Ratio & Set & Learned & Canonical & $\Delta$ & Ratio & Learned & Canonical \\
\midrule
\multirow{2}{*}{ADD6} & \multirow{2}{*}{5{,}000} & \multirow{2}{*}{0.986} & \multirow{2}{*}{0.822} & \multirow{2}{*}{0.164} & \multirow{2}{*}{13.0} & 9 & 0.992 (PT+ICM, 4{,}968) & 0.985 (PT+ICM, 4{,}968) & 0.006 & 1.78 & 0.813--0.992 & 0.361--0.985 \\
 &  &  &  &  &  & 6 & 0.988 (SSA, 5{,}000) & 0.967 (PT, 4{,}968) & 0.021 & 2.87 & 0.978--0.988 & 0.361--0.967 \\
\addlinespace[2pt]
\multirow{4}{*}{ADD8} & \multirow{2}{*}{5{,}000} & \multirow{2}{*}{0.972} & \multirow{2}{*}{0.782} & \multirow{2}{*}{0.190} & \multirow{2}{*}{7.7} & 9 & 0.982 (APT-C, 4{,}992) & 0.976 (PT+ICM, 4{,}992) & 0.006 & 1.32 & 0.965--0.982 & 0.303--0.976 \\
 &  &  &  &  &  & 6 & 0.982 (APT-C, 4{,}992) & 0.933 (PT, 5{,}040) & 0.049 & 3.77 & 0.965--0.982 & 0.303--0.933 \\
 & \multirow{2}{*}{20{,}000} & \multirow{2}{*}{0.988} & \multirow{2}{*}{0.868} & \multirow{2}{*}{0.120} & \multirow{2}{*}{10.9} & 9 & 0.996 (PT+ICM, 19{,}872) & 0.994 (PT+ICM, 19{,}968) & 0.002 & 1.39 & 0.988--0.996 & 0.303--0.994 \\
 &  &  &  &  &  & 6 & 0.996 (SSA, 20{,}000) & 0.984 (PT, 20{,}016) & 0.012 & 3.82 & 0.988--0.996 & 0.303--0.984 \\
\addlinespace[2pt]
\multirow{6}{*}{ADD12} & \multirow{2}{*}{5{,}000} & \multirow{2}{*}{0.948} & \multirow{2}{*}{0.683} & \multirow{2}{*}{0.265} & \multirow{2}{*}{6.1} & 9 & 0.948 (SA-M, 5{,}000) & 0.893 (PT+ICM-4, 5{,}040) & 0.056 & 2.07 & 0.911--0.948 & 0.201--0.893 \\
 &  &  &  &  &  & 6 & 0.948 (SA-M, 5{,}000) & 0.791 (PT, 4{,}896) & 0.157 & 4.04 & 0.911--0.948 & 0.201--0.791 \\
 & \multirow{2}{*}{20{,}000} & \multirow{2}{*}{0.969} & \multirow{2}{*}{0.807} & \multirow{2}{*}{0.163} & \multirow{2}{*}{6.3} & 9 & 0.978 (PT+ICM-4, 20{,}160) & 0.978 (PT+ICM-4, 20{,}160) & 0.000 & 0.99 & 0.957--0.978 & 0.210--0.978 \\
 &  &  &  &  &  & 6 & 0.971 (SSA, 20{,}000) & 0.942 (PT, 19{,}872) & 0.028 & 1.96 & 0.957--0.971 & 0.210--0.942 \\
 & \multirow{2}{*}{80{,}000} & \multirow{2}{*}{0.983} & \multirow{2}{*}{0.878} & \multirow{2}{*}{0.105} & \multirow{2}{*}{7.1} & 9 & 0.990 (PT+ICM-4, 79{,}920) & 0.991 (PT+ICM-4, 79{,}920) & $-$0.001 & 0.90 & 0.973--0.990 & 0.207--0.991 \\
 &  &  &  &  &  & 6 & 0.986 (SSA, 80{,}000) & 0.969 (PT, 80{,}064) & 0.017 & 2.22 & 0.973--0.986 & 0.207--0.969 \\
\addlinespace[2pt]
\multirow{6}{*}{ADD16} & \multirow{2}{*}{5{,}000} & \multirow{2}{*}{0.914} & \multirow{2}{*}{0.586} & \multirow{2}{*}{0.328} & \multirow{2}{*}{4.8} & 9 & 0.914 (SA-M, 5{,}000) & 0.732 (APT-C, 4{,}992) & 0.182 & 3.12 & 0.814--0.914 & 0.129--0.732 \\
 &  &  &  &  &  & 6 & 0.914 (SA-M, 5{,}000) & 0.732 (APT-C, 4{,}992) & 0.182 & 3.12 & 0.827--0.914 & 0.129--0.732 \\
 & \multirow{2}{*}{20{,}000} & \multirow{2}{*}{0.940} & \multirow{2}{*}{0.752} & \multirow{2}{*}{0.188} & \multirow{2}{*}{4.1} & 9 & 0.942 (PT+ICM-4, 19{,}968) & 0.939 (PT+ICM-4, 19{,}584) & 0.003 & 1.06 & 0.926--0.942 & 0.133--0.939 \\
 &  &  &  &  &  & 6 & 0.941 (SSA, 20{,}000) & 0.902 (APT-C, 19{,}968) & 0.039 & 1.67 & 0.926--0.941 & 0.133--0.902 \\
 & \multirow{2}{*}{80{,}000} & \multirow{2}{*}{0.956} & \multirow{2}{*}{0.836} & \multirow{2}{*}{0.120} & \multirow{2}{*}{3.7} & 9 & 0.963 (PT+ICM, 79{,}872) & 0.972 (PT+ICM-4, 79{,}488) & $-$0.009 & 0.76 & 0.945--0.963 & 0.159--0.972 \\
 &  &  &  &  &  & 6 & 0.961 (SSA, 80{,}000) & 0.941 (PT, 80{,}064) & 0.020 & 1.50 & 0.945--0.961 & 0.159--0.941 \\
\addlinespace[2pt]
\multirow{6}{*}{ADD24} & \multirow{2}{*}{5{,}000} & \multirow{2}{*}{0.857} & \multirow{2}{*}{0.387} & \multirow{2}{*}{0.470} & \multirow{2}{*}{4.3} & 9 & 0.857 (SA-M, 5{,}000) & 0.435 (SSA, 5{,}000) & 0.422 & 3.95 & 0.054--0.857 & 0.055--0.435 \\
 &  &  &  &  &  & 6 & 0.857 (SA-M, 5{,}000) & 0.435 (SSA, 5{,}000) & 0.422 & 3.95 & 0.054--0.857 & 0.055--0.435 \\
 & \multirow{2}{*}{20{,}000} & \multirow{2}{*}{0.885} & \multirow{2}{*}{0.518} & \multirow{2}{*}{0.368} & \multirow{2}{*}{4.2} & 9 & 0.885 (SA-M, 20{,}000) & 0.774 (PT+ICM, 20{,}736) & 0.111 & 1.97 & 0.853--0.885 & 0.058--0.774 \\
 &  &  &  &  &  & 6 & 0.885 (SA-M, 20{,}000) & 0.764 (APT-C, 19{,}872) & 0.121 & 2.06 & 0.853--0.885 & 0.058--0.764 \\
 & \multirow{2}{*}{80{,}000} & \multirow{2}{*}{0.899} & \multirow{2}{*}{0.579} & \multirow{2}{*}{0.320} & \multirow{2}{*}{4.2} & 9 & 0.908 (PT+ICM-4, 80{,}064) & 0.915 (PT+ICM-4, 80{,}640) & $-$0.007 & 0.92 & 0.890--0.908 & 0.068--0.915 \\
 &  &  &  &  &  & 6 & 0.904 (SSA, 80{,}000) & 0.872 (APT-C, 80{,}064) & 0.033 & 1.34 & 0.890--0.904 & 0.068--0.872 \\
\addlinespace[2pt]
\multirow{6}{*}{ADD32} & \multirow{2}{*}{5{,}000} & \multirow{2}{*}{0.790} & \multirow{2}{*}{0.264} & \multirow{2}{*}{0.526} & \multirow{2}{*}{3.5} & 9 & 0.790 (SA-M, 5{,}000) & 0.428 (SSA, 5{,}000) & 0.362 & 2.72 & 0.001--0.790 & 0.027--0.428 \\
 &  &  &  &  &  & 6 & 0.790 (SA-M, 5{,}000) & 0.428 (SSA, 5{,}000) & 0.362 & 2.72 & 0.022--0.790 & 0.050--0.428 \\
 & \multirow{2}{*}{20{,}000} & \multirow{2}{*}{0.833} & \multirow{2}{*}{0.411} & \multirow{2}{*}{0.422} & \multirow{2}{*}{3.5} & 9 & 0.833 (SA-M, 20{,}000) & 0.651 (APT-C, 19{,}968) & 0.182 & 2.09 & 0.737--0.833 & 0.132--0.651 \\
 &  &  &  &  &  & 6 & 0.833 (SA-M, 20{,}000) & 0.651 (APT-C, 19{,}968) & 0.182 & 2.09 & 0.789--0.833 & 0.239--0.651 \\
 & \multirow{2}{*}{80{,}000} & \multirow{2}{*}{0.847} & \multirow{2}{*}{0.486} & \multirow{2}{*}{0.361} & \multirow{2}{*}{3.4} & 9 & 0.859 (PT+ICM, 80{,}640) & 0.840 (PT+ICM-4, 79{,}872) & 0.019 & 1.14 & 0.838--0.859 & 0.430--0.840 \\
 &  &  &  &  &  & 6 & 0.854 (SSA, 80{,}000) & 0.825 (APT-C, 79{,}872) & 0.029 & 1.20 & 0.838--0.854 & 0.430--0.825 \\
\bottomrule
\end{tabular}
\end{table}

The main text states the difference, the canonical encoding's failure minus the learned energy's (equal to the
difference in success). At 20{,}000 updates, at 8, 12, 16, 24 and 32 bits, it is 0.120, 0.163, 0.188, 0.368 and 0.422
under SA-M, with failure ratios, canonical over learned, of 11, 6.3, 4.1, 4.2 and 3.5; 0.002, 0.000, 0.003, 0.111 and
0.182 between each energy's best of nine at that length, with ratios of 1.39, 0.99, 1.06, 1.97 and 2.09; and 0.012,
0.028, 0.039, 0.121 and 0.182 between the best of the six, with ratios of 3.82, 1.96, 1.67, 2.06 and 2.09. No trend in
width is read from these ratios: the last two sets differ only through the comparator, PT+ICM and PT+ICM-4 closing the
gap on the canonical encoding at up to 16 bits, and at 8, 12 and 16 bits the same family is the best on both energies.
Within each of the nine tuned families near 20{,}000 updates the learned energy leads at every width, except PT+ICM and
PT+ICM-4 at 8 and 16 bits and PT+ICM-4 at 12 bits, which are near parity (at most 0.004 either way).

At 5{,}000 updates the differences at 8, 16, 24 and 32 bits are 0.190, 0.328, 0.470 and 0.526 under SA-M, 0.006, 0.182,
0.422 and 0.362 between the best of nine, and 0.049, 0.182, 0.422 and 0.362 between the best of six.

At 80{,}000 updates the canonical encoding leads at 16 and 24 bits: its best, PT+ICM-4, succeeds at 0.972 and 0.915
against the learned energy's 0.963 (PT+ICM) and 0.908 (PT+ICM-4), differences of 0.009 and 0.007. The two are near
parity at 12 bits (0.991 and 0.990, PT+ICM-4 on both), and the learned energy leads by 0.019 at 32 bits (0.859 against
0.840). Over the best of six the learned energy leads at 12, 16, 24 and 32 bits by 0.017, 0.020, 0.033 and 0.029. Within
PT+ICM and PT+ICM-4 the canonical encoding leads at 16 and 24 bits, by 0.007 to 0.012. The 6-bit adder was evaluated only up to about 5{,}000
updates. In Fig.~\ref{fig:failure}e each family runs at its evaluated length nearest 20{,}000 updates (19{,}584 to
21{,}504); the best of nine is the smallest failure (the largest success) over the nine tuned configurations, its family
named; $\Delta$ is the SA-M difference, the canonical encoding's failure minus the learned energy's; the axis is one
minus the success, and the Wilson bars \citep{wilson1927probable}, one minus the success's bounds, mostly lie within the
markers.

\paragraph{Why the best-of curves mislead at long runs.} The best-of form of the curves (Supplementary
Fig.~\ref{fig:si-failure-best}e--h) extends each energy's $\TTS$-selected configuration to long runs, and those
configurations are selected for restarts, not for one long run. The canonical encoding's selections in the reverse mode are
APT-C with $\lambda=8$ at 8 and 12 bits, where neighboring replicas spaced $\lambda$ apart in thermodynamic length
swap with a probability of about $\mathrm{erfc}(\lambda/2)=1.5\times10^{-8}$ in the Gaussian approximation of the
ladder's design \citep{earl2005parallel}, so that its replicas practically never exchange; APT-C with $\lambda=4$ at 16 bits, a swap probability
of about 0.5\%; and SSA at 24 and 32 bits (SA-M at 6 bits). Only the 16-bit selection exchanges replicas, and its
success keeps rising to 0.939 at 80{,}000 updates, while the others level off: 0.479 at 20{,}000 updates at 8 bits (the
longest length evaluated there), and 0.305, 0.438 and 0.430 at about 80{,}000 updates at 12, 24 and 32 bits. The same
canonical encoding reaches 0.984, 0.969 and 0.941 at 8, 12 and 16 bits with exchange tempering (PT), and 0.872 and 0.825
at 24 and 32 bits with APT-C, at the longest lengths, and 0.994, 0.991, 0.972, 0.915 and 0.840 at 8, 12, 16, 24 and 32
bits with PT+ICM-4, at about 20{,}000 updates at 8 bits and about 80{,}000 elsewhere. The best-of form therefore
overstates the single-run gap against the per-length best of nine at every width from 8 to 32 bits;
Fig.~\ref{fig:failure}e compares one fixed sampler, and each energy's best sampler, at one length instead.

\paragraph{A common ceiling.} We read the canonical reverse mode as holding metastable traps that only replica exchange
leaves, at a cost that does not pay off at the $\TTS$-optimal restart lengths: at the longest length evaluated on both
energies the learned energy's nine samplers end within about 0.02 of one another from 8 to 32 bits, and the canonical
encoding's range from 0.07 to 0.99 (Supplementary Table~\ref{tab:si-adder-single}; at 6 bits, evaluated only to about
5{,}000 updates, the learned energy's PT-SA ends at 0.81 and its other eight within 0.02). At about 80{,}000 updates,
with every one of the learned energy's nine samplers, about a tenth of the sums at 24 bits (14 or 15 of 145) and about a
sixth at 32 bits (22 to 25 of 153) stay below a success of one half. The canonical encoding's PT and APT-C leave similar
shares (18 of 145 each, 28 and 25 of 153), as do its PT+ICM and PT+ICM-4 (14 and 13 of 145, 24 of 153 each); both
energies approach a common ceiling set by a minority of hard sums, with the canonical encoding's best ahead by 0.009 and
0.007 at 16 and 24 bits and the learned energy's by 0.019 at 32 bits.

\paragraph{Work.} The learned adder's work per update in the reverse mode falls from 14.4 incident terms at 8 bits to 12.1
at 32 bits (nonzero count) while the canonical encoding's stays at 5.2 to 5.3, moving the work ratio from a reversal at
6 and 8 bits (2.35 and 1.97) through 1.41 at 12 bits to 0.74, 0.64 and 0.56 from 16 to 32 bits. Expectation (f), that the
learned/canonical work per update stays within 0.8 to 1.25 times its 6-bit value, fails at 12, 16, 24 and 32 bits. In significant work, the count the main text reads, the reverse-mode ratio is 1.91, 0.73, 0.62 and 0.55 at 8, 16, 24 and 32 bits with the same readings; at 32 bits the determined modes read 4.82 to 4.92 in significant work and 4.95 to 5.04 in nonzero work.

\section*{Supplementary Note S5: coefficients and support at 4\texorpdfstring{$\times$}{x}4 in full}
\phantomsection\label{sn:e5}

The equal-workload comparison refits chosen supports of the 4$\times$4 multiplier with the exact engine's recipe, audits each
fit exactly (no case broken in any), and compares the best of nine families under the width protocol on the 90 products
(Supplementary Table~\ref{tab:si-e5}); work is counted over the factor-mode free nodes, and every term of every structure
in the comparison is significant. On the canonical encoding's own 308 terms, which the prune leaves at 289 (MUL4-canonical-refit), the fitted
coefficients need 0.55 [0.47, 0.69] of the canonical encoding's best updates and 0.51 [0.44, 0.64] of its work, and the
two medians' intervals do not overlap (3{,}449 [2{,}952, 4{,}082] against 6{,}317 [5{,}125, 7{,}165]). Adding terms in
the order of MUL4-P4-sig's magnitudes buys fewer updates at a roughly proportional cost in work up to about 400 terms (the largest
308 and 400 terms, refitted as MUL4-top308 and MUL4-top400: 0.45 and 0.41 of the canonical updates at work of 49.5 and 63.8) and little beyond it.
Choosing the support forward by the objective's first-order score, with a refit after each batch, is slower than
truncation at equal size (MUL4-forward400, 400 terms: 1.26 [1.06, 1.51] times the updates) and shares only 33\% and 45\% of MUL4-P4-sig's
largest 400 and 588 terms. Pool P3 alone (MUL4-P3-exact: column-local terms to interaction order three plus the canonical terms, among them
the canonical encoding's 14 of order four; 449 terms after the prune) needs 0.57 [0.45, 0.66] of the canonical updates
at 0.92 [0.74, 1.07] of its work and 1.54 [1.26, 1.76] times MUL4-P4-sig's updates, showing that P4's column-local
fourth-order terms matter for updates at 4$\times$4; forward selection to 588 terms (MUL4-forward588, 466 after the prune) needs 0.41 [0.34, 0.51] of
the canonical updates at 0.75 [0.63, 0.94] of its work.

\begin{table}[!tbp]
\centering
\scriptsize
\caption{\textbf{Equal-workload frontier on the 4$\times$4 multiplier.} Factor mode in the equal-workload comparison under the width protocol (90
products), best of nine families, re-selected in every bootstrap replicate. Terms: the support after the prune; Work:
incident terms per update over the factor-mode free nodes (every term significant in every row); $\rhostar$ against the
canonical encoding's best of nine with 95\% product-bootstrap intervals; Work $\times$ $\TTS$ in thousands; Frontier:
no structure of the comparison, including the two arms given only in Supplementary Note~S5 (MUL4-P3-exact and MUL4-forward588), has both a work per update and a
median no larger, with one of them smaller (point estimates).}
\label{tab:si-e5}
\setlength{\tabcolsep}{2pt}
\begin{tabularx}{\textwidth}{@{}>{\raggedright\arraybackslash}p{2.65cm}>{\raggedright\arraybackslash}X>{\raggedleft\arraybackslash}p{0.82cm}>{\raggedleft\arraybackslash}p{0.7cm}>{\raggedright\arraybackslash}p{3.35cm}>{\raggedright\arraybackslash}p{2.4cm}>{\raggedright\arraybackslash}p{2.4cm}>{\raggedleft\arraybackslash}p{0.95cm}>{\raggedright\arraybackslash}p{1.05cm}@{}}
\toprule
Structure & Support & Terms & Work & Best of nine (family, $\TTS$ [95\% CI]) & $\rhostar$, updates & $\rhostar$, work & Work $\times$ $\TTS$ & Frontier \\
\midrule
\mbox{MUL4-canonical} & hand-written & 308 & 38.1 & SA-M, 6{,}317 [5{,}125, 7{,}165] & 1 & 1 & 241k & no \\
\mbox{MUL4-canonical-refit} & canonical's, refitted & 289 & 35.5 & HBA, 3{,}449 [2{,}952, 4{,}082] & 0.55 [0.47, 0.69] S & 0.51 [0.44, 0.64] S & 122k & yes \\
\mbox{MUL4-top308} & largest 308, refitted & 299 & 49.5 & SA-M, 2{,}866 [2{,}383, 3{,}340] & 0.45 [0.37, 0.56] S & 0.59 [0.49, 0.73] S & 142k & yes \\
\mbox{MUL4-forward400} & forward to 400, refitted & 349 & 46.5 & HBA, 3{,}232 [2{,}711, 3{,}552] & 0.51 [0.42, 0.60] S & 0.62 [0.52, 0.73] S & 150k & yes \\
\mbox{MUL4-top400} & largest 400, refitted & 386 & 63.8 & SA-M, 2{,}570 [2{,}051, 2{,}815] & 0.41 [0.32, 0.48] S & 0.68 [0.54, 0.81] S & 164k & yes \\
\mbox{MUL4-iterated} & iterated prune, refit & 483 & 81.6 & SA-M, 2{,}324 [1{,}944, 2{,}718] & 0.37 [0.31, 0.45] S & 0.79 [0.66, 0.97] S & 190k & yes \\
\mbox{MUL4-P4-sig} & pool P4, pruned & 588 & 100.2 & SA-M, 2{,}325 [1{,}865, 2{,}711] & 0.37 [0.30, 0.46] S & 0.97 [0.78, 1.20] U & 233k & no \\
\bottomrule
\end{tabularx}
\end{table}

\paragraph{Iterated pruning.} Every coefficient below half the step of MUL4-P4-sig's 8-bit word at its own scale
($1.14\times10^{-3}$) is removed, the remaining support is refitted with the exact engine's recipe and audited exactly, and the
step is repeated until no coefficient falls below the threshold. From MUL4-P4-sig the support shrinks $588\to498\to484\to483$
in three refits, and from MUL4-P4-v1 it shrinks $596\to500\to491\to488\to482$ in four. No truncation or
refit breaks a case or falls below the floor 0.02 (minimum case gap 0.043 to 0.052); at 4$\times$4 the small
coefficients carry likelihood and gap margin rather than correctness: removing 90 to 96 of them raises the objective at
$\beta$ 40 from 0.064 to 0.075, the first refit recovers almost all of it, and the fixed points end at 0.064657 and
0.064658. The two lineages meet: their fixed points share 472 terms (98\% of each support), with coefficient cosine
0.99956 on them and no sign change. The fixed point from MUL4-P4-sig enters the comparison as MUL4-iterated: as many updates as
MUL4-P4-sig (1.00 [0.87, 1.17]) at 0.81 [0.71, 0.95] of its work.

\section*{Supplementary Note S6: hardware-facing cost in full}
\phantomsection\label{sn:hw}

The hardware
evidence is one campaign, registered before any certification, build or board run, targeting single-replica color-sweep
hard-threshold SSA \citep{onizawa2023fast,shin2023memory,gonzalez2011parallel} as the compact integer-only FPGA kernel
(avoiding the nonlinear activation tables, multi-bit uniform comparators and replicated state arrays of Gibbs, Metropolis
and parallel-tempering samplers \citep{pervaiz2019weighted,chowdhury2023full,nikhar2024all}) and measuring only the
practical deployment structures of the main text:
the multiplier series in P3 at 4$\times$4, 5$\times$5 and 6$\times$6 (MUL4-P3, MUL5-P3 and
MUL6-P3-seed1) and at 7$\times$7 MUL7-P3-seed1, and the adder ladder's learned structures at 8, 16, 24 and 32 bits (ADD8-learned, ADD16-learned-seed1, ADD24-learned-seed2 and ADD32-learned-seed2), each with its canonical structure. MUL7-P3-seed1, the structure of seed~1 (20261026) and the panel's other-seed arm, was fixed for the campaign before the seed rule chose seed~2 (20261027, MUL7-P3-seed2) for the panel's headline; its audit of every case is the GPU one, and its frozen audit covers an 80-case subset. Supplementary
Table~\ref{tab:si-hw} lists every network and every pair; every value here is read from the campaign's final
analysis record and its final report.

\begin{table}[!tbp]
\centering
\scriptsize
\setlength{\tabcolsep}{2.2pt}
\caption{\textbf{Hardware on the main-text structures.} (a) Every network (learned: MUL4-P3, MUL5-P3,
MUL6-P3-seed1, MUL7-P3-seed1 and the ladder's adders): word (own, certified; common, the pair's; both,
the two equal), terms kept, largest integer code, incident entries, routed lookup tables (LUT) and flip-flops (FF) at 10\,ns, Fmax (MHz) at 4\,ns
(xck26, out of context), colors per sweep $k$ and parallel steps per trial $P$, and on the KR260 at 99.999\,MHz the
median $\TTS$ in cycles, the main run's mean validity $\bar p$, cases never valid and loaded-idle power (W). The
canonical integers at 6$\times$6 and 7$\times$7 are equal at 9 and 13 bits: one network, two board runs. (b) Every
pair: learned over canonical median $\TTS$ in cycles and wall time at the common word and in cycles at the own words,
with paired case-bootstrap 95\% intervals (2{,}000 resamples), and the registered float ratio within SSA (Float).}
\label{tab:si-hw}
\textit{(a) Networks}\\[2pt]
\begin{tabular}{@{}lllrrrrrrrrrrr@{}}
\toprule
Width & Energy & Word & Terms & Code & Incid. & LUT & FF & Fmax & $k$ / $P$ & $\TTS$ & $\bar p$ & Never & Idle \\
\midrule
4$\times$4 & canonical & W7 both & 308 & 56 & 729 & 4{,}235 & 2{,}868 & 195 & 12 / 48 & 5{,}139 & 0.055 & 0 & 3.16 \\
4$\times$4 & learned & W7 both & 409 & 58 & 1{,}137 & 5{,}705 & 2{,}902 & 167 & 12 / 48 & 2{,}094 & 0.105 & 0 & 3.16 \\
5$\times$5 & canonical & W9 own & 565 & 216 & 1{,}372 & 7{,}593 & 4{,}423 & 148 & 15 / 600 & 161{,}097 & 0.021 & 0 & 3.16 \\
5$\times$5 & canonical & W13 common & 565 & 864 & 1{,}372 & 7{,}906 & 4{,}772 & 149 & 15 / 600 & 161{,}097 & 0.021 & 0 & 3.16 \\
5$\times$5 & learned & W13 both & 862 & 4{,}095 & 2{,}453 & 18{,}198 & 5{,}384 & 121 & 15 / 600 & 17{,}596 & 0.160 & 0 & 3.17 \\
6$\times$6 & canonical & W9 own & 906 & 224 & 2{,}245 & 10{,}919 & 5{,}703 & 138 & 18 / 10{,}944 & 5{,}137{,}023 & 0.012 & 0 & 3.17 \\
6$\times$6 & canonical & W13 common & 906 & 224 & 2{,}245 & 10{,}919 & 5{,}703 & 138 & 18 / 10{,}944 & 5{,}490{,}001 & 0.012 & 0 & 3.17 \\
6$\times$6 & learned & W13 both & 1{,}359 & 4{,}095 & 3{,}900 & 26{,}282 & 7{,}000 & 112 & 18 / 576 & 108{,}980 & 0.035 & 0 & 3.18 \\
7$\times$7 & canonical & W9 own & 1{,}312 & 224 & 3{,}312 & 14{,}867 & 7{,}006 & 129 & 20 / 39{,}040 & 11{,}508{,}371 & 0.017 & 0 & 3.19 \\
7$\times$7 & canonical & W13 common & 1{,}312 & 224 & 3{,}312 & 14{,}867 & 7{,}006 & 129 & 20 / 39{,}040 & 11{,}242{,}688 & 0.017 & 0 & 3.19 \\
7$\times$7 & learned & W13 both & 1{,}957 & 4{,}095 & 5{,}545 & 33{,}148 & 8{,}344 & 102 & 20 / 2{,}400 & 637{,}423 & 0.033 & 0 & 3.19 \\
\midrule
8 bits & canonical & W3 both & 81 & 2 & 158 & 3{,}614 & 2{,}982 & 272 & 4 / 16 & 238 & 0.315 & 0 & 3.16 \\
8 bits & learned & W3 both & 81 & 2 & 245 & 3{,}786 & 3{,}147 & 275 & 4 / 64 & 139 & 0.817 & 0 & 3.16 \\
16 bits & canonical & W3 both & 161 & 2 & 318 & 7{,}023 & 5{,}929 & 268 & 4 / 64 & 468 & 0.471 & 1 & 3.16 \\
16 bits & learned & W3 both & 153 & 3 & 469 & 6{,}902 & 6{,}254 & 257 & 4 / 160 & 208 & 0.880 & 0 & 3.16 \\
24 bits & canonical & W3 both & 241 & 2 & 478 & 10{,}030 & 8{,}841 & 266 & 4 / 48 & 440 & 0.380 & 9 & 3.17 \\
24 bits & learned & W3 both & 224 & 3 & 689 & 10{,}641 & 9{,}326 & 258 & 4 / 112 & 259 & 0.738 & 0 & 3.17 \\
32 bits & canonical & W3 both & 321 & 2 & 638 & 13{,}399 & 11{,}887 & 263 & 4 / 80 & 591 & 0.412 & 17 & 3.18 \\
32 bits & learned & W3 both & 297 & 3 & 917 & 14{,}287 & 12{,}532 & 253 & 4 / 208 & 273 & 0.792 & 1 & 3.18 \\
\bottomrule
\end{tabular}
\\[6pt]
\textit{(b) Pairs, learned / canonical}\\[2pt]
\begin{tabular}{@{}llrlllll@{}}
\toprule
Width & Mode & Cases & Common & Cycles, common word & Wall time, common word & Cycles, own words & Float \\
\midrule
4$\times$4 & factor & 90 & W7 & 0.408 [0.317, 0.524] S & 0.408 [0.317, 0.524] S & 0.408 [0.317, 0.524] S & 0.49 \\
5$\times$5 & factor & 340 & W13 & 0.109 [0.102, 0.121] S & 0.109 [0.102, 0.121] S & 0.109 [0.101, 0.121] S & 0.11 \\
6$\times$6 & factor & 64 & W13 & 0.0199 [0.0171, 0.0268] S & 0.0198 [0.0171, 0.0268] S & 0.0212 [0.0186, 0.0263] S & 0.025 \\
7$\times$7 & factor & 64 & W13 & 0.0567 [0.0449, 0.068] S & 0.0567 [0.0449, 0.068] S & 0.0554 [0.0443, 0.0669] S & 0.0625 \\
\midrule
8 bits & reverse & 129 & W3 & 0.583 [0.523, 0.633] S & 0.586 [0.526, 0.636] S & 0.583 [0.523, 0.633] S & 0.68 \\
16 bits & reverse & 137 & W3 & 0.444 [0.406, 0.484] S & 0.445 [0.406, 0.485] S & 0.444 [0.406, 0.484] S & 0.41 \\
24 bits & reverse & 145 & W3 & 0.588 [0.527, 0.633] S & 0.589 [0.528, 0.633] S & 0.588 [0.527, 0.633] S & 0.39 \\
32 bits & reverse & 153 & W3 & 0.462 [0.400, 0.521] S & 0.463 [0.401, 0.521] S & 0.462 [0.400, 0.521] S & 0.33 \\
\bottomrule
\end{tabular}
\end{table}

\paragraph{Coefficient words.} Each coefficient $\theta$ is rounded to a $W$-bit two's-complement integer
$n=\mathrm{round}(\theta/s)$, ties away from zero, clipped to $\pm(2^{W-1}-1)$, with one scale $s$ per structure;
terms that round to zero are dropped. The scale is the designer's: among the candidate scales, the one with the largest
minimum case gap over every case of every mode divided by the largest incident L1 norm of the rounded coefficients. The
candidates are the full-range scale (the largest $|\theta|$ over $2^{W-1}-1$), the lattice $2^{-(W-1)}$ and, where one
exists, a common rational scale of the coefficients, and on the adders and at 4$\times$4 also a grid of 32 scales per
octave from the full-range scale to the largest $|\theta|$. A
structure is correct at a word when every case of every mode keeps every legal state below every illegal one (minimum case
gap above $10^{-9}$), and its certified word is the smallest word of the ladder $W\in\{3,4,5,7,8,9,13,17\}$ at which it
is correct with every larger ladder word correct too. Correctness is decided exactly: on the adders by a min-plus
recursion along the column chain in every mode, at 4$\times$4 over the full energy cube, at 5$\times$5 by the exact
chain recursion, and at 6$\times$6 and 7$\times$7 by the GPU exact audit, cross-checked on a fixed 80-case subset by the frozen CPU
recursion to $10^{-9}$ relative. No refit at the word is used. Every
structure has a certified word: both energies at 3 bits at every adder width, both at 7 bits at 4$\times$4, and the
learned structures at 13 bits against the canonical encoding's 9 from 5$\times$5 to 7$\times$7, as expected before any
run (the learned word at least the canonical one on the multiplier, both adders at 3 bits at 8 and 16 bits and at most 5
at 24 and 32). At 16 bits both adders are correct at 3 bits, with minimum case gaps of 0.127 for the canonical encoding and
0.508 for the learned structure. Rounding drops small terms of the learned structures: at their
words 409 of 571 terms remain at 4$\times$4, 862 of 1{,}072 at 5$\times$5, 1{,}359 of 1{,}692 at 6$\times$6 and 1{,}957 of 2{,}309 at 7$\times$7, and
81 of 169, 153 of 287, 224 of 462 and 297 of 557 on the adders from 8 to 32 bits; the canonical structures keep every
term. At 7$\times$7 the learned structure's minimum case gap at 13 bits is about 8\% of its floating-point floor; the
certification asks for correctness, not the floor. Each pair runs at its common word, the larger of its two certified words, and each structure also at its own word
where the two differ.

\paragraph{Synthesis.} Every network was synthesized alone as a term-table network on the KR260's FPGA (xck26; Vivado,
out of context, routed): one hard-threshold SSA node per variable with an accumulator of $1+\lceil\log_2(L_{1,i}/s+1)
\rceil$ bits for node $i$ of incident L1 norm $L_{1,i}$, the integers fixed in the logic, one color class updated per
clock cycle, at a 10\,ns pass for area and a 4\,ns pass for the highest clock; every network passed the check of its
arithmetic against its integer reference in register-transfer-level (RTL) simulation. The learned networks have more incident entries than the
canonical ones at every width and need more lookup tables at every width but 16 bits, where the routed learned network
is the smaller (6{,}902 against 7{,}023 lookup tables) and the synthesized one is not (7{,}257 against 7{,}255); the
expectation that the learned network needs more lookup tables at every width is therefore met before routing and not
after it. At 6$\times$6 the learned network takes 26{,}282 against 10{,}919 routed lookup tables at 112 against
138\,MHz, and at 7$\times$7 33{,}148 against 14{,}867 at 102 against 129\,MHz; on the adder the learned networks reach 253 to 275\,MHz and the canonical ones 263 to 272\,MHz. The two
networks of every width need the same number of colors per sweep (4 on the adder; 12, 15, 18 and 20 on the multiplier from
4$\times$4 to 7$\times$7), giving both the same number of cycles per sweep. A parallel step is a count, not a latency:
a step's latency follows the largest fan-in it updates. How many nodes may update at once, and at what input precision,
is mapped for parallel $p$-bit Ising machines by \citet{onizawa2026unified}; the color sweep used here is the
conflict-free end of that range \citep{gonzalez2011parallel,nikhar2024all}, and the precision studied here is that of the coefficients.

\paragraph{Board.} Each network ran on the KR260 board in a sampling core that runs the panel's cases and trials without
the processor, every build at 99.999\,MHz: one core for the multipliers to 6$\times$6 and the 8-bit adder, and a wider
core, equal to the first bit for bit on structures of up to 64 nodes, for 7$\times$7 and the 16- to 32-bit adders. The
core therefore changes between the 8- and the 16-bit adder, and every ratio is formed within one width. Each structure ran the color-sweep SSA at its registered panel's
color-sweep setting, or at the SSA family's tuned setting where the panel has no color-sweep arm (4$\times$4 and
5$\times$5), unchanged at every word and without re-tuning, on its panel's evaluation cases (90, 340, 64 and 64 factor cases
on the multiplier; 129, 137, 145 and 153 sums on the adder). A replay run of 256 trials per case checked every final
state and case hash bit for bit against the core's reference, and a main run of 4{,}096 trials per case gave each
case's validity $p_c$ and $\TTS$ in cycles, $(P+3)\max\{1,\ln 0.01/\ln(1-p_c)\}$ for $P$ parallel steps per trial, so that
a case with a validity of at least 0.99 counts one trial. At 7$\times$7 the setting is the panel's color-sweep
arm for the hardware structure (MUL7-P3-seed1), run at MUL7-P3-seed2's SSA tuning selection: $T$ 5{,}000 for the learned
network and 80{,}000 for the canonical one, both best cells at two successes in 128 trajectories on the median product,
leaving the lengths chosen at the panel's counting resolution; nothing was re-tuned. Every one of the 38 board runs,
replay and main, reproduces its reference bit for bit. Learned over
canonical, the median $\TTS$ in cycles at the common word, with paired case-bootstrap 95\% intervals (2{,}000
resamples), is 0.408 [0.317, 0.524] at 4$\times$4, 0.109 [0.102, 0.121] at 5$\times$5, 0.0199 [0.0171, 0.0268] at
6$\times$6 and 0.0567 [0.0449, 0.0680] at 7$\times$7 in factor mode, and 0.583 [0.523, 0.633], 0.444 [0.406, 0.484], 0.588 [0.527, 0.633] and 0.462 [0.400,
0.521] at 8, 16, 24 and 32 bits in the reverse mode, all S, the headline expectation met: every interval below 1, in the
direction of the registered floating-point ratio within the SSA family. The board ratios compare the two
energies on one SSA core. In software, single-replica random-scan SSA trails color-sweep tempering on either energy at 6$\times$6 and 7$\times$7: 4.03 and 4.10 times on the learned energy (MUL6-P3-seed1 and MUL7-P3-seed2) and 22.0 and 20.5 times on the canonical encoding. Each ratio takes the energy's SSA median at its best evaluated length, re-measured (descriptive) at 1{,}024 trajectories per product (322{,}147 and 1{,}342{,}731 updates on the learned energy, 3{,}356{,}827 and 14{,}550{,}096 on the canonical), over its tempering median in Table~\ref{tab:series}; the SSA rows of Table~\ref{tab:si-family-medians} are at the tuning-selected lengths. The learned 6$\times$6 tempering median sits at its length floor, 79{,}926 updates at $T$ 80{,}000, so its fourfold is a lower bound. At the own words the points differ only at 6$\times$6 and
7$\times$7, 0.0212 [0.0186, 0.0263] and 0.0554 [0.0443, 0.0669]; at 5$\times$5 the canonical medians at 9 and 13 bits
are equal (161{,}097 cycles) and only the lower bound moves, to 0.101. The floating-point
ratios differ in magnitude from the board's at the adders and at 4$\times$4 (at 32 bits 0.462 against 0.33): they
count parallel steps or updates, the board counts cycles with a per-trial overhead, and the difference is not
attributed. At 7$\times$7 the floating-point ratio is the panel's color-sweep ratio in updates, 0.0625, which with
both cells at the counting resolution equals the ratio of the two lengths. Some sums are never solved in 4{,}096 trials (1, 9 and 17 of the canonical encoding's at 16, 24 and 32
bits, 1 of the learned structure's at 32 bits); their $\TTS$ is infinite, and every median is finite. Wall time
follows the cycles within 0.5\% in every pair. Loaded with any network, the idle board draws 3.16 to 3.19\,W, the two
networks of a pair within one step of the sensor. We infer that the energy per solution follows the cycle ratio; the
board-level sensor cannot separate the core's dynamic power from the platform's static draw, so this is not measured.

\paragraph{Reading Fig.~\ref{fig:hw}.} The gray diamonds are the registered floating-point ratio within the SSA family that the
campaign's expectation names: SSA in updates at 4$\times$4 and 5$\times$5, the color sweep in parallel steps at
6$\times$6 and on the adder, and in updates at 7$\times$7; their units differ from the board's clock cycles, which
include a per-trial overhead. Panel (c) counts routed lookup tables at the 10\,ns pass (xck26, out of context).

\paragraph{Board against the floating-point model.} The paired floating-point model of each network runs the replay
cases on the same random streams. A case's board validity is compared with the model's by a two-sample test, Newcombe's
hybrid score interval \citep{newcombe1998interval} for the difference of the two validities at 95\%, a case departing when the interval excludes 0.
Four of the 2{,}712 replay comparisons depart, all in the canonical 7$\times$7 network's replay runs and in both
directions (one case at 9 bits, three at 13), where a 5\% test would flag about 136 by chance for independent draws;
the learned 7$\times$7 network has none. Over the 16 rows before 7$\times$7,
1{,}852 of the 2{,}520 comparisons are equal and the largest
difference is 15 of 256 trials (the canonical 16-bit adder), with mean signed differences within 1.3 counts for every
network: the shared streams keep the two sides close and the integer rounding moves some trajectories without a
direction.

\paragraph{Earlier measurements.} Before this campaign the two 25-node circuits were measured on the same board with an
earlier core, at each energy's own word and at an 8-bit word. The main text cites one of those runs: in the multiply
mode of the 4$\times$4 multiplier, under each energy's own color-sweep SSA configuration, MUL4-P4-sig needed 0.86 [0.68,
0.99] of the canonical encoding's cycles at 8 bits and 0.73 [0.55, 0.88] at the own words. The
closest silicon reference, the 65\,nm 5-bit by 5-bit invertible multiplier of \citet{smithson2019efficient}, built from
designed gate configurations with noise weights chosen by experimental sweeps, names automatic noise adjustment as its
next step; here the noise is chosen by the tuned protocol and the energy itself is learned.

\section*{Supplementary Note S7: further discussion}
\phantomsection\label{sn:discussion}

\paragraph{Answer coverage at finite length.} In the underdetermined modes a solver must also reach the different
legal answers, and faster solving and wide answer coverage are different properties; this paper measures the first.
Large learned supports concentrate the distribution over multiple valid answers at finite run length. Write $\bar D_\chi$ for the mean over sums of the collision distance from uniform, $n\sum_k p_k^2-1$, of a sum's answer distribution $p$ over its $n$ legal answers, estimated without bias from the sampled counts $c_k$ of $N$ answers as $n\sum_k c_k(c_k-1)/[N(N-1)]-1$. In the adder's reverse mode the effective answer fraction $1/(1+\bar D_\chi)$ is 0.64 for ADD6-large against 0.99 for the
canonical encoding at $T$ 1{,}000, falling to 0.45 at $T$ 5{,}000; over factor pairs MUL4-P4-full reaches 0.82 against 1.00. Every
adder structure here is fitted with equal legal energies, giving an exactly uniform Gibbs distribution over the legal completions of a sum
at every $\beta$ (computed exactly for ADD6-compact: 1 to within $1.4\times10^{-13}$ for all 127 sums);
the observed concentration is a property of the finite-length SSA output. The
compact fit keeps 0.93 under random-scan updates (non-inferiority confirmed), while under a color sweep the compact
fits concentrate as well. MUL4-P4-full's legal energies within a product differ by at most 0.04, and the exact Gibbs distribution
over factor pairs at $\beta=40$ is nearly uniform (effective answer fraction 0.998), against 0.82 at $T$ 20{,}000.

\paragraph{Repeated patterns.} A term's pattern is its set of nodes written as input, product and carry positions
relative to the term's lowest input indices, so one pattern can occur at several places in the array. In the first
sampled engine's multiplier points (MUL5-P3-v1 and MUL6-P3-v1) the patterns that recur at two or more places hold 89\% of MUL5-P3-v1's terms and 86\% of its coefficient magnitude, and 94\% and 93\% of
MUL6-P3-v1's. The two widths share 220 recurring patterns, a Jaccard overlap of 0.58. The connection pattern
carries over between widths; the coefficients do not: a 6$\times$6 start built by tiling the 5$\times$5 patterns at their
mean coefficients breaks correctness (minimum case gap $-2.7$) and was not used. The analysis is descriptive and enters
no result.

\paragraph{Higher-order terms or pairwise structures with auxiliary nodes.} A pairwise machine reaches the same relations
only through auxiliary nodes whose couplings keep the legal set's gap once they are minimized over
\citep{biamonte2008nonperturbative,bian2014discrete}. Three properties of the method rest on the higher-order form: a
term of interaction order $k$ penalizes an illegal pattern of $k$ nodes directly; no node is hidden, making the fitted distribution
the Gibbs distribution itself rather than a marginal; and on a fixed pool the objective is convex, as for any exponential family
with every variable observed \citep{wainwright2008graphical}. With every node visible and every term column-local, the
column-chain dynamic program audits each adder fit exactly. The costs of the form are
measured here: a term of interaction order $k$ makes its $k$ nodes mutually conflicting, setting the coloring by the cliques the
terms induce (18 colors at 6$\times$6 for the canonical encoding and for MUL6-P3-v1, the clique bound), and the learned points on their floor need
wider words.

\paragraph{Prior circuit scales and comparisons.} Additional numerical detail accompanies the related-work overview in the
Introduction. Hardware Boltzmann training has been demonstrated on a full adder \citep{kaiser2022hardware,jhonsa2026cmos},
and the generalized probabilistic approximate optimization algorithm checks its sampled optimization against the exact
Markov flow on a four-node majority gate and a five-node full adder \citep{abdelrahman2025generalized}. In the modular
restricted Boltzmann machine work of \citet{patel2022logically}, first hits on one product appear about four times earlier
than the most frequently sampled state settles, motivating sample-verification checks, while their reported speed factors
of about $10^4$ over a CPU and $10^3$ over a GPU compare hardware platforms executing one algorithm rather than two
energies under equal sampler tuning. Among excited-spectrum constructions, \citet{he2023many} merge linear-programming
four-body gates to cut a 2$\times$2 factorizer from 32 energy levels to 9 and raise the sampled share of correct factors
on one product of a 108-$p$-bit 6$\times$6 factorizer from about 4\% to above 80\%, falling to about 28\% at
10$\times$10; \citet{he2024direct} work their direct two-level truth-table expansion through an eight-node 2$\times$2
factorizer, and on 4$\times$4 to 7$\times$7 multipliers its gap scaled to the incident-weight budget is 0.16, 0.080,
0.040 and 0.020 against floors of 0.02, 0.005, 0.0048 and 0.0047. Earlier gate-level invertible encodings ran in hardware
emulation on a 48-$p$-bit 4-bit adder and a 46-$p$-bit 2$\times$2 multiplier coupled unidirectionally
\citep{pervaiz2017hardware}, in software and FPGA simulation across three organizations of an 8$\times$8 multiplier of
192 to 208 nodes \citep{nishino2018study}, and in simulation on a 448-$p$-bit 32-bit adder with unidirectional carry
couplings \citep{camsari2017stochastic}, while IsingFormer tested learned proposals on a 10$\times$10 multiplier and
products up to 24 bits on hand-written energies \citep{bunaiyan2025isingformer}. In silicon, beside the 65\,nm
5$\times$5 factorizer \citep{smithson2019efficient}, a 130\,nm chip of 1{,}143 $p$-bits with voltage-controlled magnetic
tunnel junctions factored 35 on a 27-$p$-bit pairwise circuit, with board-link bandwidth limiting measurement and 20-bit
products simulated on a modified 275-$p$-bit design \citep{duffee2025integrated}. Outside gate-level encodings,
factor-bit cost functions factored 945 on eight $p$-bits \citep{borders2019integer} and 16- to 30-bit semiprimes in
simulation \citep{xu2025probabilistic}, and variable-only three-body 3-SAT energies reached about 78\% average success on
ten 100-variable instances in 5{,}000 annealing steps where a 631-node pairwise gate-level circuit solved none
\citep{he2026solving}.

\section*{Supplementary Note S8: protocol and experiment records}
\phantomsection\label{sn:records}

Every run of this paper except the multiply-mode check below and the descriptive term and incident counts of the direct design in the related work was fixed in a dated note before it ran; recorded
entries are never edited, and a correction is an appended note. The 25-node studies Confirmation~A, Confirmation~B and Confirmation~C were pre-registered with their
hypotheses, testing sequences and margins. The equal-tuning comparison and its extensions, the equal-workload comparison,
the adder ladder from 6 to 32 bits, the first sampled engine's multiplier points, the hardware measurements (the first synthesis study, the first board batch,
and the hardware campaign on the main-text structures) and the multiplier series each have a pre-run note, tuning and evaluation freezes and a completion note in the
project's experiment record. The 7$\times$7 fit's completion note follows the frozen audit of every case of the chosen seed. The three added families ran as supplements under one uniform
amendment, each width's tuning and evaluation frozen before its run, and each circuit family's nine-family analysis ran once and first reproduced the registered
analysis, except that the 6$\times$6 analysis was rerun once to correct its previous-width argument, which changed only the
cross-width fields and opened no new result. The descriptive re-measurement of low-count cells and the fresh read of Appendix~\ref{app:protocol} have their own pre-run note and analysis record. The maximal-gap comparison of Supplementary Note~S9 has a fit recipe, notes on its solver, a pre-run note, tuning and evaluation freezes, an interim note and a close note. One
descriptive check has no entry in that record: the multiply-mode check at 4$\times$4 (Supplementary Note~S3), which enters
no comparison or expectation; its run on the learned structures had its inputs fixed by checksum before it started, and
the canonical encoding's values come from an earlier run of the same code. Every number of this paper is read
from these records and the check's own.

\section*{Supplementary Note S9: likelihood against maximal gap on one pool}
\phantomsection\label{sn:maxgap}

\paragraph{The comparison.} With pool P3, the node budget, each width's floor and the rules for the term budget and the
fallback held fixed, the likelihood objective is replaced by a maximal-gap linear program, and the two run on identical
products, pilots, seed namespaces and streams, with all nine families tuned on the same grids (Supplementary
Table~\ref{tab:si-maxgap}; Fig.~\ref{fig:width}a, b of the main text draws the arm). The program also holds all legal energies equal, as the gap-maximizing construction of
\citet{onizawa2021high} does and the likelihood fits on the multiplier do not. The term budget is a rule and not a count:
MUL5-P3 keeps 848 terms and the fallback takes it to 1{,}072, while MUL5-P3-maxgap's natural count, 821, lies below
$K=848$, so that it needs neither prune nor fallback and delivers 820; at 4$\times$4 the counts are 571 and 457.

\begin{table}[!tbp]
\centering
\scriptsize
\caption{\textbf{Likelihood against maximal gap on one pool.} Multiplier, factor mode, pool P3, on the same products and
streams; readings S, U, R as in Table~\ref{tab:series}; the best of six gives the same value in every best-of row. Same
family: each structure at its own tuning-selected configuration and length. 5$\times$5's reading is the confirmatory one,
evaluated at the stricter first step (97.5\% interval) of a pre-registered two-hypothesis Holm correction. 4$\times$4 is
outside that family: its direction had been seen before the comparison was fixed, so neither its row nor the ratio across
the two widths is prospective, and two widths alone do not establish a trend. Within the families at
4$\times$4, the maximal-gap PT+ICM-4 selection sits at the top of its $\lambda$ grid, which is not widened at that width.}
\label{tab:si-maxgap}
\setlength{\tabcolsep}{4pt}
\begin{tabular}{@{}lll@{}}
\toprule
 & 4$\times$4 & 5$\times$5 \\
\midrule
\multicolumn{3}{@{}l}{\textit{Structures}} \\
Likelihood fit & \mbox{MUL4-P3} & \mbox{MUL5-P3} \\
Maximal-gap fit & \mbox{MUL4-P3-maxgap} & \mbox{MUL5-P3-maxgap} \\
Terms (significant), likelihood; maximal gap & 571 (452); 457 (424) & 1{,}072 (847); 820 (776) \\
Minimum case gap, likelihood; maximal gap (floor) & 0.0200053; 0.0672238 (0.02) & 0.0050008; 0.0337553 (0.005) \\
Exact objective at $\beta$ 40, likelihood; maximal gap & 0.267; 0.881 & 1.584; 5.160 \\
\midrule
\multicolumn{3}{@{}l}{\textit{Likelihood over maximal gap, each at its best of nine}} \\
Updates & 0.68 [0.56, 0.79] S & 0.55 [0.46, 0.58] S \\
Updates, 97.5\% interval (first step of the correction) & --- & [0.45, 0.59] \\
Nonzero work & 0.82 [0.67, 0.96] S & 0.67 [0.56, 0.71] S \\
Significant work & 0.72 [0.59, 0.84] S & 0.58 [0.48, 0.61] S \\
Same family, updates: range; S/U/R & 0.58 to 0.70; 9/0/0 & 0.28 to 0.62; 9/0/0 \\
Same family, nonzero work & 0.70 to 0.84; 6/3/0 & 0.33 to 0.75; 9/0/0 \\
Same family, significant work & 0.62 to 0.74; 9/0/0 & 0.29 to 0.65; 9/0/0 \\
Single-run success at a nominal length, difference (pp) & $T$ 1{,}000: $+22.0$ [18.4, 25.6] S & $T$ 5{,}000: $+30.4$ [28.6, 32.1] S \\
The same, within each family & 8 S, 1 R (PT+ICM-4, $-12.9$) & 9 S \\
5$\times$5 over 4$\times$4, updates & \multicolumn{2}{l}{0.81 [0.64, 0.97] S: one ratio, not a trend reading} \\
\midrule
\multicolumn{3}{@{}l}{\textit{Fresh read, descriptive: 1{,}024 trials per product, the best cells fixed}} \\
Updates, products resampled & 0.75 [0.62, 0.88] S & 0.56 [0.49, 0.61] S \\
Updates, products fixed & [0.69, 0.79] S & [0.53, 0.57] S \\
Nonzero work, products resampled & 0.90 [0.75, 1.06] U & 0.68 [0.59, 0.73] S \\
Nonzero work, products fixed & [0.84, 0.95] S & [0.64, 0.69] S \\
Significant work, products resampled & 0.80 [0.66, 0.94] S & 0.59 [0.51, 0.64] S \\
Significant work, products fixed & [0.74, 0.84] S & [0.55, 0.60] S \\
Fresh over registered median, likelihood; maximal gap & 1.01; 0.91 & 1.04; 1.02 \\
\midrule
\multicolumn{3}{@{}l}{\textit{Maximal gap over the canonical encoding's best of nine}} \\
Updates & 0.81 [0.66, 1.000] U & 0.79 [0.73, 0.88] S \\
Nonzero work & 1.35 [1.09, 1.67] R & 1.32 [1.22, 1.47] R \\
Significant work & 1.25 [1.01, 1.54] R & 1.26 [1.16, 1.41] R \\
\midrule
\multicolumn{3}{@{}l}{\textit{Best cells: family, restart length, median $\TTS$ in updates}} \\
Likelihood & SA-M, $T$ 250: 3{,}486 & HBA, $T$ 500: 19{,}879 \\
Maximal gap & SA-M, $T$ 100: 5{,}125 & APT-C, $T$ 20{,}000: 36{,}037 \\
Canonical encoding & SA-M, $T$ 100: 6{,}317 & APT-C, $T$ 20{,}000: 45{,}508 \\
\bottomrule
\end{tabular}
\end{table}

\paragraph{The maximal-gap recipe.} On the pool, the program maximizes $\gamma-10^{-4}\sum_S|\theta_S|$ subject to one
energy for all legal states, every illegal state at least $\gamma$ above it and Eq.~\ref{eq:budget}, so that $\gamma$ is
the minimum case gap. Cuts come from exact separation, the series' column-chain search for the lowest illegal state of
every visible slice, with at most 20{,}000 new violators per round, starting from the legal states' single flips and
20{,}000 random states. A round with no new violator certifies a solve over the whole state cube; a support takes at most
31 solves. After a solve the coefficients are projected onto the equal-energy subspace, those at or below $10^{-9}$ in
magnitude are dropped and the result is normalized to $\max_i\ell_i=1$. If more than $K$ terms remain, the $K$ largest are
kept and the program is solved again on them, and the canonical terms would be added if the kept support's certified gap
lay at or below the floor. MUL5-P3-maxgap comes from Gurobi 13.0.3's barrier without crossover, at tolerances of $10^{-9}$,
only an optimal status being taken: 12 solves on the pool and 3 on the kept terms. MUL4-P3-maxgap is an earlier structure,
stored and not refitted, from the same solver's barrier without crossover, needing no prune and no fallback. The program's
optimum is unique in its objective and not necessarily in its coefficients, so a kept support can depend on the point the
solver returns; the solver was fixed before the fit ended.

\paragraph{Audits.} The certificate does not rest on the solver, and the coefficients can be checked by the exact audit
without it. The frozen exact audit of Supplementary Note~S1 covers every case: at 5$\times$5, 340 factor and 1{,}024
multiply cases, none at or below the floor, its minimum case gap within $5\times10^{-14}$ relative of the separation's; at
4$\times$4, 90 and 256 cases with the same outcome, within $10^{-6}$ of the stored gap. A fit's gates ask for a certificate on
every solve and for a projection change and a legal spread of at most $10^{-8}$. On MUL5-P3-maxgap the stage on the whole
pool has a legal spread of $2.99\times10^{-8}$, above that limit, and the delivered stage on the kept terms has
$1.29\times10^{-9}$; reading the gate on the delivered stage was chosen after the value was seen, and the limit was not
relaxed. That spread comes from dropping 838 small terms after the projection, and the kept stage solves its own program
with the equality rows.

\paragraph{Readings.} The primary reading is the best-of-nine median $\TTS$ in updates of the likelihood structure over
that of the maximal-gap structure, with the products resampled jointly for all arms and the best family and length
re-selected in every replicate, the maximal-gap cells drawing their counts from a stream of their own. Evaluated at the first step of the pre-registered two-hypothesis Holm correction (95\% family-wise), the 5$\times$5
comparison has a 97.5\% upper bound of 0.59, below 1, surviving at either step of the correction. The single-run difference is a registered secondary reading outside the correction: the paired difference in mean
single-run success, each arm at its best of nine at a nominal length fixed before any read, near a quarter of the
likelihood structure's best median. APT-C is the best family at that length on both arms at both widths, and the difference
in updates between the two cells accounts for 2.3 pp at 4$\times$4 and none at 5$\times$5. The fresh read is descriptive:
each arm's evaluation-selected best cell, read once on a new namespace at 1{,}024 trials per product with its family and
length held fixed, with one interval resampling the products and one holding them fixed, the second a statement about every
product of that width and no others.

\end{document}